\documentclass[twocolumn]{aastex631}

\usepackage{booktabs}
\usepackage{amsmath}

\newcommand{\pharo}{PHARO}
\newcommand{\palmartwo}{Hale 5.1 m telescope}

\begin{document}

\title{\Large Stellar companions sculpt hot Jupiter formation and spin-orbit evolution}

\shorttitle{Stellar Companions Sculpt Hot Jupiters}
\shortauthors{Shariat et al.}

\author[0000-0003-1247-9349]{Cheyanne Shariat}
\affiliation{Department of Astronomy, California Institute of Technology, 1200 East California Boulevard, Pasadena, CA 91125, USA}
\email{cshariat@caltech.edu}

\author[0000-0002-6871-1752]{Kareem El-Badry}
\affiliation{Department of Astronomy, California Institute of Technology, 1200 East California Boulevard, Pasadena, CA 91125, USA}

\author[0000-0002-7846-6981]{Songhu Wang} 
\affiliation{Department of Astronomy, Indiana University, 727 East 3rd Street, Bloomington, IN 47405-7105, USA}

\author[0000-0002-0376-6365]{Xian-Yu Wang} 
\altaffiliation{Sullivan Prize Postdoctoral Fellow}
\affiliation{Department of Astronomy, Indiana University, 727 East 3rd Street, Bloomington, IN 47405-7105, USA}

\author[0000-0002-7670-670X]{Malena Rice}
\affiliation{Department of Astronomy, Yale University, 219 Prospect Street, New Haven, CT 06511, USA}

\author[0000-0002-6618-1137]{Jerry W. Xuan}
\altaffiliation{51 Pegasi b Fellow}
\affiliation{Department of Earth, Planetary, and Space Sciences, University of California, Los Angeles, Los Angeles, CA 90095, USA}
\affiliation{Department of Physics, University of California, Santa Barbara, Santa Barbara, CA 93106, USA}

\author[0000-0002-5741-3047]{David R. Ciardi}
\affiliation{NASA Exoplanet Science Institute, Caltech/IPAC, Pasadena, CA 91125, USA}

\correspondingauthor{Cheyanne Shariat}

\begin{abstract}
Stellar companions can drive hot-Jupiter (HJ) migration and spin-orbit misalignment, but their role in HJ formation remains uncertain. We construct a homogeneous census of resolved stellar companions to $147$ northern HJs with measured projected obliquities. We obtain uniform adaptive-optics imaging and combine these observations with {\it Gaia} common proper-motion pairs to identify 8 new companion candidates, bringing the \textit{observed} companion fraction to $71/147=48\%$.  Modeling the full survey selection function yields an \textit{intrinsic} companion fraction of $62\pm5\%$ for mass ratios $q_\star=0.1$--$1$ and projected separations $s=50$--$50{,}000$~au, roughly 3--4$\times$ enhanced relative to field stars. Including white-dwarf companions would increase this fraction further. HJs with resolved companions at $50$--$2{,}000$~au are nearly twice as likely to be misaligned compared to systems without detected companions: $46\%$ compared to $24\%$ ($p=0.009$). The misaligned fraction rises steadily from $5\%$ among the coolest hosts to $80\%$ among the hottest, without a sharp transition at the Kraft Break, while the intrinsic companion fraction remains roughly constant across the temperature range. These trends are consistent with HJs beginning with a broad obliquity distribution, followed by progressively weaker tidal realignment at higher stellar temperatures. Contrary to previous work, we find that most stellar companions in our sample are capable of driving eccentric Kozai--Lidov (EKL) oscillations to the tidal limit under suitable orbital configurations, making high-eccentricity migration dynamically promising. Taken together, these results indicate that stellar companions sculpt HJ formation and spin--orbit architectures.
\end{abstract}


\section{Introduction}
\label{sec:introduction}

The first exoplanet discovered around a main-sequence star was a hot Jupiter (HJ).  The 51 Pegasi system packs a Jupiter-mass planet into a $4.2$-day orbit, only $\sim10~{\rm R_\odot}$ from its host star \citep{Mayor1995}.  Its compact orbit posed a formation problem because giant planets are expected to assemble farther from their host stars \citep{Lin1996,DawsonJohnson2018}, as observed in our Solar System.  Disk-driven inward migration was proposed within a year of the discovery \citep{Lin1996}.  Three decades later, the processes that deliver HJs to their present-day orbits are still debated \citep{DawsonJohnson2018}.

Formation models for HJs group into three broad classes:  in-situ formation, disk migration, and high-eccentricity migration. 
In-situ formation assembles a giant planet close to the star from solids that drift inward through the disk \citep[e.g.,][]{Batygin2016}.
In disk migration, a giant planet forms several au from its host star and migrates inward through torques from the gas disk \citep[e.g.,][]{Lin1996}. In high-eccentricity migration, gravitational interactions with another planet or star drive a giant planet to high eccentricity in an initially wide orbit; tides then shrink and circularize its orbit \citep[e.g.,][]{RasioFord1996,WuMurray2003,FabryckyTremaine2007,Naoz2011HotJupiters}.  The observed HJ population may contain contributions from all three routes \citep{DawsonJohnson2018}.  Their predictions differ in planetary eccentricity, spin--orbit angle, presence of nearby planets, and presence of distant stars.  Testing these predictions requires measuring the properties for a large sample of Hot Jupiters and their environments using different techniques.

The occurrence of stellar companions offers one observational test of these formation routes. 
A distant star can tilt the protoplanetary disk or drive a giant planet onto a highly eccentric orbit \citep[e.g.,][]{Batygin2012,Lai2018,WuMurray2003,FabryckyTremaine2007,NaozFarrRasio2012,Naoz2016}. If these interactions influence the formation of HJs, their companion frequency or architecture should differ from those of comparable field stars.
A prediction of dynamical migration is that HJs in these systems should also exhibit a broader distribution of spin--orbit angles \citep[e.g.,][]{FabryckyTremaine2007, WeldonNaozHansen2025}.  Radial velocity and high-resolution imaging surveys have tested these predictions by searching for companions around HJ hosts and comparing aligned and misaligned systems \citep[e.g.,][]{Wollert2015,Evans2016,Ngo2015,Ngo2016,Bohn2020}.

The Friends of Hot Jupiters (FOHJ) survey made one of the first population-wide measurements of stellar companions and their association with HJ obliquities \citep{Knutson2014,Ngo2015,Piskorz2015,Ngo2016}.  
From $35$ systems with Rossiter--McLaughlin measurements, 
\citet{Ngo2016} find only a $1.4\sigma$ excess of companions around misaligned systems.
The expanded $77$-system survey found that $\sim50\%$ of HJs had stellar companions at $50$--$2000$ au, $\sim3\times$ larger than the occurrence rate of such companions for normal field stars, while companions inside $50$ au were comparatively rare \citep{Ngo2016}.  
Their simplified dynamical analysis compared Kozai--Lidov and
general-relativistic precession timescales for planets initially at 1--5~au and excluded systems with additional planetary companions, estimating that stellar companions could have driven migration for only a minority of HJs.
\citet{Ngo2016} therefore argued that while wide binaries are strongly correlated with HJ formation, only a small fraction of these stellar companions are capable of driving high-e migration.

Later studies continued to find high stellar multiplicity among HJ hosts, but revised how that multiplicity should be interpreted for HJ formation. VLT/SPHERE imaging found companions around $\sim55\%$ of HJ hosts, with similarly high rates for systems containing eccentric and circular planets \citep{Bohn2020}.  
A recent {\it Gaia}-based search likewise found a $6\times$ enhancement of wide companions around M dwarfs hosting short-period giant planets \citep{Gan2026}, suggesting that this trend extends to lower-mass stars.
Population modeling similarly showed that close-binary suppression and other selection effects could account for much of the wide-companion excess reported by FOHJ \citep{MoeKratter2021}.  These studies left two central questions: whether the intrinsic excess of stellar companions around HJ hosts is suggestive of their formation mechanism, and whether those companions are associated with stellar obliquity excitation. Prior to this work, the association between stellar companions and HJ obliquities had been tested only with the original 35-system FOHJ sample.

These open questions can now be revisited with a much larger obliquity sample and greater sensitivity to stellar companions.  The most recent obliquity catalogs contain more than $300$ planetary systems \citep{Knudstrup2024,WangWangBatygin2026}, with the majority being HJs.  The misaligned fraction is found to increase with host-star effective temperature, including among stars above the Kraft Break, where changes in stellar structure are thought to make tidal realignment inefficient \citep{Winn2010,Schlaufman2010,Lai2012Tides,LinOgilvie2017,AndersonWinnPenev2021,Knudstrup2024,Zanazzi2024,Zanazzi2025,Wang2026}.  Intriguingly, the stellar binary fraction also increases toward hotter (more massive) primaries \citep{Duchene2013,Offner2023}, 
potentially complicating the interpretation of observed trends of obliquity with host star temperature.

At the same time, \textit{Gaia} has transformed companion searches by providing homogeneous, all-sky parallaxes and proper motions for more than a billion stars \citep{GaiaEDR3Summary,ElBadry2021}.  Bound companions can now be readily distinguished from chance alignments using astrometry, and parallaxes now provide better estimates for host star properties and projected physical separations.  The all-sky catalog also extends companion searches beyond the small fields of adaptive-optics images, allowing bound companions to be identified out to $\theta>100\arcsec$. 
However, \textit{Gaia} DR3 is incomplete for companions within $\theta\lesssim2\arcsec$, especially when the companion is much fainter than the host, so high-resolution imaging is required to probe these small separations \citep[e.g.,][]{Ziegler2018,ElBadry2024Renaissance}.
Combining \textit{Gaia} astrometry and adaptive-optics imaging now has the ability to provide a homogeneous census of stellar companions across close and wide separations, and most importantly, with an explicitly defined selection function.

In this work, we construct a census of stellar companions for $147$ northern HJs with measured obliquities by combining \textit{Gaia} common-proper-motion pairs with uniform high-resolution imaging.  We leverage our homogeneous survey to infer the {\it intrinsic} companion fraction and properties over $50$--$50,000$ au, test whether resolved companions are associated with spin--orbit misalignment, and assess the implications for HJ formation through dynamical formation channels.

The remainder of this paper is organized as follows. Section~\ref{sec:observations} defines the parent sample and describes the imaging observations, Section~\ref{sec:companion_framework} presents the companion census, and Section~\ref{sec:contrast_completeness} describes the selection function.  Section~\ref{sec:descriptive_results} presents the demographic and obliquity results, Section~\ref{sec:discussion} discusses their implications, and Section~\ref{sec:ekl_capability} evaluates the capability of the observed companions for EKL-induced migration. Section~\ref{sec:conclusions} summarizes the conclusions. Appendices \ref{app:wd_comps}, \ref{app:kraft_break}, \ref{app:planet_mass_ratio_obliquity}, \ref{app:ekl_method}, \ref{app:companion_tables}, and \ref{app:image_atlas} provide additional details and data.

\section{Observations and Data Reduction}\label{sec:observations}

\subsection{Parent sample}\label{subsec:parent_sample}

We begin with $251$ planetary systems with published stellar-obliquity
constraints from \citet{WangWangBatygin2026}\footnote{\url{www.stellarobliquity.com}}.
The catalog contains each planet's class, orbital period, mass,
sky position, host effective temperature, and projected spin--orbit angle $|\lambda|$.
We select planets classified as hot Jupiters -- generally defined as those with
$0.3 < M_p/M_{\rm J}<13$ and $P < 10$ days -- that are accessible from the Palomar Observatory in the northern hemisphere ($\delta>-30\arcdeg$). This leaves $147$ HJs with measured spin--orbit angles, which constitute our parent sample.

Among these $147$ systems, $63$ were already known to host resolved stellar companions: $46$ common-proper-motion pairs from {\it Gaia} and $17$ from dedicated imaging.  We obtained PHARO images for $78$ systems, including $10$ of the previously known companion hosts.  The PHARO sample identifies $8$ new stellar companion candidates, bringing the total to $71$ companion hosts.
Of the $69$ systems not imaged by PHARO, $53$ have a previously known companion, and $16$ have no detected companion from previous high-resolution imaging.
Table~\ref{tab:census_summary} summarizes the parent and imaging samples.
For the obliquity analysis (Section \ref{subsec:projected_obliquity}), 
we exclude CoRoT-18, CoRoT-19, HATS-14, TOI-2524, and WASP-49, whose projected-obliquity uncertainties are $\sigma_\lambda\geq50\arcdeg$. This leaves $142$ hosts. Throughout this work, we define misalignment as systems satisfying $|\lambda|>10\arcdeg$ and $|\lambda|-2\sigma_\lambda>0$.

\begin{figure*}
    \centering
    \includegraphics[width=0.99\textwidth]{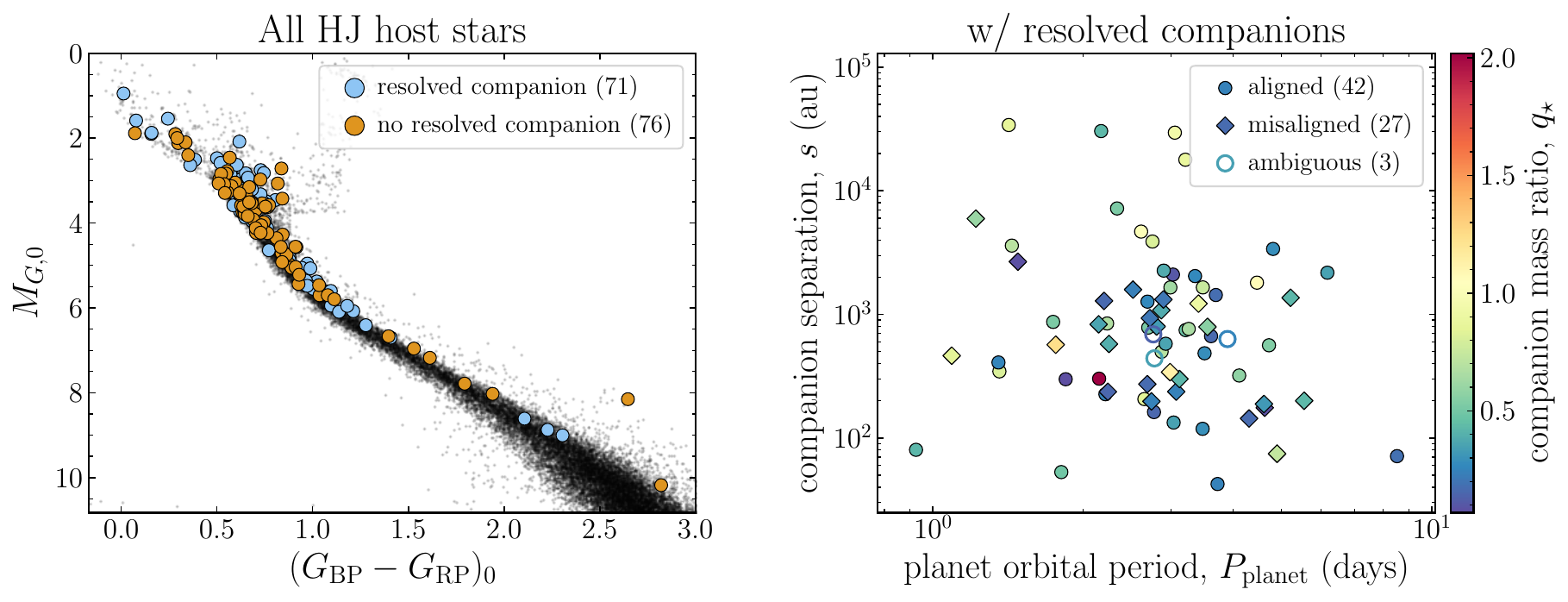}
    \caption{
    Summary of the HJ host-star parent sample. 
    {\bf Left:} Dereddened {\it Gaia} color--magnitude diagram for the $147$ HJ hosts, separated by those with (blue) and without (orange) a resolved companion. Black points show {\it Gaia} stars within $100$~pc for context. 
    {\bf Right:} Planet orbital period versus projected separation of the resolved companion for the $71$ companion hosts. Circles and diamonds denote aligned and misaligned HJs, respectively, using ($|\lambda|>10^\circ$ and $|\lambda|-2\sigma_\lambda>0$) as the misalignment criteria. The color denotes the companion-to-host mass ratio, $q_\star=M_{\rm comp}/M_\star$. HJ hosts are typically FGK main-sequence stars and span a broad range of planet periods, projected separations, and mass ratios.
    }
    \label{fig:sample_cmd_architectures}
\end{figure*}

Figure~\ref{fig:sample_cmd_architectures} shows all $147$ HJ hosts on the dereddened {\it Gaia} color--magnitude diagram and summarizes the planet and companion properties of the $71$ systems with resolved companions.
Most hosts are FGK stars on or near
the main sequence, with a smaller number of hotter A stars,
cooler M dwarfs, and the pre-main-sequence host K2-33
\citep{Mann2016}.

\subsection{Observations}

We obtained near-infrared adaptive optics (AO) images with \pharo\ on the \palmartwo\ at Palomar Observatory.  \pharo\ uses a $1024\times1024$ HgCdTe detector and provides 25 and 40 mas pixel scales \citep{Hayward2001}.  All observations use the $25~{\rm mas}\ {\rm pixel}^{-1}$ mode, providing a $25.6\arcsec\times25.6\arcsec$ field of view \citep{Hayward2001}.

We observed $78$ northern HJ targets over six nights between
May 2025 and July 2026.
The general observing strategy was to obtain one image for every target, and if a companion was detected, obtain a second image in a different filter to constrain its spectral type and verify that it was not an image artifact.
This resulted in $66$ images in
$K_s$, $13$ in $K$, $30$ in  $K$-continuum, $33$ in Br$\gamma$, $23$ in $H$, and $5$ in $J$.
For each observation, we use a five-point dither pattern.
The image quality changes across targets and nights due to variations in AO performance and atmospheric conditions (e.g., Appendix \ref{app:image_atlas}).

\begin{deluxetable}{lr}
\tablecaption{Sample selection\label{tab:census_summary}}
\tablehead{
\colhead{Category} &
\colhead{Number}
}
\startdata
Northern hot-Jupiter sample & 147 \\
known stellar companions & 63 \\
no known stellar companions & 84 \\
PHARO imaging sample & 78 \\
new PHARO companion & 8 \\
total companions & 71 \\
\enddata
\end{deluxetable}

\subsection{Reduction and stacking}

We reduced the PHARO images using the SImMER package \citep{Savel2022}.  PHARO raw files
store the detector in four quadrants, which we first assembled into one image. For each observing block, SImMER constructs median darks, subtracts the matched dark from each flat, normalizes and median-combines the flats, and constructs a flat-corrected median sky image.  Science frames are divided by the flat, sky-subtracted, corrected for bad pixels, and shifted to place the host star at
the detector center.

\subsection{Image quality}
Image quality affects point spread function (PSF) subtraction, source localization, and relative photometry. For each stacked image, we estimate an effective full width at half maximum (FWHM) by creating an azimuthally averaged radial profile about the centered peak.
 The median FWHM is $0.108\arcsec$, with a 16th--84th percentile range of $0.096$--$0.174\arcsec$. This sets the effective angular resolution of our survey ($\sim0.1 \arcsec$) and is used throughout the analysis.
 We use $160\times160$ pixel ($4\arcsec$) stacked images for visual inspection.

\section{Companion Identification}
\label{sec:companion_framework}
We create a census of resolved stellar companions to all northern HJs by combining published stellar companions, {\it Gaia} common-proper-motion pairs, and new PHARO companions from our AO campaign. We specify the discovery channel for each system because the three searches cover different separations and contrasts. The sensitivity of each survey is quantified and incorporated into the unified selection function (Section \ref{sec:contrast_completeness}).

The PHARO imaging sample contains $78$ systems, and $8$ host newly discovered companions, while $10$ additional systems already had known companions.
The remaining $60$ have no reported companion, and no clear companion is observed in the PHARO image.
The new PHARO companions are CoRoT-19, HAT-P-49, HAT-P-50, HAT-P-70, HATS-02, HATS-14, WASP-60, and WASP-084.  

\subsection{Previously known companions}

We classify $63$ systems as {\it previously known} companion hosts: $46$ appear in the {\it Gaia} EDR3 wide-binary catalog of \citet{ElBadry2021}, and $17$ were established through published high-resolution imaging. The {\it Gaia} binary catalog selects pairs with consistent parallaxes and proper motions, and for each pair, estimates their chance-alignment probabilities empirically. The median chance-alignment probability is $2.4\times10^{-5}$, and all are below $2\%$, indicating that these {\it Gaia} pairs are highly likely to be bound.

Among the $17$ systems identified through previous imaging, $14$ have consistent proper motions, establishing them as bound companions \citep{Ngo2015,Ngo2016,Mugrauer2019,Temple2019,Schlagenhauf2024,Wollert2015,Johnson2018}.
Three others, WASP-103 B, WASP-020 B, and KELT-21 B/C, do not have published epoch astrometry, but have empirically measured chance alignment probabilities of $<1\%$ \citep{Southworth2016,Johnson2018, Bohn2020}.

\subsection{New PHARO companions}\label{subsec:new_pharo_comps}

We review every stacked PHARO image for resolved sources. We inspect the science stacks and residual images produced by reference differential imaging (RDI), which models and subtracts the primary PSF using reference-star images. The reference stars are generally other science targets from the same night with images taken in the same filter. We performed principle component analysis (PCA)-based RDI with {\tt VIP} \citep{GomezGonzalez2017}. We classify sources as companion candidates if they are detached from the primary PSF and persist at the same location across filters, epochs, or reference PSFs. These checks reject wavelength-dependent speckles and artifacts tied to a single reference PSF.

\begin{figure*}
    \centering
    \includegraphics[width=0.95\textwidth]{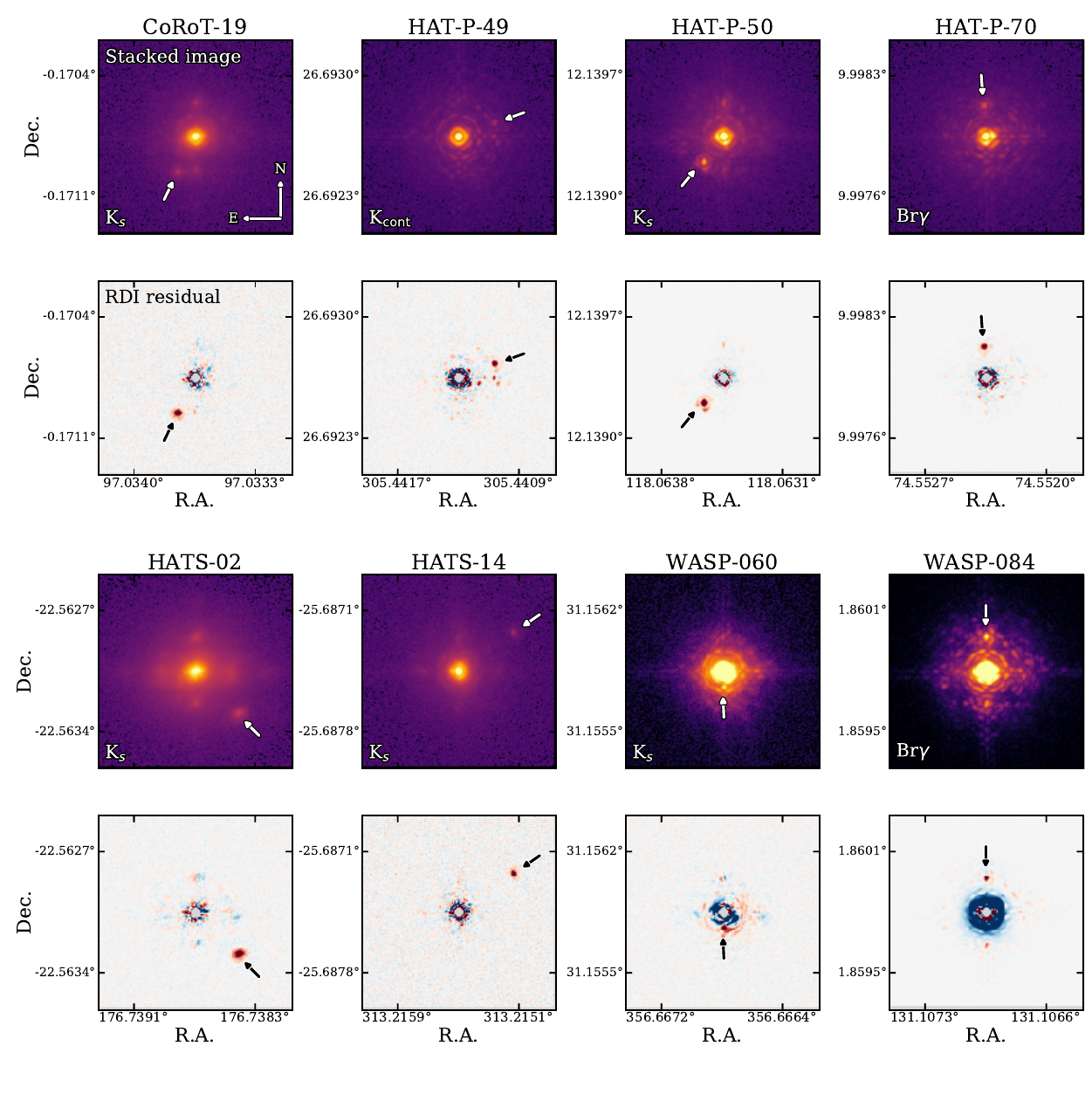}
    \caption{
    $4\arcsec\times4\arcsec$ PHARO images of the eight newly detected companions. 
    For each system, the upper panel shows the stacked science image, and the lower panel shows the corresponding RDI PSF-subtracted residual with the filter labeled in the bottom left. Arrows mark the companion positions. 
    Each companion candidate, besides WASP-084, is identified at a consistent location in multiple independent images and across at least two different filters.}
    \label{fig:new_pharo_detections}
\end{figure*}

Eight new sources meet these criteria: CoRoT-19, HAT-P-49, HAT-P-50, HAT-P-70, HATS-02, HATS-14, WASP-60, and WASP-084.  
Figure~\ref{fig:new_pharo_detections} shows all new PHARO companions, which can be readily distinguished both in the stacked images and PSF-subtracted images.
WASP-84 has the least secure candidate among the eight, as it is recovered in multiple image stacks but only in the Br$\gamma$ filter. Excluding WASP-084's candidate companion does not change our population-wide conclusions.

\subsubsection{Chance alignment probability}\label{subsubsec:pharo_chance_align}

We estimate the chance alignment probability of each new PHARO companion candidate using \citep{Ngo2015,Ginski2016,Evans2016}:
\begin{equation}\label{eq:chance_alignment}
    P_{{\rm chance},i}=1-\exp\left[-\pi\rho_i^2
    \Sigma_i(<m_{{\rm cand},i})\right],
\end{equation}
where $\pi\rho_i^2$ is the area enclosed by the candidate's angular separation and $\Sigma_i(<m_{{\rm cand},i})$ is the surface density of field stars brighter than the candidate. Their product is therefore the expected number of unrelated stars this bright that would fall at least this close to the host. Assuming that field stars follow Poisson statistics, $P_{{\rm chance},i}$ is the probability of finding one or more such stars by chance.

For each candidate, we calculate its $K_s$ magnitude from the unresolved 2MASS magnitude of the system and the PHARO contrast. We then use the Besan\c{c}on Galaxy Model to estimate the cumulative number of stars at least this bright at the target's Galactic coordinates \citep{Robin2003,Skrutskie2006}. The resulting probabilities range from $0.005\%$--$0.320\%$.
The probability that \emph{at least one} source is a chance alignment is
\begin{equation}
P(N_{\rm chance}\geq 1)=1-\prod_{i=1}^{8}\left(1-P_{{\rm chance},i}\right)=0.95\%.
\end{equation}

As an independent check, we repeat the calculation using observed 2MASS source counts \citep[e.g.,][]{Oberst2017,Johnson2018}. For each target, we count 2MASS sources brighter than the candidate within a $30\arcmin$ radius, divide by the searched area to estimate the local surface density, $\Sigma_i(<m_{{\rm cand},i})$, and insert this value into Equation \eqref{eq:chance_alignment}. The $30\arcmin$ region provides sufficient sources while remaining representative of the target's local Galactic field. This calculation gives a $0.85\%$ probability that at least one of the eight detections is a chance alignment, consistent with the Besan\c{c}on estimate. Because 2MASS becomes incomplete at faint magnitudes, this value is a lower bound for candidates below its completeness limit \citep{Skrutskie2006}.
Overall, the low probabilities indicate that the new PHARO sources are unlikely to be chance alignments.

\subsection{Relative astrometry and photometry}

We report approximate positions and contrasts for the eight PHARO companions (Table~\ref{tab:pharo_candidate_measurements}).  From an initial guess of the source position, we refine this estimate with a flux-weighted centroid in a $13\times13$ pixel box.  From the angular separations, we identify projected separations between the two stars with $s_{\rm proj}=\theta \times D$, where $D$ is the {\it Gaia} parallax distance in parsecs.

We estimate contrast using equal circular apertures on the primary and companion, with a radius of one FWHM and a minimum of $2.5$ pixels. We estimate the image background around the primary using a sigma-clipped annulus spanning $3$--$5$ aperture radii. 
At the companion position, the dominant background is typically light from the extended PSF of the primary.
We measure this light using identical apertures placed around the primary at the same separation as the companion, but at up to $16$ other position angles.  We exclude apertures within $18\arcdeg$ of the companion to avoid including its flux.  The median flux in these reference apertures is subtracted from the companion measurement, and their scatter sets the formal uncertainty.  We also measure the contrast after subtracting a reference PSF from the same image.  The difference between the two contrast measurements estimates the systematic uncertainty from the extraction method.

For partially blended sources, the primary and companion overlap too strongly for aperture photometry.  We therefore construct an empirical PSF profile from the target image, subtract it from the primary, and measure the companion flux in the residual image.  The spread across different reference PSFs is included in the contrast uncertainty. WASP-60 is the only target where the companion is marginally resolved, and this method was applied (e.g., Figure \ref{fig:new_pharo_detections}).

WASP-60 is recovered at the same location in the $K_s$, $K_{\rm cont}$, and Br$\gamma$ PHARO images and their PSF-subtracted residuals (Figure \ref{fig:new_pharo_detections} and \ref{fig:pharo_atlas_05}). Keck/NIRC2 imaging obtained $10$ days later also shows a $5\sigma$ point source at the same sky position after subtraction (Mehla et al., private communication)\footnote{Following discussions with the Indiana University team, A. Kraus independently detected the companion to WASP-60 in high-resolution near-infrared imaging (private communication).}. These independent detections support its classification as a companion.

\begin{table}
\centering
\caption{New PHARO companion candidates.}
\label{tab:pharo_candidate_measurements}
\begin{tabular}{llrrr}
\toprule
System & Filter & $\theta$ ($\arcsec$) & $s_{\rm proj}$ (au) & $\Delta m$ \\
\midrule
CoRoT-19  & $K_s$          & 0.792 & 632  & $4.29^{+0.18}_{-0.21}$ \\
HAT-P-49  & $K_s$          & 0.798 & 273  & $5.56^{+0.11}_{-0.23}$ \\
HAT-P-50  & $K_s$          & 0.659 & 300  & $2.97^{+0.07}_{-0.32}$ \\
HAT-P-70  & $K_s$          & 0.628 & 198  & $4.21^{+0.10}_{-0.12}$ \\
HATS-02   & $K_s$          & 1.225 & 409  & $3.71^{+0.11}_{-0.19}$ \\
HATS-14   & $K_s$          & 1.373 & 690  & $5.20^{+0.06}_{-0.06}$ \\
WASP-060  & $K_{\rm cont}$ & 0.313 & 144  & $4.98^{+0.34}_{-0.32}$ \\
WASP-084 & Br$\gamma$     & 0.714 & 71.3 & $4.34^{+0.11}_{-0.12}$ \\
\bottomrule
\end{tabular}
\tablecomments{Projected separations use \textit{Gaia} parallaxes; angular separations use the nominal $0.025\arcsec$ pixel scale.}
\end{table}

\section{Selection Function}
\label{sec:contrast_completeness}
To infer intrinsic companion demographics, we model the probability of detecting a companion around each HJ host through PHARO imaging, previous high-resolution imaging, or {\it Gaia}. The imaging sensitivity is described by contrast curves, while the {\it Gaia} sensitivity depends on angular separation and magnitude contrast.

\subsection{Contrast curves}

We calculate the contrast curves with VIP version 1.4.2 \citep{GomezGonzalez2017}.  We subtract the stellar PSF with one-component reference differential imaging (RDI) and run VIP's \texttt{contrast\_curve} routine.  
To measure the flux lost during PSF subtraction, VIP injects artificial companions along four radial spokes around the star and measures their recovered flux. The curve is evaluated at radial intervals of one FWHM using a $5\sigma$ threshold and the small-sample correction from \citet{Mawet2014}.
We then test the resulting limit by injecting companions at eight evenly spaced position angles and retain the contrast curve when at least seven are recovered with ${\rm S/N}\geq5$. 
The resulting target-specific contrast curves define the selection function for high-resolution imaging: for each host, they identify the faintest detectable companion as a function of angular separation and capture differences in image quality among the observations.

\begin{figure*}
    \centering
    \includegraphics[width=0.98\textwidth]{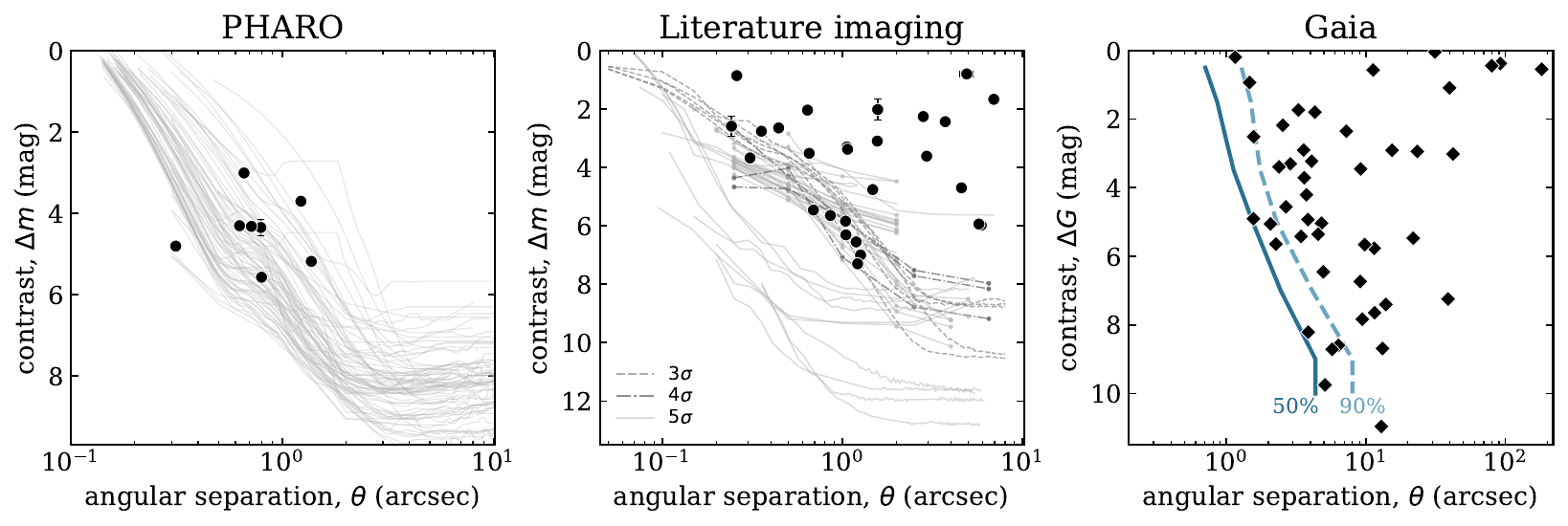}
    \caption{Sensitivity to resolved stellar companions from PHARO (left), previous imaging, and {\it Gaia} (right).  
    {\bf Left:} $5\sigma$ contrast curve for each PHARO image. 
    {\bf Middle:} $47$ contrast curves for literature sources, where line styles mark $3\sigma$, $4\sigma$, and $5\sigma$ curves, depending on which is reported.
    {\bf Right:} $50\%$ and $90\%$ {\it Gaia} recovery limits from \citet{ElBadry2024Renaissance}. Black points mark detected companions.}
    \label{fig:contrast_curves}
\end{figure*}

We also compile the published contrast curves for each of the $17$ companions identified from previous imaging campaigns (Figure~\ref{fig:contrast_curves}), as well as $13$ with previous imaging and no detected companion. Since most contrast curve data are not publicly available, we digitize published contrast curves for all sources from the discovery papers.
{\it Gaia} sensitivity is evaluated for all $147$ hosts; $77$ hosts have PHARO curves, and all $17$ literature hosts have published contrast curves. 

Five companions identified through previous imaging---HAT-P-20 B, HAT-P-32 B, K2-267 B, WASP-1 B, and WASP-180 B---are resolved in {\it Gaia} DR3 but absent from the common-proper-motion catalog of \citet{ElBadry2021}. HAT-P-32 B is blended with its host and has no reliable {\it Gaia} astrometry, while the other four fail the catalog's parallax or proper-motion quality cuts.
These cuts can reject genuine companions if they are unresolved binaries because orbital motion distorts the single-source {\it Gaia} astrometric solution \citep{ElBadry2021}. HAT-P-20 B may be such a case: its {\it Gaia} DR3 single-source fit is poor, with ${\rm RUWE}=23.2$ \citealt{Gaia2023}. We therefore retain all five systems based on their published confirmations \citep{Ngo2015,Temple2019,Schlagenhauf2024}. Manual inspection of each candidate pair also shows proper motions consistent with physical association.

Figure~\ref{fig:contrast_curves} compares the $8$ PHARO companions, $17$ previous literature companions, and $46$ resolved {\it Gaia} companions with the corresponding survey sensitivity.  The new PHARO companions span $0.313$--$1.37\arcsec$ and $\Delta m=3.0$--$5.6$ mag.  
{\it Gaia} DR3 extends the search from $1.16$ to $182\arcsec$.  The faintest plotted source is the KELT-9 companion at $12.9\arcsec$ and $\Delta G=10.96$ mag.  Its photometric mass ratio is $q_\star=0.075$, below the $q_\star\geq0.1$ population modeled here.

The new PHARO companions are faint enough that their effect on published host-star atmospheric properties should be small. Simple two-component spectral estimates using the measured near-IR contrasts predict optical contamination of typically $<1\%$ (up to a few percent for HAT-P-50), corresponding to $T_{\rm eff}$ shifts of only $\sim10--40$~K for most systems and $\lesssim 70$~K in the most affected case; expected $[\mathrm{Fe/H}]$ biases are at the level of a few hundredths of a dex.

\subsection{Binary population model}
\label{sec:population_model}

We infer the {\it intrinsic} companion fraction of HJ hosts over mass ratios $0.1\leq q_\star\leq1$ and separations $50\leq s/{\rm au}\leq50{,}000$, where our survey is sensitive. The $50$~au inner limit is approximately the smallest separation to which the AO imaging is sensitive for nearby stars and matches previous work \citep{Ngo2016}. Differences in distance and inner working angle are accounted for on a target-by-target basis by the selection function.

The mass ratio is defined as $q_\star=M_2/M_1$, where $M_1$ is the host and $M_2$ is the companion. We derive companion mass ratios from the measured contrast using the main-sequence relations of \citet{PecautMamajek2013}. For {\it Gaia} companions, we use the absolute $G$-band magnitude relation; for imaging companions, we use the corresponding relation in the available $K_s$, $K$, $J$, or $H$ bandpass.
Only one source (KELT-9, $q_\star=0.075$) is below the mass ratio threshold, five sources have $q_\star>1$, and one source has a white dwarf companion (WASP-71). These sources are excluded from the demographic analysis.

For each host, we draw $10^3$ companion orbits in every cell of a $50\times50$ grid uniform in $q_\star$ and $\log s$, giving $10^6$ simulated companions per target.
We weight each cell according to the expected frequency of companions with that mass ratio and separation, assuming the wide-binary properties of solar-type stars.
We adopt the {\it intrinsic} mass-ratio distributions derived by \citet{ElBadry2019Twins} for wide binaries with primary masses 
$0.1 < M_1/{\rm M_\odot} < 2.5$, mass ratios $0.1 < q  <1$, and separations $50 < s/{\rm au} < 50{,}000$. 
This mass-ratio distribution, $p_{\rm EB19}(q_\star\mid M_{1,i},s)$, depends on both $M_1$ and $s$ \citep[][see their Figure 9 and Table G1]{ElBadry2019Twins}. We evaluate this distribution at the measured mass of each HJ host,
$M_{1,i}$.

For the separation distribution, we adopt: 
\begin{align}
 p(s)\equiv\frac{{\rm d}P}{{\rm d}\log s}&\propto
 \begin{cases}
 1, & 50\leq s/{\rm au}<500,\\
 s^{-0.6}, & 500\leq s/{\rm au}\leq50{,}000,
 \end{cases}
 \label{eq:EB_weight_s}
\end{align}
which is log-uniform from $50$--$500$ au \citep{ElBadry2019} and follows ${\rm d}P/{\rm d}s\propto s^{-1.6}$ from $500$--$50{,}000$ au \citep{ElBadry2018}. The total weight of a cell is then
\begin{equation}\label{eq:EB_weight_q}
w_i(q_\star,s)\propto
p_{\rm EB19}(q_\star\mid M_{1,i},s)\,p(s).
\end{equation}
We estimate each $M_{1,i}$ from its measured $T_{{\rm eff}}$ and map the component masses to {\it Gaia} and near-infrared absolute magnitudes using the main-sequence relations of \citet{PecautMamajek2013}. The {\it Gaia} parallax distance then gives the apparent contrast and angular separation, $\theta=s/D_i$.

Our imaging census excludes companions that were initially more massive than the HJ host and have since become a faint white dwarf (WD) or were unbound during WD formation\footnote{Rapid mass loss during WD formation is predicted to produce `WD kicks' of $v_{\rm kick}\sim0.5$--$2$~km~s$^{-1}$ \citep[e.g.,][]{Hwang2025,OConnor2026,Fuller2026}. The signature of such kicks is evident in {\it Gaia} wide binaries \citep{ElBadry2018,Hwang2025}, {\it Gaia} triples \citep{Shariat2023, Shariat2025a}, and the properties of WDs in clusters \citep[][]{Weidemann1977,Fellhauer2003,Heyl2007,Davis2008}.}.
Appendix~\ref{app:wd_comps} presents a separate calculation for initially more massive companions that have evolved into WDs. These systems are not included in the main analysis, which focuses on the present-day main-sequence companion fraction.

\subsection{PHARO selection function}
\label{sec:pharo_selection}

We calculate a target-specific $5\sigma$ contrast curve for each PHARO-imaged host using VIP (Figure~\ref{fig:contrast_curves}). We count a simulated companion as detected when its predicted contrast is brighter than the contrast limit at its angular separation, $\theta$. The recovery probability is zero outside the radial range covered by the contrast curve.

\subsection{Previous-imaging selection function}
\label{sec:literature_selection}

For the $29$ hosts with contrast curves published from previous imaging, including $16$ detections and $13$ non-detections, we assemble all available published contrast curves in their reported filters. We calculate the corresponding contrasts using $1$~Gyr, solar-metallicity MIST isochrones \citep{Dotter2016,Choi2016} and count a companion as detected if it is brighter than the limit of any available curve. We set the recovery probability $P_{\rm lit}=0$ outside the published angular ranges and for all other hosts (Figure \ref{fig:contrast_curves}).

\subsection{{\it Gaia} selection function}
\label{sec:gaia_selection}

For each simulated companion, we calculate its apparent magnitude $G_2$ and contrast $\Delta G=G_2-G_1$ in the {\it Gaia} $G$ filter. We evaluate the empirical {\it Gaia} DR3 recovery probability, $S_{\rm Gaia}(\theta,\Delta G)$, from the ``no cuts'' curves in Figure~2 of \citet{ElBadry2024Renaissance}. These curves give the recovery of resolved companions in bins of $\Delta G$. We interpolate linearly with angular separation along each curve and between the curves in $\Delta G$. For similar-brightness pairs, the effective angular resolution is $\approx1\arcsec$ (Figure \ref{fig:contrast_curves}).

We also require $G_2\leq21$ and an expected parallax signal-to-noise ratio $\varpi_i/\sigma_\varpi(G_2)>5$. These cuts approximate the faint limit of {\it Gaia} EDR3 and the minimum parallax significance used in wide-binary catalogs \citep{GaiaEDR3Summary,ElBadry2021}. The resulting {\it Gaia} recovery probability is
\begin{equation}
 P_{{\rm Gaia},i}=
 S_{\rm Gaia}(\theta,\Delta G)\,
 \mathbb{H}(G_2\leq21)\,
 \mathbb{H}\!\left[\frac{\varpi_i}{\sigma_\varpi(G_2)}>5\right],
 \label{eq:gaia_selection}
\end{equation}
where $\mathbb{H}$ is a step function that equals $1$ when the condition is satisfied and $0$ otherwise.

\subsection{Total selection function}
\label{sec:combined_selection}

For simulated companion $k$ around HJ host $i$, the probability of recovery by {\it Gaia}, PHARO, or previous imaging is
\begin{equation}
 P_{{\rm total},i,k}
 =
 1-(1-P_{{\rm Gaia},i,k})
 (1-P_{{\rm PHARO},i,k})
 (1-P_{{\rm lit},i,k}).
 \label{eq:total_selection}
\end{equation}
We set $P_{{\rm PHARO},i,k}=0$ for targets without PHARO imaging and $P_{{\rm lit},i,k}=0$ for targets without a published contrast curve. 
We define the sensitivity for host $i$ as the weighted mean recovery probability over all simulated companions:
\begin{equation}
S_i
=
\frac{\displaystyle\sum_k w_{i,k}P_{{\rm total},i,k}}
{\displaystyle\sum_k w_{i,k}},
\label{eq:target_sensitivity}
\end{equation}
where $w_{i,k}$ is the field-binary weight from Equation~\eqref{eq:EB_weight_q} and \eqref{eq:EB_weight_s}. Thus, $S_i$ is the fraction of companions in the modeled mass-ratio and separation range that our observations would detect around host $i$.

\begin{figure}
    \centering
    \includegraphics[width=0.99\linewidth]{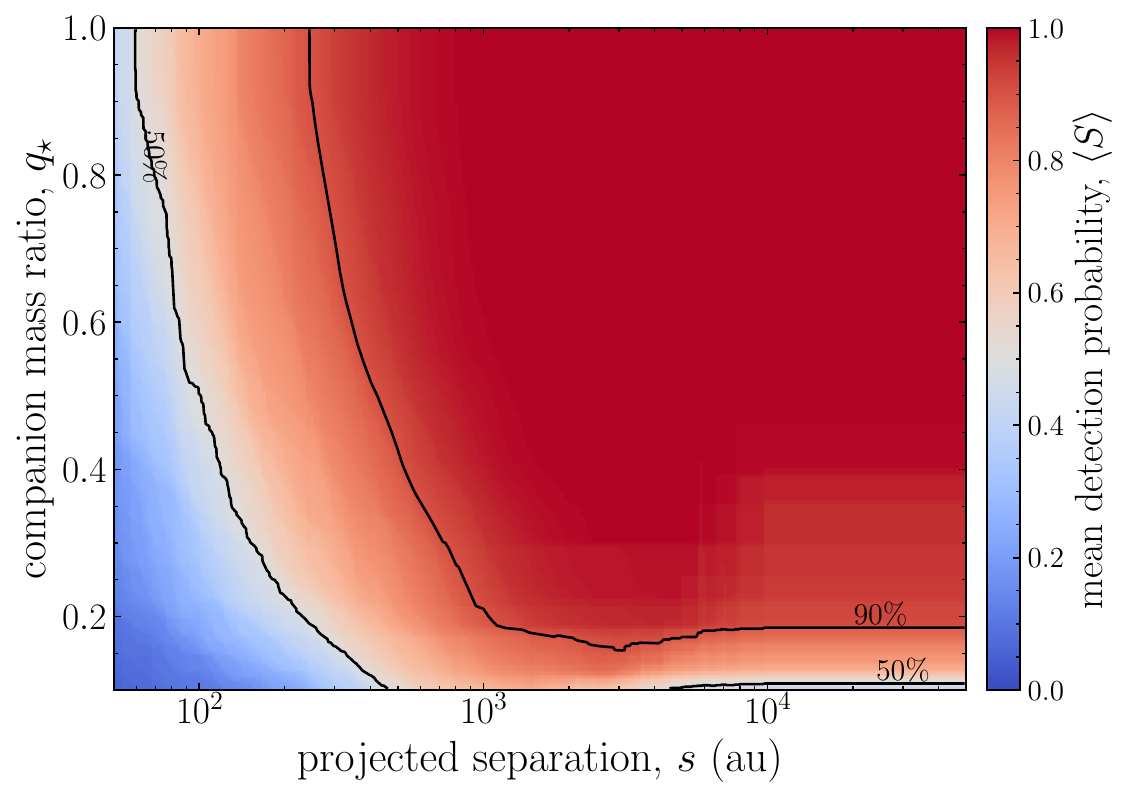}
    \caption{
    Mean probability of detecting a main-sequence companion across the 147 HJ hosts as a function of projected separation $(s)$ and companion mass ratio $(q_\star)$. At each $(s,q_\star)$, we combine the Gaia, PHARO, and literature-imaging probabilities for each host and then average equally over the sample. Black curves mark mean $50\%$ and $90\%$ detection probabilities.
    }
    \label{fig:mean_selection_surface}
\end{figure}

Figure~\ref{fig:mean_selection_surface} shows the combined sensitivity as a function of the separation and mass ratio parameter space. At each $(s,q_\star)$, we calculate the Gaia, PHARO, and literature-imaging detection probability for each of the $147$ hosts using its mass, distance, and available contrast curves, and then average equally over the sample. The mean recovery exceeds $90\%$ for $q_\star\gtrsim0.2$ at $s\gtrsim10^3$~au, but decreases toward closer separations and lower-mass companions. This averaged surface summarizes the survey coverage; the population inference retains each host's individual selection function and integrates it over the adopted companion distribution to obtain $S_i$.

Note that our selection function does not include the probability that an HJ was first detected, confirmed by follow-up spectroscopy, and selected for an obliquity measurement. Each of these steps can introduce an additional bias.
At the first transit-detection stage,
a close stellar companion can dilute the transit signal, making the planet harder to detect. For HJs, this selection bias is likely negligible 
because their transit signal is large. \citet{Littlefield2026} found a $\sim100\%$ {\it TESS} sensitivity to close-in gas giants around primaries with comparably bright unresolved companions ($\Delta m\geq4$), similar to the completeness around apparently single stars. 
By contrast, only $\sim50\%$ of comparable planets would be detected if they orbited the fainter secondary. Spectral blending from close, bright companions may also disfavor radial-velocity confirmation or obliquity measurements \citep{Santerne2015}, although this effect has not been quantified uniformly for our sample.
Both selection effects preferentially \emph{remove} binaries from the observed sample, so our inferred binary fractions are conservative lower limits.

\subsection{Inferring intrinsic demographics}
\label{sec:binary_fraction_likelihood}

The process for deriving intrinsic demographics from the observed data can be summarized as follows:

\begin{enumerate}
    \item Draw companions with mass ratios and projected separations $(q_\star,s)$ and map them to a filter-specific contrast and projected angular separation ($\Delta m$,$\theta$).
    
    \item Evaluate the PHARO, previous-imaging, and {\it Gaia} recovery probabilities (Figure \ref{fig:contrast_curves}), combine them using Equation~\eqref{eq:total_selection}, and average using Equation~\eqref{eq:target_sensitivity} to obtain $S_i$.
    
    \item Calculate the observed binary fraction from the $62$ previously identified companions and $8$ new PHARO candidates ($70/147$).
    
    \item Infer the intrinsic binary fraction $f_{\rm bin}$ with
    Equation~\eqref{eq:binary_fraction_likelihood}.
\end{enumerate}

Given $f_{\rm bin}$ and the target sensitivities $S_i$, the likelihood is 
\begin{equation}
 \mathcal{L}(f_{\rm bin})=
 \prod_{i\in\mathcal{K}} S_i f_{\rm bin}
 \prod_{j\in\mathcal{U}}(1-f_{\rm bin}S_j),
 \label{eq:binary_fraction_likelihood}
\end{equation}
where $\mathcal{K}$ and $\mathcal{U}$ iterate over the systems with and without detected companions, respectively. Each target is treated as an independent Bernoulli trial, with probability $f_{\rm bin}S_i$ of yielding a detection. Following \citet{Ngo2016}, we set $S_i=1$ for systems with a detected companion. We adopt a uniform prior over $0<f_{\rm bin}<1$.

\subsection{Companion fractions versus $q_\star$, $s$, and $T_{\rm eff}$}
\label{sec:binned_fraction_method}

To measure the intrinsic mass-ratio and projected-separation distributions, we repeat the inference in six mass-ratio bins with edges $q_\star=(0.1,0.2,0.3,0.4,0.55,0.75,1)$ and six equal-width bins in $\log s$ from $50$--$2{,}000$~au. 
For each HJ host $i$ in $(q_\star,s)$ bin $b$, we calculate $S_{i,b}$ by restricting the sums in Equation~\eqref{eq:target_sensitivity} to simulated companions in that bin. We then replace $S_i$ with $S_{i,b}$ in Equation~\eqref{eq:binary_fraction_likelihood} and classify detections according to whether the observed companion lies in the same bin. This gives the selection-corrected companion fraction in each mass-ratio or projected-separation interval.  Although the bins are fitted independently and are not constrained to reproduce the global result, their summed fractions approximately match the intrinsic companion fraction inferred over $50$--$2{,}000$~au.

To measure the companion fraction as a function of host temperature ($T_{\rm eff}$), we divide the $147$ systems into $T_{\rm eff}$ bins of roughly equal $N_{\rm bin}$ and repeat Equation~\eqref{eq:binary_fraction_likelihood} in each bin using each target's full $50$--$50{,}000$~au selection function. For comparison in Section~\ref{sec:companion_obliquity_discussion} (Figure~\ref{fig:teff_fractions}), we calculate wide-binary fractions as a function of $T_{\rm eff}$ for field stars using the results aggregated by \citet{Offner2023} over $100$--$10{,}000$~au. The binary fractions for early-M, FGK, and A stars are based on \citet{Winters2019}, \citet{Raghavan2010}, and \citet{DeRosa2014}, respectively. We scale these fractions to $50$--$50{,}000$~au using the corresponding lognormal separation distributions from each study, assuming $a=s$ for this comparison, and map primary mass to $T_{\rm eff}$ with the main-sequence relation of \citet{PecautMamajek2013}.

\section{Results}
\label{sec:descriptive_results}

\subsection{Binary fraction}\label{sec:results_binary_fraction}

We detect companions around 70 of 147 hosts, giving an observed binary
fraction of $48\pm4\%$. Correcting for the survey selection function gives
\begin{equation}
 f_{\rm bin,50-50000}=62.0\pm4.9\%,
 \label{eq:primary_fbin}
\end{equation}
or $3$--$4\times$ larger than the field-star fraction over the same mass ratio and separation range \citep{Raghavan2010,Moe2017}.
The \emph{birth} companion fraction is even larger than this, since initially more massive companions that evolved into WDs are not included in this calculation (see Appendix~\ref{app:wd_comps}).

As a conservative check, we repeat the inference using only the $46$ {\it Gaia} common-proper-motion companions, whose consistent parallaxes and proper motions establish the companions as bound with high confidence. This gives $f_{\rm bin}= 60.9 \pm 7.0\%$ over the same range, which is consistent with our total result. See Table \ref{tab:binary_fraction_inference} for the inferred binary fractions over all ranges.

\begin{figure}
    \centering
    \includegraphics[width=\columnwidth]{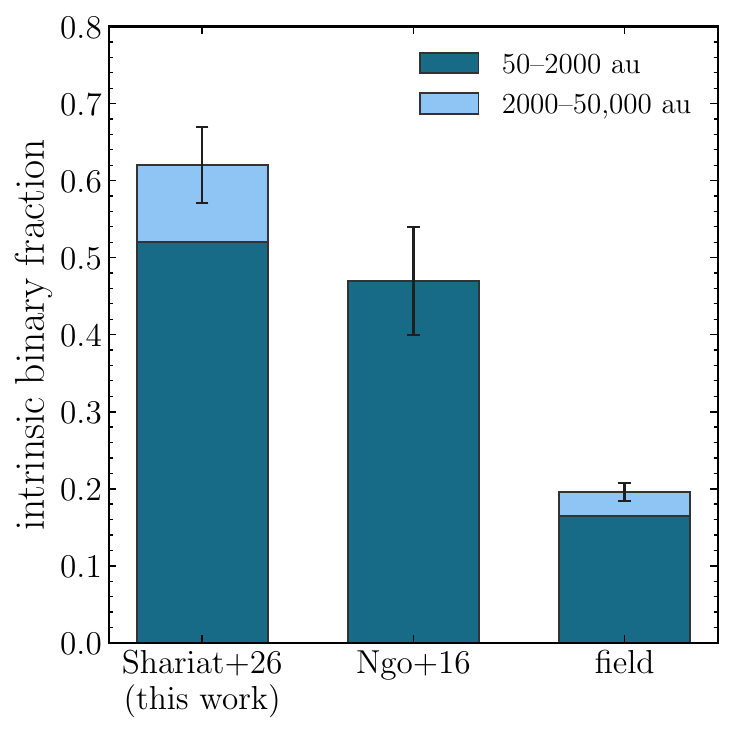}
    \caption{Intrinsic binary fractions of HJ hosts and field stars.  The dark bars show the 50--2000 au fraction, and the light extension gives the total fraction through 50,000 au. We compare our finding to the hot-Jupiter result of \citet{Ngo2016} over 50--2000 au and field stars. The HJ stellar binary fraction is significantly higher than that of field stars.}
    \label{fig:binary_fraction_comparison}
\end{figure}

Figure~\ref{fig:binary_fraction_comparison} compares our result with previous measurements. Over $s=50$--$2{,}000$~au, we infer $f_{\rm bin}=52.0 \pm 5.2\%$, consistent with the $47\pm7\%$ reported by \citet{Ngo2016}. Restricting the calculation to the $31$ {\it Gaia} common proper-motion companions in this range gives $f_{\rm bin}=49.6 \pm 7.5\%$ (Table~\ref{tab:binary_fraction_inference}), again consistent with the fiducial estimate.

\vspace{-0.5cm}
\begin{deluxetable}{lcc}
\tablecaption{Intrinsic binary fractions\label{tab:binary_fraction_inference}}
\tablehead{
\colhead{Range (au)} &
\colhead{\textit{Gaia} only} &
\colhead{All companions}
}
\startdata
$50$--2000~au & $0.496 \pm 0.075$ & $0.520 \pm 0.052$ \\
$50$--50,000~au & $0.609 \pm 0.070$ & $0.620 \pm 0.050$
\enddata
\end{deluxetable}

\subsection{Mass ratio and separation}
\label{sec:companion_distribution_results}

\begin{figure*}
    \centering
    \includegraphics[width=0.98\textwidth]{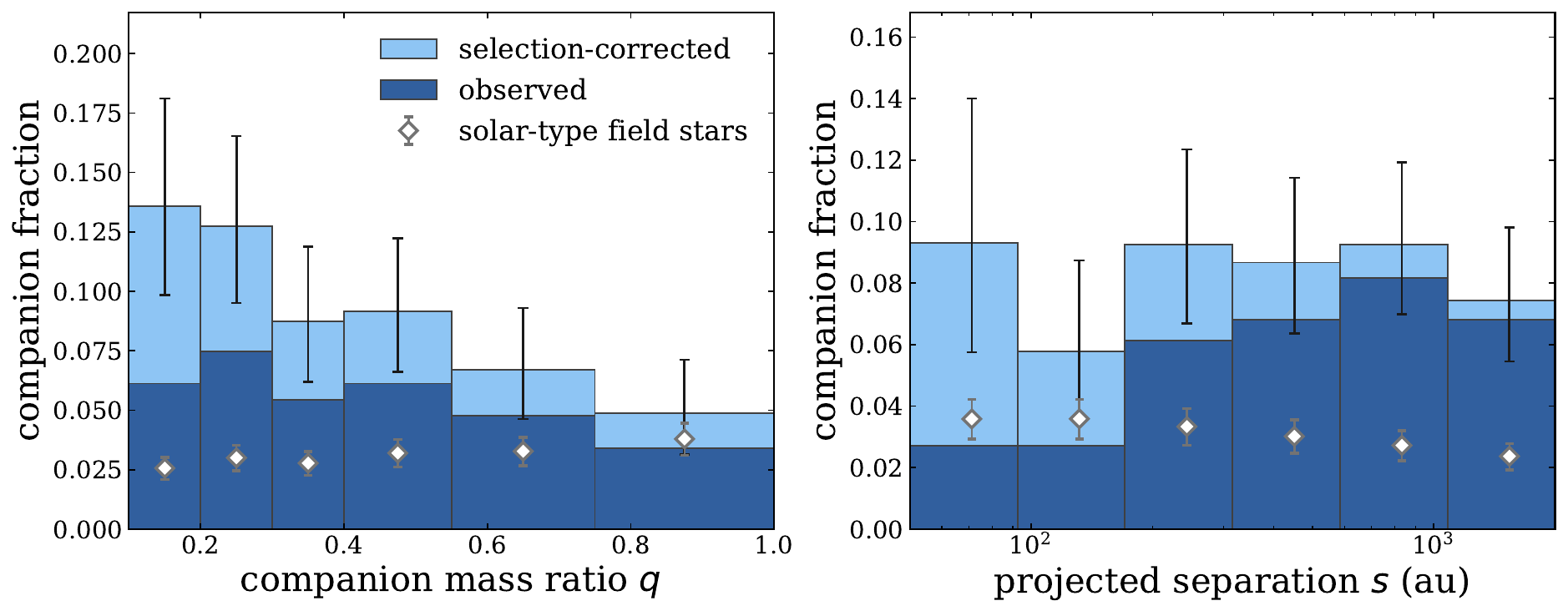}
    \caption{Companion fractions over 50--2000 au, binned by
    stellar-companion mass ratio (left) and projected separation (right). Light bars and black intervals show the selection-corrected posterior medians and 16th--84th percentiles.  Dark bars show the observed fractions. White diamonds show the solar-type field distribution from \citet{Moe2017}.}
    \label{fig:intrinsic_q_s_dist}
\end{figure*}

Figure~\ref{fig:intrinsic_q_s_dist} shows the intrinsic companion fraction as a function of companion mass ratio ($q_\star$) and projected separation ($s$).
The mass-ratio distribution declines toward higher $q_\star$. The two bins spanning $0.1<q_\star<0.3$ contain $20$ detected companions despite sensitivities of only $50\%$--$60\%$, so correcting for incompleteness strengthens the preference for low-mass companions. In the lowest-$q_\star$ bin, for example, the correction raises the companion fraction from approximately $6\%$ to $\sim14\%$.

The companion excess around HJ hosts is strongest at low mass ratios. Although wide binaries in the field also favor unequal-mass companions, the HJ-host distribution declines more steeply toward higher $q_\star$ \citep{Moe2017,ElBadry2019Twins}.
The normalization, however, is substantially higher around HJ hosts: the field-star companion fraction is $\approx3\%$ in each of the two lowest-$q_\star$ bins, compared with $\sim14\%$ for HJ hosts. At $0.75<q_\star<1$, the two populations have similar companion fractions of $\sim5\%$. 

The {\it observed} companion fraction rises toward wider separations, but this trend largely reflects the survey sensitivity.
The completeness correction is largest in the innermost bin, where it raises the inferred fraction from $2.7\%$ to $9.3\%$.
After correction, all six separation bins have overlapping fractions between $6\%$ and $10\%$. 
We quantify the distribution with the power law
\begin{equation}
    \frac{{\rm d}N}{{\rm d}\log s}\propto s^\pi,
\end{equation}
where $\pi=0$ corresponds to a log-uniform distribution. We find $\pi=0.02\pm0.13$, demonstrating that the projected-separation distribution is consistent with being log-uniform from $50$ to $2{,}000$~au. The inferred HJ-host fraction exceeds the field-star fraction in every separation bin \citep{Moe2017}, indicating that the wide-companion excess spans the full separation range.

\subsection{Projected obliquity}\label{subsec:projected_obliquity}

\begin{figure}
    \centering
    \includegraphics[width=\columnwidth]{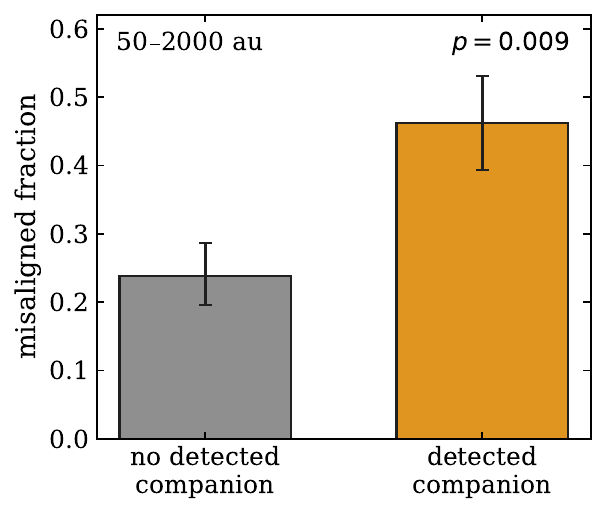}
    \caption{Misaligned fractions for HJ hosts without (left) and with (right) a detected resolved companion at projected separations $50$--$2{,}000$~au. Misalignment is defined as $|\lambda|>10^\circ$ and $|\lambda|-2\sigma_\lambda>0$. HJs with stellar companions have a  $\sim2\times$ larger misaligned fraction ($p=0.009$).}
    \label{fig:misalignment_single_panel}
\end{figure}

Figure~\ref{fig:misalignment_single_panel} compares the fraction of misaligned HJs among systems with and without detected resolved stellar companions at projected separations of $50$--$2000$~au. We consider only HJs with $\sigma_\lambda<50^\circ$ ($142/147$), and classify them as misaligned when $|\lambda| > 10\arcdeg$ and $|\lambda|-2\sigma_\lambda>0$.

Among the $52$ HJs with detected companions in this separation range, $24$ are misaligned, giving a misaligned fraction of $46 \pm 7\%$. Among the $88$ systems without a detected companion in this separation range, $21$ are misaligned, giving $24 \pm 4.5\%$. Companion hosts are therefore $\sim2\times$ more likely to be misaligned. A two-sided Fisher exact test gives $p=0.009$, indicating an association between resolved stellar companions and projected spin--orbit misalignment at the $95\%$ level.

\begin{figure*}
    \centering
     \includegraphics[width=\textwidth]{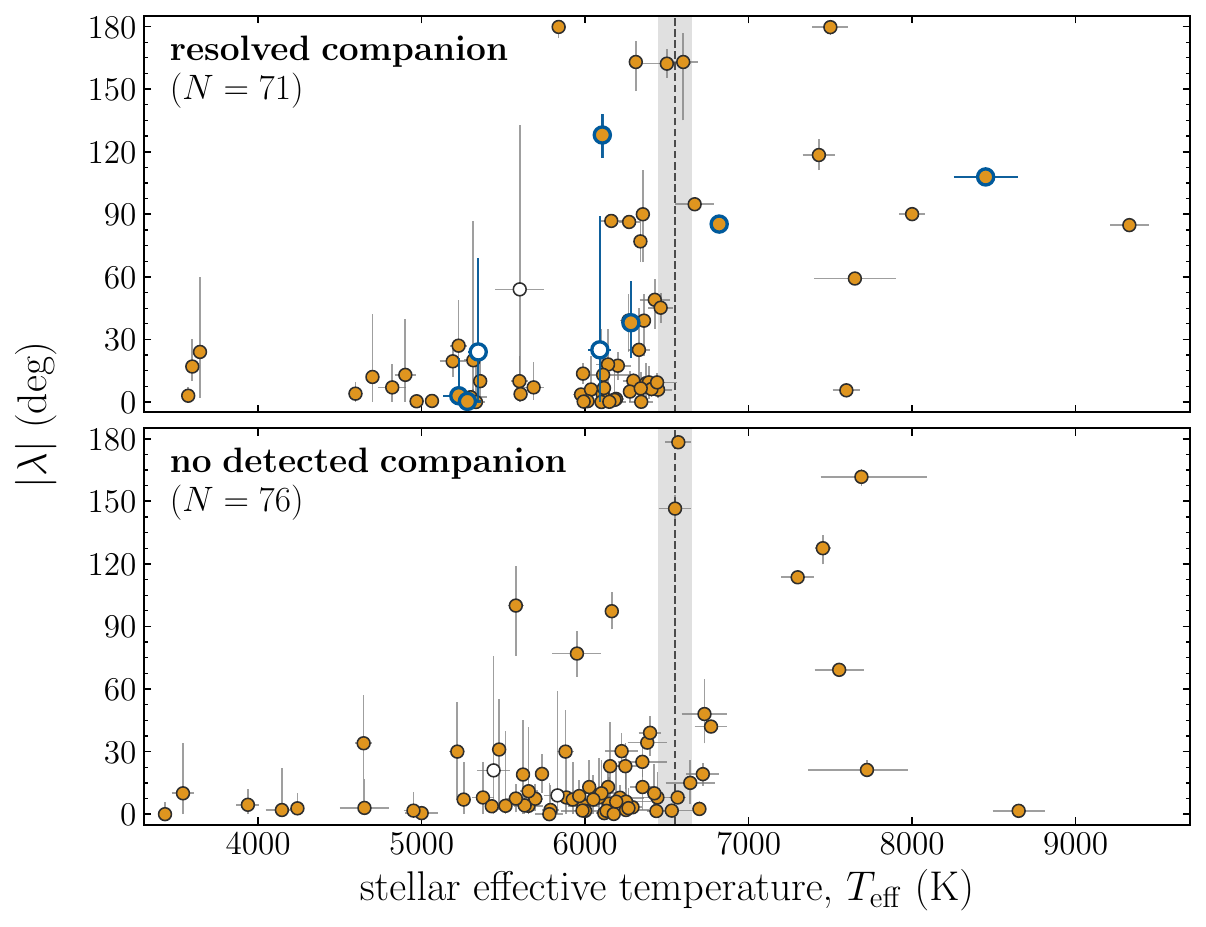}
    \caption{Projected obliquities as a function of host-star temperature for systems with (top) and without (bottom) known resolved companions.  Blue outlines mark the eight PHARO companions identified in this work; gray bands show the empirical transitions from \citet{Beyer2024}. Open circles show systems with poorly measured obliquities ($\sigma_\lambda \geq 50^\circ$). For a version of this plot scaled by the metallicity-dependent Kraft Break, see Figure \ref{fig:teff_lambda_kraft}.
    Four of the new PHARO detections are confidently misaligned.}
    \label{fig:teff_lambda}
\end{figure*}

We highlight that the new PHARO detections provide much of the statistical leverage for this result, explaining why we find stronger evidence for an association than previous works \citep[e.g.,][]{Ngo2015}. 
Before including the eight new detections, we find that the null hypothesis -- i.e., no association between stellar companions and misalignment --  cannot be rejected at the $95\%$ level ($p=0.08$).  Four out of the six new PHARO companions with well-measured obliquities are around misaligned HJs.  The other two systems, CoRoT-19 and HATS-14, have poorly measured $\lambda$ (Figure \ref{fig:teff_lambda}). Thus, the statistical association is driven in large part by the new PHARO detections around securely misaligned systems.

Figure~\ref{fig:teff_lambda} shows projected obliquity against host-star effective temperature for HJs with and without an accepted resolved companion. In both samples, misaligned HJs become more common above the $\sim6500$~K Kraft Break, consistent with previous measurements of the temperature--obliquity relation \citep{Winn2010,Albrecht2012,Knudstrup2024,Wang2026}. On both sides of this transition, the misaligned fraction is higher among companion hosts. Below $6500$~K, $16/57$ companion hosts are misaligned, compared with $9/59$ systems without a detected companion ($28\%$ versus $15\%$; $p=0.116$). Above $6500$~K, the corresponding counts are $10/11$ and $10/15$ ($91\%$ versus $67\%$; $p=0.197$). We discuss the association between companions and stellar obliquity further in Section~\ref{sec:companion_obliquity_discussion}.

The eight new PHARO companion hosts span $T_{\rm eff}=5227$--$8450$~K, with six below the Kraft Break and two above it (Figure~\ref{fig:teff_lambda_kraft}). Four cooler hosts have retained obliquity constraints, of which two are misaligned; both hotter hosts are misaligned. The other two cooler hosts, CoRoT-19 and HATS-14, are excluded from the obliquity analysis. The PHARO detections therefore contribute to the companion--obliquity comparison on both sides of the Kraft Break.

\section{Discussion}
\label{sec:discussion}

\subsection{Comparison with previous surveys}
\label{sec:ngo_subset_results}
Over $50$--$2000$ au, we infer a companion fraction of $52\pm5\%$ (Table \ref{tab:binary_fraction_inference}), consistent with the $47\pm7\%$ measured for $77$ HJ hosts by \citet{Ngo2016}.  \citet{Bohn2020} similarly measured a multiplicity rate of $54.8^{+6.3}_{-9.9}\%$ for $40$ HJ hosts.  
Neither survey had access to {\it Gaia} DR3 astrometry.
Our census expands the HJ host sample to $147$ systems and uses DR3 parallaxes and proper motions to identify bound companions out to $50{,}000$~au.
Within the $s=50$--$2000$~au interval, our companion fraction is $\sim3\times$ larger than the $16.4\pm1.0\%$ field-star fraction from \citet{Raghavan2010}, corresponding to a difference of $\approx6.6\sigma$.  The excess appears across all separations (Figure \ref{fig:intrinsic_q_s_dist}), establishing a large excess of wide stellar companions around HJ hosts.

A complementary \textit{Gaia}-based search by \citet{Gan2026} found co-moving companions around $13/38$ ($34.2\pm9.5\%$) M dwarfs hosting short-period giant planets over projected separations of $100$--$10{,}000$~au, compared with $5.3\pm3.7\%$ for their field sample.  Because this observed fraction was not corrected for \textit{Gaia} incompleteness, we reanalyze their data using our \textit{Gaia} selection function and target-specific likelihood described in Sections~\ref{sec:gaia_selection} and~\ref{sec:binary_fraction_likelihood}. 
For the selection correction, we adopt $q_\star$ distributions from \citet{ElBadry2019Twins} and separation distributions from M dwarf planet hosts \citep{Clark2024}.
We infer an intrinsic companion fraction of $42.8 \pm 9.3\%$. The result ranges from $40\%$--$45\%$ assuming different choices for the intrinsic separation distribution.  Compared over the same separation interval with the $7.2^{+1.7}_{-1.5}\%$ field M-dwarf multiplicity measured by \citet{Ward2015}, this corresponds to an enhancement of $6 \pm 1.8$.
We caution that this comparison is likely metallicity-dependent because planet hosts are typically more metal-rich, but M-dwarf multiplicity decreases toward higher metallicity \citep{Moe2019, Gan2026}.

At face value, the strong excess of wide companions could suggest that they are dynamically important for HJ formation, potentially through high-eccentricity migration driven by the stellar companion.  In this channel, an inclined and eccentric outer star drives the eccentric Kozai--Lidov (EKL) effect, exciting the planet's eccentricity and inclination until tides shrink and circularize its orbit \citep{WuMurray2003,FabryckyTremaine2007,NaozFarrRasio2012,Naoz2016}.  This mechanism naturally connects a wide stellar companion with both the formation of an HJ and a broad distribution of spin--orbit angles \citep{Naoz2011HotJupiters,AndersonStorchLai2016}.  Whether it operates in a particular system, however, depends on the companion's mass and orbit, the mutual inclination, and competing precession from general relativity and additional planets \citep{PuLai2018,Denham2019,Wei2021,Faridani2022}.

Using proto-HJ semimajor axes of $1$--$5$~au and accounting statistically for additional planetary companions, \citet{Ngo2016} estimated that stellar EKL could produce at most $16\pm5\%$ of HJs. They therefore argued that wide binaries may instead trace environments favorable for giant-planet formation, with stellar EKL contributing to only a subset of systems. Because our companion excess is strongest at low mass ratios, it remains important to test whether the measured companions can drive the required eccentricity growth. 
In Section~\ref{sec:ekl_capability}, we revisit this calculation using the measured companion architectures and observationally motivated proto-HJ orbits, finding that stellar EKL {\it is} dynamically feasible in most systems.

\subsection{White dwarf companions}
\label{sec:white_dwarf_companions}

The above demographic results assume only main-sequence companions. In cases where the HJ host began as the initially lower-mass star, it is possible that the companion is now a white dwarf (WD).
We consider WD companions in Appendix~\ref{app:wd_comps}, and find that the inferred companion fraction increases to $69\%$ when considering faint WD companions, and to $71\%$ when WD mass loss and kicks are also included.
In this model, $7\%$ of initial companions become unbound WDs, while another $8.5\%$ remain bound but evade detection because they are too faint or close to the host star.

The evolution of a stellar companion into a WD impacts the dynamical formation of HJs. 
Mass loss expands the outer orbit, while an asymmetric kick changes its semimajor axis, eccentricity, and mutual inclination \citep{Stephan2024,Fuller2026}. 
A surviving binary can therefore be dynamically \emph{rejuvenated}, where an initially dormant system can become EKL-active due to a reshuffling of the orbits and masses \citep{Stephan2024}. A strong kick can also unbind the companion entirely \citep{ElBadry2018,Fuller2026}. 
In the simulations of \citet{Stephan2024}, $162$ HJs formed before the companion became a WD and another $83$ formed only after, \emph{and because of}, the WD kick. Among the systems that formed HJs before the kick, $\sim50\%$ had their companions subsequently unbound. 

Our estimated birth rate $f_{\rm bin, birth} \approx71\%$ (Appendix~\ref{app:wd_comps}) describes the birth companion fraction of HJs in our sample, without considering these dynamics. If WD evolution and kicks increase the probability of forming an HJ \citep[e.g.,][]{Stephan2024}, binaries containing WD progenitors would be overrepresented among HJ hosts relative to our assumed field population, and thus the true birth companion fraction could be larger than reported.

\subsection{Close binary fraction}
\label{sec:close_companions}

Our imaging analysis is sensitive to companions $\gtrsim50$~au and therefore does not measure the close stellar companion population. Detecting companions inside $\sim50$~au generally requires long-term radial velocity monitoring or astrometric accelerations.  Surveys with this sensitivity find a strong \emph{deficit} of close companions around planet hosts.  Combining radial velocities with adaptive-optics imaging, \citet{Ngo2016} inferred that only $3.9^{+4.5}_{-2.0}\%$ of HJ hosts have stellar companions between $1$ and $50$~au, $\sim4\times$ lower than the field-star fraction.  \citet{Kraus2016} similarly detected only $23$ companions with projected separations below $50$ au and $q_\star>0.4$ among Kepler planet hosts, compared with $58$ expected from the field population, a $4.6\sigma$ deficit.  For Kepler gas-giant hosts, \citet{Wang2015Multiplicity} found no companions inside $20$ au, compared with a field-star fraction of $18\pm2\%$.  These independent measurements show that the wide-companion excess around HJ hosts is accompanied by a dearth of companions within tens of au.

The close-binary deficit most likely reflects suppressed planet formation. A nearby stellar companion can truncate or perturb the circumstellar disk, reducing its mass and lifetime and increasing the collision velocities of growing planetary bodies \citep{Kraus2016,MoeKratter2021,Ziegler2026}. Combining several planet surveys, \citet{MoeKratter2021} inferred nearly complete suppression of circumstellar planets at binary separations below $1$~au. At a binary separation of $10$~au, planet occurrence is only $15\pm15\%$ of that around single stars, whereas the suppression becomes negligible beyond approximately $200$~au. Close binaries can therefore inhibit planet formation, while wider companions coexist with giant planets and possibly influence their orbital evolution.

\subsection{Outer planetary companions}
\label{sec:total_companions}

Outer giant planets are also common in HJ systems, although their frequency is uncertain.  In the Friends of Hot Jupiters sample, \citet{Knutson2014} inferred that $51\pm10\%$ of HJs host at least one $1$--$13~{\rm M_{\rm J}}$ companion at 1--20 au. Using the same dataset, \citet{Bryan2016} obtained $70\pm8\%$ with a revised population likelihood.  These estimates rely largely on linear RV trends, making it hard to distinguish these signals from other effects, such as stellar activity. Subsequent observations indeed find RV--activity correlations in HAT-P-4, HAT-P-22, and HAT-P-32 and question the trend in XO-2 \citep[p.~145]{Rosenthal2022Thesis}.
The inferred fraction is also sensitive to the assumed (unknown) stellar jitter, as discussed in \citet{Bryan2016}. Given these substantial uncertainties \citep[see also][]{Li2026}, we adopt the original $51\pm10\%$ estimate below only to perform an illustrative calculation.

By combining our stellar binary fraction with the fraction of outer giant planets, we estimate the fraction of HJs with {\it any} massive outer companion.  If outer giant planets and resolved stellar companions occur independently (a highly simplified assumption, but see \citealt{Ngo2015,Ngo2017}), the fraction hosting at least one of the two is
\begin{equation}
f_{\rm companion}
 = 1-(1-f_{\rm bin})(1-f_{\rm p,out}).
\label{eq:massive_companion_fraction}
\end{equation}
Substituting $f_{\rm bin}=0.620\pm0.049$ (Figure \ref{fig:binary_fraction_comparison}) and $f_{\rm p,out}=0.51\pm0.10$ \citep{Knutson2014} gives 
\begin{equation}
    f_{\rm companion}=0.81\pm0.05.
\end{equation}
The original Friends of Hot Jupiters sample contained four systems with both outer giant and stellar companions, consistent with independent occurrence, and yielded a combined fraction of $72\pm16\%$ using the earlier stellar-companion measurement \citep{Ngo2015}.  Our updated estimate therefore suggests that the {\it vast majority} of HJ systems host a massive outer companion: either a $1$--$13~{\rm M_{\rm J}}$ planet at $1$--$20$ au or a stellar companion with $0.1\leq q_\star\leq1$ at $50$--$50{,}000$ au or both. 

An independent Bayesian analysis of $N=11$ HJs from the California Legacy Survey also found a high outer-planet occurrence. \citet{Zink2023} measured $1.3^{+1.0}_{-0.6}$ outer planets per HJ over $M\sin i=0.3$--$30~{\rm M_{\rm J}}$ and periods of $10$--$40,000$ days. If the number of outer giants follows a Poisson distribution, this corresponds to $f_{\rm p,out}=73^{+17}_{-23}\%$. They measured a statistically consistent rate of $1.0\pm0.3$ outer giants per warm- or cold-Jupiter system, and found that the companions to HJs tend to be more massive and eccentric. They argued that this architecture is consistent with coplanar high-eccentricity migration, in which an eccentric outer planet drives the inner giant to a small periastron before tides circularize its orbit \citep{Petrovich2015}. The high frequencies of both outer planets and stellar companions indicate that most HJs reside in dynamically active systems, although the role these companions played in HJ formation remains uncertain. We investigate this further in Section \ref{sec:ekl_capability}.

\subsection{Companions and spin--orbit misalignment}\label{sec:companion_obliquity_discussion}

\begin{figure*}
    \centering
    \includegraphics[width=\linewidth]{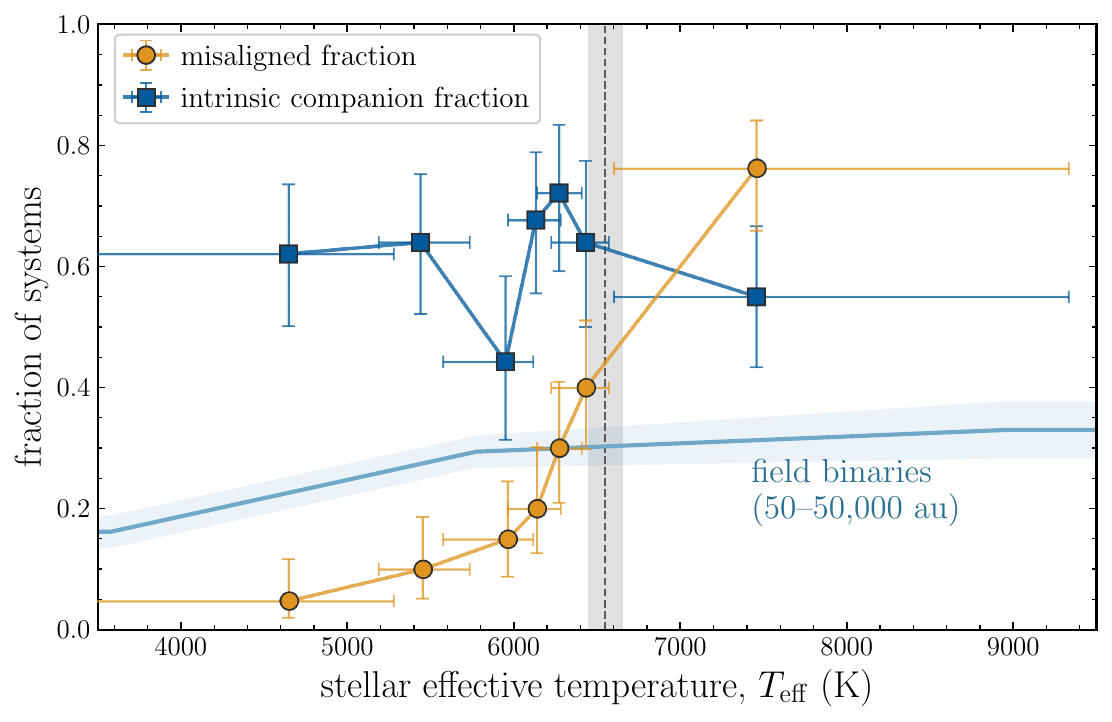}
    \caption{Misalignment fraction as a function of host-star temperature.
    We compare the misaligned fraction (orange), selection-corrected companion fraction (blue), and expected field-star binary fraction (light blue band).  The dashed line and shaded region mark the Kraft Break and its transition width \citep{Beyer2024}. For a version of this plot scaled by the metallicity-dependent Kraft Break, see Figure \ref{fig:teff_fractions_kraft}.
    The misaligned fraction rises steadily with host-star temperature, while the intrinsic companion fraction remains roughly constant across the same temperature range.}
    \label{fig:teff_fractions}
\end{figure*}

HJs with resolved stellar companions at $50$--$2{,}000$~au are nearly twice as likely to be misaligned as those without a known companion (Figure \ref{fig:misalignment_single_panel}): $45\%$ versus $24\%$ ($p=0.009$). The 35-system survey of \citet{Ngo2015} found the same trend at only $1.4\sigma$, leading them to conclude that there was {\it ``no correspondence between hot-Jupiter spin--orbit misalignment and the incidence of directly imaged stellar companions.''} 
Although they recognized that tides could erase the obliquities of cool-star systems, their sample was too small to separate this temperature dependence from an association with stellar companions. Our larger census, including the $6$ new PHARO detections with precise obliquities, reveals a statistically significant association between resolved companions at $50$--$2{,}000$~au and misalignment at $95\%$ confidence.

One possible interpretation is that HJs form through at least two pathways: one that commonly produces misaligned initial orbits and another that generally produces initial alignment. 
Companion-driven high-eccentricity migration (e.g., EKL) excites orbital inclinations and should therefore contribute disproportionately to the misaligned population \citep{FabryckyTremaine2007,NaozFarrRasio2012,AndersonStorchLai2016}.  Systems \emph{without} a detected companion would then be a mixture. Misaligned systems in this group may have an unseen companion or may have lost one during WD formation (Appendix~\ref{app:wd_comps}), while some aligned systems may have formed through pathways that generally predict initial alignment, such as disk migration \citep{Lin1996,DawsonJohnson2018,Polanski2026}. 
Other aligned systems may have undergone dynamical migration but were subsequently realigned by tides, particularly around cool stars \citep{Lai2012Tides,LinOgilvie2017,AndersonWinnPenev2021}.
This mixture could explain why the sample without detected companions below the Kraft Break is not completely aligned but has a lower misalignment fraction than the companion sample. 
Our present-day companion fraction of $62\pm5\%$ over $0.1\leq q_\star\leq1$ and $50$--$50{,}000$~au leaves room for multiple pathways, although it predicts that companion-driven EKL migration could be the dominant mechanism for HJ formation (Section \ref{sec:ekl_capability}).

Figure~\ref{fig:teff_fractions} shows that the misaligned fraction increases strongly with $T_{\rm eff}$, whereas the companion fraction shows no obvious temperature dependence.  The misaligned fraction rises from $5\%$ in the coolest bin to $40\%$ right below the Kraft Break to $76\%$ in the hottest bin. Across the same temperature bins, the inferred companion fraction does \emph{not} show a significant trend. Across the temperature range, the intrinsic companion fraction remains consistently $3$--$4\times$ elevated relative to the field-star fraction.
Thus, the increase in misalignment toward the Kraft Break is not caused by a higher incidence of stellar companions around hotter hosts.
Instead, these trends are consistent with companions exciting a broad initial obliquity distribution, followed by tidal realignment that preferentially damps misalignments around cooler stars \citep[e.g.,][for a recent discussion]{Rice2022}. We discuss this interpretation further in Section~\ref{sec:tides_kraft_break}.

Stellar companions can produce a broad initial HJ obliquity distribution through several mechanisms.
Before disk dispersal, an inclined companion can torque the protoplanetary disk, tilting the planetary orbits relative to the stellar spin while leaving them mutually aligned \citep{Batygin2012,Spalding2014,Lai2018}.
Subsequent disk migration could then potentially produce an HJ.
After disk dispersal, a companion can excite a planet's eccentricity and inclination through the eccentric Kozai--Lidov mechanism, allowing tides to shrink and circularize the orbit \citep{FabryckyTremaine2007,NaozFarrRasio2012,AndersonStorchLai2016,MuNozLaiLiu2016}.
These processes can also act together. Disk-induced misalignment followed by companion-driven high-eccentricity migration can produce an obliquity distribution with a broad peak near $90^\circ$ \citep{Vick2023}.
The evidence for such a polar peak among misaligned systems is not presently observed in the data \citep{Dong2023}.
After an HJ has already formed, an inclined stellar companion can change the obliquity by making its orbital axis precess faster than the stellar spin can follow \citep{Lai2018}.

A stellar companion could also trigger planet--planet scattering or secular chaos, which can tilt planetary orbits even without a stellar companion \citep{RasioFord1996,Wu2011,DawsonJohnson2018}.
The higher misaligned fraction among HJs with stellar companions could therefore reflect either direct EKL excitation or interactions between planets triggered by a companion.
Our current data cannot distinguish these possibilities.

Stellar binaries hosting aligned and misaligned HJs
overlap in companion mass ratio and projected separation (Figure~\ref{fig:ekl_q_separation}).
Their median $q_\star$ values are $0.484$ and $0.347$, and their median separations are $773$ and $573$~au, respectively. 
Systems with and without resolved companions also have similar planet-to-star mass ratios ($9.52\times10^{-4}$ versus $1.05\times10^{-3}$). Companion hosts have smaller $a/R_\star$,
which should strengthen tidal realignment rather than explain their higher misaligned fraction (Appendix~\ref{app:planet_mass_ratio_obliquity}).
Thus, while excess misalignment is associated with resolved companions, there is not a measurable difference in system architecture: companion mass, separation, or planet mass.
Stellar companion eccentricities and mutual inclinations remain largely unknown, and future astrometric monitoring will be required to constrain their orbits and dynamical implications.

\subsection{Companions across the Kraft Break}\label{sec:companions_kraft_break}

The Kraft Break is a transition from slowly rotating cooler stars to rapidly rotating hotter stars near mid-F spectral types.  \citet{Beyer2024} show that this transition is sharp, with a center at $6550$~K and a width of only $200$~K, as shown by the gray band in Figure~\ref{fig:teff_lambda}. 
Above the break, stellar convective envelopes become thin or disappear, weakening magnetic braking and allowing hotter stars to retain faster rotation. 
Early obliquity studies found that HJ misalignment depends on the host star's temperature.  \citet{Schlaufman2010} find ten systems with possible line-of-sight misalignment, all around stars above the Kraft Break (masses $1.2$--$1.5~{\rm M_\odot}$), and \citet{Winn2010} showed that large sky-projected obliquities occurred preferentially above $T_{\rm eff}\approx6250$~K.  Subsequent measurements confirmed that cool HJ hosts are predominantly aligned, whereas hot hosts span a broad range of obliquities \citep{Albrecht2012,AlbrechtDawsonWinn2022,WangWangBatygin2026}.

This temperature dependence motivated models in which HJs are initially misaligned around both cool and hot stars, but tides preferentially realign the cool-star systems.
In proposed tidal models, deeper convective envelopes in cool stars allow more efficient obliquity damping, while the thin or absent convective envelopes of hot stars preserve more of the obliquity produced during formation or migration \citep[][but see also the latter two works for a resonance locking mechanism]{Lai2012Tides,RogersLin2013,Xue2014,LiWinn2016,LinOgilvie2017,AndersonWinnPenev2021,SpaldingWinn2022,Zanazzi2024, Zanazzi2025}. Using homogeneous single-star samples, \citet{Wang2026} measured a rotation break at $6510^{+97}_{-127}$~K and an HJ obliquity transition at $6447^{+85}_{-119}$~K.  The agreement between these temperatures and the $6550$~K break measured by \citet{Beyer2024} motivates a division at $6500$~K associated with the stellar Kraft Break\footnote{Adopting instead the $6105$~K obliquity transition reported by \citet{Wang2026} for binaries, companion hosts remain more frequently misaligned on both sides of the division}.

On both sides of the rotational Kraft Break, HJ hosts with resolved companions have a higher misaligned fraction (e.g., Figure \ref{fig:teff_lambda}), although neither discrepancy is statistically significant at the current sample size.  
Among stars cooler than $6500$~K, $15/56$ HJs with companions are misaligned, compared with only $10/60$ systems without a known companion ($27\%$ versus $17\%$; $p=0.259$).
Above 6500~K, the corresponding counts are $10/11$ and $10/15$ ($91\%$ versus $67\%$; $p=0.197$), again showing the same trend.

Because tides can erase misalignment excited by companions, differences in planet mass or orbital separation should contribute to the observed association \citep{Albrecht2012,Lai2012Tides,AndersonWinnPenev2021,SpaldingWinn2022}. We examine this possibility in Appendix~\ref{app:planet_mass_ratio_obliquity} (Figure~\ref{fig:planet_tidal_parameters}). Below each host's metallicity-dependent Kraft break, misaligned HJs tend to have lower planet-to-star mass ratios, consistent with weaker tidal realignment by lower-mass planets \citep{AlbrechtDawsonWinn2022,Rusznak2025}. However, this trend is driven by the $N=4$ misaligned systems without detected companions. HJs in binaries also have smaller $a/R_\star$ on average, which should strengthen tidal realignment and produce more alignment, opposite to the observed trend. Thus, the full cool-star sample shows the expected tendency for higher-mass planets to be aligned, but neither $q_p$ nor $a/R_\star$ alone explains why companion hosts are more often misaligned.

\subsection{Tidal realignment across the Kraft Break}\label{sec:tides_kraft_break}

The misaligned fraction increases gradually toward hotter hosts, whereas the companion fraction remains high without a comparable temperature dependence (Figure~\ref{fig:teff_fractions}). This behavior is consistent with stellar companions contributing a broad \textit{initial} obliquity distribution and tides preferentially erasing misalignments around cool stars \citep[e.g.,][]{Rice2022,SpaldingWinn2022}. 
A gradual transition is not surprising. 
The rotational Kraft break marks a sharp change in magnetic braking \citep{Matt2015, Beyer2024}, whereas the obliquity distribution is shaped by tidal dissipation and could have a different sensitivity to $T_{\rm eff}$. Because the relation between stellar structure and $T_{\rm eff}$ also depends on metallicity, any corresponding obliquity transition should occur at different $T_{\rm eff}$ for different hosts \citep{SpaldingWinn2022}. 
In Appendix~\ref{app:kraft_break}, we show that the increase in misalignment persists when each host is considered relative to its metallicity-dependent Kraft break.

We compare the observed misalignment-$T_{\rm eff}$ trend with predictions from five tidal models. 
In the equilibrium-tide model, turbulent convection dissipates the lagging tidal bulge more efficiently in cool stars with substantial convective envelopes, but the same torque damps the obliquity and shrinks the orbit on comparable timescales \citep{Zahn1977,Barker2009,Vidal2020}. In the inertial-wave model, an obliquity-specific component of the tidal forcing excites waves in the stellar convective envelope. Dissipating these waves changes the stellar spin direction without directly shrinking the orbit \citep{Lai2012Tides}. In the frequency-dependent-$Q$ model, tidal dissipation becomes much weaker at high forcing frequencies, allowing the star to realign after only moderate orbital decay before further tidal evolution slows \citep{Penev2018}. The decoupled-envelope model assumes that tides reorient only a low-inertia outer layer rather than the whole star, so less orbital angular momentum is needed for alignment \citep{Winn2010,Dawson2014}. In the resonance-locking model, the tidal forcing is coupled to the evolving stellar gravity modes. Because these modes evolve more rapidly in the hydrogen-burning radiative cores of cool stars, their obliquities are damped more strongly than those of hot stars \citep{Witte1999,Witte2001,Burkart2014,Fuller2017,Ma2021,Zanazzi2024,Zanazzi2025}.

\begin{figure}
    \centering
    \includegraphics[width=0.82\linewidth]{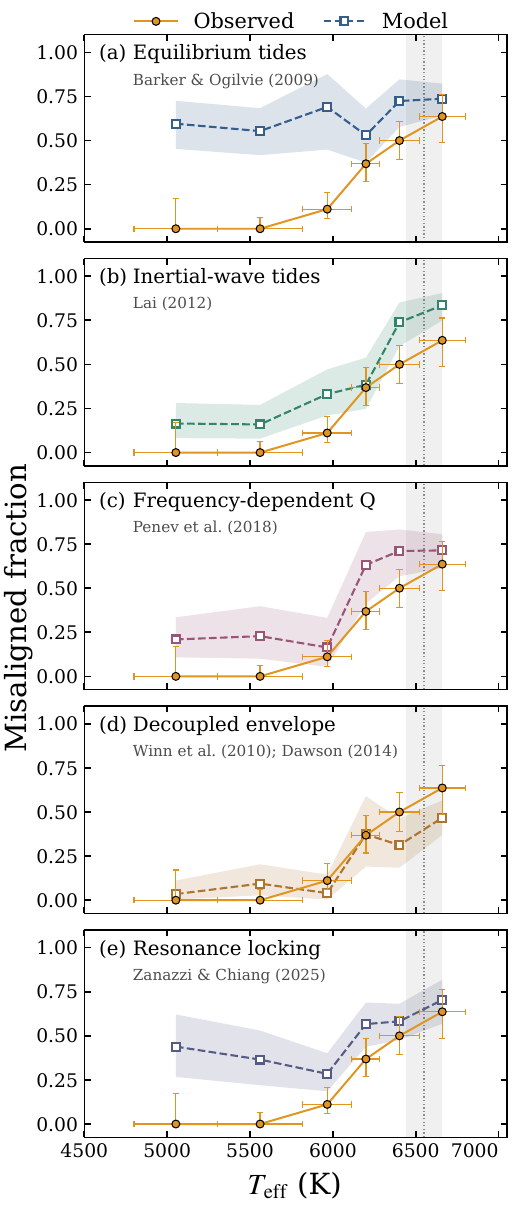}
    \caption{
    Misaligned fractions for observed HJs (orange circles) and predictions for different tidal models (open squares). The tidal models show population synthesis simulations from \citet{AlbrechtDawsonWinn2022} in panels (a)--(d) and \citet{Zanazzi2025} in panel (e).
    For the simulations, we show the median and central $68\%$ interval after incorporating uncertainties due to measurement and sample counts. The dotted line and gray band mark the rotational Kraft Break \citep{Beyer2024}.}
\label{fig:tidal_models}
\end{figure}

Figure \ref{fig:tidal_models} compares the observed rise in misalignment toward hotter stars with predictions from the above tidal models. 
For panels (a)--(d), we use the results from the toy population syntheses performed by \citet{AlbrechtDawsonWinn2022}, described in their Appendix A. Panel (e) uses the resonance-locking population from Figure~7 of \citet{Zanazzi2025}, who consider both axisymmetric and non-axisymmetric gravity modes. 
In every panel, we compare the model with the same $88$ HJ hosts satisfying $4800\leq T_{\rm eff}/{\rm K}\leq6800$, $0.5<M_p/M_{\rm J}\leq15$, and $a/R_\star<10$.

To compare the simulations with our observations, we bin both into the same temperature ranges and apply the misalignment criterion, $|\lambda|>10\arcdeg$ and $|\lambda|>2\sigma_\lambda$. For each simulated system, we assign a measurement error by drawing from uncertainties of the observed hosts in the same bin.
Repeating this $100$ times produces the model line and band in Figure \ref{fig:tidal_models}, which show the median and central $68\%$ interval after including measurement and finite-sample (binomial) uncertainty.

Most tidal models broadly reproduce the observed trend of higher misalignment towards hotter stellar hosts.
The equilibrium-tide population, however, retains a much larger misaligned fraction around cool stars than observed. The other four populations produce a stronger rise toward hotter hosts and are qualitatively closer to the data, although several still overpredict the cool-star fractions and none reproduces every bin. 
These comparisons cannot yet identify a preferred tidal model because the simulations start from different populations and use different parameters. Different tidal models can also reproduce the HJ period distribution, depending on how and when HJs form \citep{Ma2026}. A stronger test would use the same initial population and compare the predicted obliquities, stellar rotation rates, planet masses, and orbital separations with observations. Overall, the models support a picture in which tides erase misalignments more efficiently around cooler stars.

\subsection{Binarity and obliquity across planet populations}

Warm Jupiters (WJs) provide a useful comparison because their wider orbits produce weaker tidal realignment, so their present-day obliquities preserve their migration histories.
Among systems without known stellar companions, \citet{Wang2024WarmJupiters} found all $23$ WJs to be aligned, including all eight around stars hotter than 6100~K, a $3.4\sigma$ difference from HJs around similarly hot stars.
\citet{EspinozaRetamal2025} likewise inferred that $95\%$ of single-star WJs have true obliquities $\psi\lesssim30\arcdeg$. This alignment contrasts with the high misalignment rate among HJs with resolved companions in our sample.
Only a few WJs have hosts above the 6500~K Kraft Break, however, so their obliquities around hot stars remain poorly tested \citep{Wang2026}.

The orbital properties and relative occurrence of WJs compared to HJs provide another test for HJ formation channels. In the simulations of \citet{Petrovich2016}, stellar perturbers produce fewer WJs relative to HJs than planetary perturbers, making the warm-to-hot ratio an additional test of the migration mechanism. Related simulations reproduce the cold-Jupiter eccentricity
distribution through stellar EKL when the initial orbits
are modestly eccentric \citep{WeldonNaozHansen2025}.  Planet--planet scattering also predicts misaligned WJs, especially beyond $\sim0.3$~au \citep{Esposito2026}. Comparing WJ occurrence, eccentricities, and obliquities in systems with and without stellar companions would therefore help distinguish stellar-driven migration from planet--planet interactions.

Like HJs, sub-Saturns are close-in gas-rich planets, but their lower masses $0.05$--$0.3~{\rm M_{\rm J}}$ make tidal realignment of the host star less efficient.
In the single-star sample of \citet{Radzom2024}, $0/5$ sub-Saturns in compact multiplanet systems were misaligned, compared with $9/17$ apparently isolated planetary systems, a $2.6\sigma$ difference. 
Using a mass-dependent boundary in scaled orbital separation, \citet{WangWang2026WarmSubSaturns} similarly found that warm sub-Saturns are aligned, whereas close-in (``hot'') sub-Saturns are frequently misaligned ($3.2\sigma$).
This mirrors the obliquity difference between warm and hot Jupiters.
Several close-in sub-Saturns remain highly misaligned around cool stars, where their low masses make tidal realignment of the host inefficient \citep{Radzom2024,Dugan2025}. Compact systems containing smaller planets are also preferentially aligned \citep{Albrecht2013,Morton2014,Louden2021,Louden2024, Handley2026}. Together, these results associate sub-Saturn misalignment with close-in, apparently isolated planetary architectures rather than universal primordial misalignment. However, the binary classifications account only for known companions and are often heterogeneous. 

Whether stellar companions are similarly associated with misalignment among warm Jupiters and sub-Saturns remains unknown. Existing studies distinguish systems with and without known companions, but do not correct for the selection function. Moreover, planets in known binaries include both aligned and misaligned examples \citep{WuMurray2003,Hjorth2021,Gupta2024,Veldhuis2026}, so individual systems do not establish a population-wide association.  Our HJ census does reveal such an association between obliquity and companions (Figure \ref{fig:misalignment_single_panel}). A comparable companion census of warm Jupiters and sub-Saturns is needed to determine whether this result extends beyond HJs.

\section{Can eccentric Kozai-Lidov migration explain the observed HJ population?}
\label{sec:ekl_capability}

\begin{figure*}
    \centering
    \includegraphics[width=0.95\linewidth]{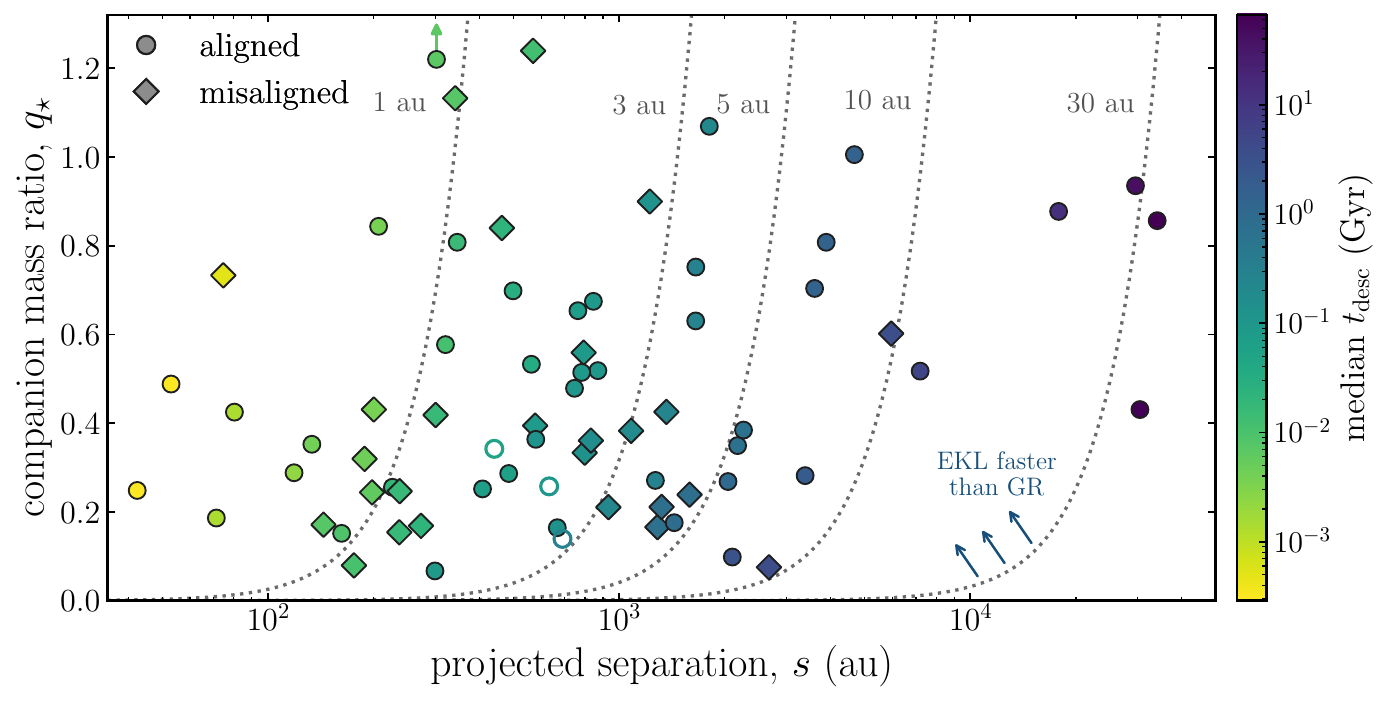}
    \caption{Capacity for eccentric Kozai--Lidov (EKL) oscillations among proto-HJs.
    We show the companion mass ratio and projected separation for the northern HJ sample. Circles and diamonds denote aligned and misaligned HJs, respectively. Color denotes the median descent time to the planet's present-day pericenter after marginalizing over orbital configurations. Hollow points mark systems with poorly measured obliquities ($\sigma_\lambda \geq 50^\circ$). 
    TOI-3714 ($q_\star = 2.02$) is shown at its measured separation with its mass ratio displaced downward.
    Gray curves mark where the quadrupole EKL timescale is comparable to the initial general-relativistic apsidal precession rate ($t_{\rm quad}\dot{\omega}_{\rm GR}=1$) for the labeled progenitor semimajor axes. Systems above and to the left of a curve can undergo EKL oscillations. Most observed stellar companions can drive EKL-induced migration if the proto-HJ began beyond $\sim3$--$5$ au.}
    \label{fig:ekl_q_separation}
\end{figure*}

A natural interpretation for the elevated stellar companion rate of HJs is that these companions played an active dynamical role in HJ formation, possibly by driving high-eccentricity migration.
In the eccentric Kozai--Lidov (EKL) mechanism, an inclined stellar companion exchanges angular momentum with the planetary orbit, driving eccentricity--inclination oscillations. Tides can then shrink and circularize the orbit during repeated close passages to the host star \citep{Kozai1962,Lidov1962,FabryckyTremaine2007,NaozFarrRasio2012,Petrovich2015, Naoz2016,LiuY2026}. General-relativistic apsidal precession can suppress this eccentricity growth when it is faster than the secular precession induced by the companion. 

An observable prediction of high-eccentricity migration -- specifically stellar EKL migration followed by tidal dissipation -- is a transient population of eccentric giant planets caught before their orbits have fully circularized \citep[e.g.,][]{Socrates2012}. Several candidate proto-HJs have now been identified on or near plausible tidal migration tracks, including several with stellar companions \citep{WuMurray2003,Santerne2014,Masuda2017,Dong2021,EspinozaRetamal2023,Gupta2024,Heidari2025,Bieryla2025,Thomas2026,Liveoak2026}.  Their existence demonstrates that this pathway operates, although it remains unclear whether their abundance is consistent with high-eccentricity migration models \citep[e.g.,][]{Dawson2015,Jackson2023}.

\subsection{EKL capability of the observed companions}
\label{subsec:ekl_system_capability}

We test whether the resolved stellar companions in our sample can drive HJ formation through EKL migration using two complementary calculations. First, we calculate regions where the quadrupole-level EKL precession timescale ($t_{\rm quad}$) is shorter than the initial GR apsidal-precession rate ($\dot{\omega}_{\rm GR}$): {\it i.e.,} EKL oscillations can be present when $t_{\rm quad}\dot{\omega}_{\rm GR}\lesssim1$ \citep{FabryckyTremaine2007}. We calculate this for several characteristic proto-HJ semimajor axes, shown with the gray curves in Figure~\ref{fig:ekl_q_separation}. 

Second, for each system, we estimate the EKL descent time \citep[$t_{\rm desc}$;][]{Weldon2024}: the time required for a giant planet on an initially wide orbit to reach a pericenter equal to the present-day pericenter of the HJ. Reaching this pericenter marks entry into the high-eccentricity migration regime, assuming subsequent tidal circularization is efficient, and the planet survives \citep[see, e.g.,][for a discussion]{Gupta2024,Liveoak2026,Weldon2026}.
For each system, we draw $10^4$ possible initial configurations while keeping the measured host star mass, planet mass, and companion masses fixed. Proto-HJ semimajor axes and eccentricities follow the mass-dependent distributions measured by the California Legacy Survey \citep{Fulton2021,VanZandt2026,Blunt2026}\footnote{\citet{Ngo2017} found no significant differences in the masses, eccentricities, or periods of giant planets in single- and multi-stellar systems from $0.1$--$5$~au, supporting the use of the same cold-Jupiter distributions here.}.
Stellar-companion eccentricities follow the observed separation-dependent distribution of {\it Gaia} wide binaries \citep{HwangTingZakamska2022}, which is roughly uniform at $\sim100$~au, thermal at $\sim300$~au, and super-thermal above $\sim1000$~au. Mutual inclinations are isotropic \citep[consistent with wide triples;][]{Tokovinin2022,Shariat2025}, and the measured projected separations are deprojected to orbital semi-major axes assuming random orbital phases and viewing orientations. 
We retain configurations that are dynamically stable, overcome initial GR precession, and fall within the calibrated range of the analytic expression. 
For each {\it accepted} initial configuration, we calculate $t_{\rm desc}$. 
Appendix~\ref{app:ekl_method} provides more details on this calculation.

Only $16\%$ of draws satisfy the criteria for the analytic $t_{\rm desc}$ expression, which requires $i_{\rm mut}>50\arcdeg$ \citep{Weldon2024}. The reported medians therefore measure dynamical capability {\it among EKL-favorable configurations}, not among the entire wide binary population. However, because HJs are rare \citep[$\sim3\%$ as common as cold Jupiters;][]{Petigura2018,VanZandt2026}, the low accepted fraction does not by itself rule out a substantial contribution from stellar EKL to the HJ population. We evaluate this separately in Section~\ref{subsec:ekl_population}.

Figure~\ref{fig:ekl_q_separation} summarizes both calculations, showing the capacity for HJs in wide binaries to undergo EKL migration. 
Assuming $e_{\rm out}=0.5$ and $a_{\rm out}=s$, $45/71$, $56/71$, and $64/71$ companions can undergo EKL oscillations ($t_{\rm KL}\dot{\omega}_{\rm GR}<1$) for proto-HJs at $3$, $5$, and $10$~au, respectively. If stellar companions are preferentially eccentric with $e_{\rm out}=0.9$, the corresponding counts increase to $56/71$, $63/71$, and $67/71$. A substantial fraction ($\sim65\%$) of cold Jupiters reside at $\gtrsim3$~au separations \citep{Fulton2021,VanZandt2026}.
Correspondingly, we conclude that most measured companions can induce EKL in the proto-HJ\footnote{White dwarf (WD) companions, which are challenging to observe in magnitude-limited imaging surveys, could further boost HJ production via EKL (see discussion in Section \ref{sec:white_dwarf_companions}).}.

\begin{figure}
    \centering
    \includegraphics[width=0.9\linewidth]{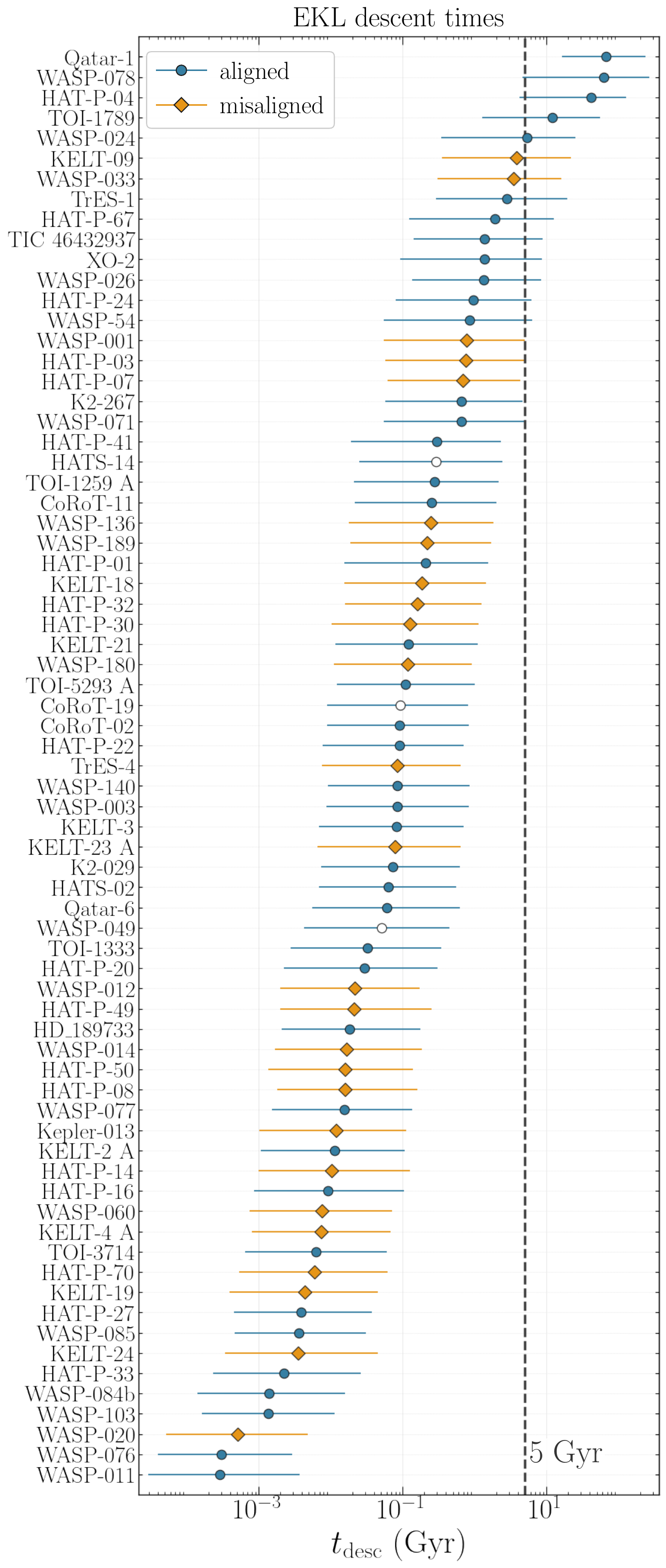}
    \caption{EKL descent times ($t_{\rm desc}$) for the $71$ HJs with resolved stellar companions. 
    Here, $t_{\rm desc}$ is the time required for octupole-level EKL evolution to decrease the proto-HJ pericenter to the observed present-day orbit \citep{Weldon2024}. 
    Markers show the median and 16th--84th percentile range for aligned (circles), misaligned (diamonds), and ambiguous (hollow) HJs. Systems are ordered by median $t_{\rm desc}$, and the dashed line marks 5~Gyr. The initial proto-HJ semi-major axis is sampled from the mass-dependent occurrence law of giant planets \citep{VanZandt2026}.
    Nearly all observed stellar companions are capable of driving high-eccentricity evolution via EKL.}
    \label{fig:ekl_tdesc_by_system}
\end{figure}

Figure~\ref{fig:ekl_tdesc_by_system} shows the descent-time distributions for the individual systems. The median $t_{\rm desc}$ is below 3~Gyr for $64/71$ systems ($90\%$), and even the {\it 84th percentile} is below 10~Gyr for $62/71$ ($87\%$). Thus, most measured companions both overcome GR for plausible proto-HJ orbits and admit stable configurations with EKL strong enough to reach the observed planetary pericenter within a few~Gyr\footnote{These ages are plausible for HJ hosts, whose vertical velocity dispersions (a kinematic age indicator) are statistically indistinguishable from that of matched field stars \citep{Hu2026}.}. Applying the same calculation to the M dwarfs with short-period giant planets from \citet{Gan2026}, we find a median $t_{\rm desc}<10$~Gyr for all $13$ systems. 
Thus, most measured companions admit EKL oscillations that can reach the observed planetary pericenter within $10$~Gyr across both FGK and M-dwarf hosts.

The four systems with median $t_{\rm desc}>10$~Gyr -- HAT-P-4, WASP-78, Qatar-1, and TOI-1789 -- have the widest companions in our sample, at projected separations of $18{,}000$--$34{,}000$~au. At these separations, EKL forcing is weak for most companion orbits. Galactic tides and stellar flybys, however, can change the eccentricity and orientation of such wide binary orbits and move them into configurations that produce stronger EKL evolution \citep[e.g.,][]{Rodet2021,Kontiainen2026,Grishin2025}. Our calculation may therefore
underestimate the influence of the widest companions.
Additional companions can also change the migration pathway: in HAT-P-7, a wide stellar companion can excite the eccentricity of a closer stellar companion, whose repeated encounters with the planet can trigger high-eccentricity migration \citep{Yang2025}.
Lastly, we remark that the similar descent times of aligned and misaligned systems in our FGK sample suggest that present-day obliquity is largely determined by subsequent tidal evolution \citep[e.g.,][]{Winn2010}.

\subsection{Contribution to the HJ population}
\label{subsec:ekl_population}

To test whether EKL can produce HJs at the observed rate, we perform a simple population synthesis calculation.
We draw cold-Jupiter masses and semimajor axes from the occurrence grid of \citet{VanZandt2026} over $0.8$--$13\,{\rm M_{\rm J}}$ and $1$--$30$~au. Stellar companion mass ratios follow \citet{ElBadry2019Twins}, binary semimajor axes follow the broken separation law in Equation~\eqref{eq:EB_weight_s}, outer eccentricities follow \citet{HwangTingZakamska2022}, and mutual inclinations are isotropic. 
We retain only systems that satisfy the stability criterion of \citet{Mardling2001} and $\epsilon_{\rm oct}<0.1$ \citep{Naoz2013Secular,MuNozLaiLiu2016}.
For each stable system, we use Equation~19 of \citet{MuNozLaiLiu2016} to determine whether its sampled inclination can drive the planet to a specified pericenter. For a pericenter threshold $r$, we set $e_{\rm req}=1-r/a_p$ and calculate the critical inclination required to reach $e_{\rm req}$. This calculation combines quadrupole and octupole torques and includes the limits imposed by GR and planetary tidal-bulge precession \citep{MuNozLaiLiu2016}. 
We repeat this calculation for three pericenter thresholds from \citet{Weldon2026}: migration, mass loss, and complete disruption.
Planets that reach the migration threshold but not the mass-loss threshold form intact HJs; those that cross the mass-loss threshold but avoid complete disruption form stripped HJs. We count both outcomes as successful HJ formation \citep[e.g.,][]{Weldon2026}.

Ignoring additional planets, which can strengthen \emph{or} suppress stellar EKL \citep{Takeda2008,PuLai2018,Denham2019}, this calculation converts $f_{\rm EKL}=14.4\%$ of cold Jupiters in wide binaries into HJs. 
For comparison, the detailed numerical integrations of \citet{Weldon2026} provide $f_{\rm EKL}=12.2\%$. 
The corresponding initial architectures and HJ-forming subset are shown in Figure~\ref{fig:ekl_simulations_initial_corner}.

Interestingly, surviving HJ progenitors are drawn from a wide range of initial planetary semimajor axes $a_{\rm in}$. This is because larger values of $a_{\rm in}$ increase the EKL torque, allowing more planets to reach extreme eccentricities and smaller pericenters \citep{AndersonStorchLai2016,MuNozLaiLiu2016}. However, the stronger torque also increases the fraction that become completely tidally disrupted, so surviving HJ progenitors are not preferentially drawn from the largest $a_{\rm in}$ values, consistent with previous results \citep[e.g.,][]{MuNozLaiLiu2016}\footnote{For a more detailed treatment of HJ migration with mass loss, see \citet{Weldon2026}.}.

We convert the EKL efficiencies from the simulations into an occurrence per FGK star using 
\begin{equation}
    \eta_{\rm HJ}=\eta_{\rm CJ}f_{\rm wide}f_{\rm EKL},
\end{equation}
where $\eta_{\rm CJ}=20\%$ is the cold-Jupiter occurrence rate and $f_{\rm wide}=23\%$ is the wide-binary fraction \citep{Raghavan2010,VanZandt2026}. This gives an HJ occurrence of $0.56\%$--$0.66\%$, comparable to the observed occurrence of $0.5$--$1\%$ \citep{Wright2012,Petigura2018}.

We emphasize that just because EKL can be a dominant contributor to the HJ population, it does not need to operate in most cold-Jupiter systems. HJs are $\approx10$--$30$ times rarer than cold Jupiters \citep[e.g.,][]{Petigura2018,VanZandt2026}, so their population can be supplied only by the subset of systems with a favorable combination of initial planetary orbit and stellar-companion orbit (shown in Figure \ref{fig:ekl_simulations_initial_corner}). 
In our model, successful HJ progenitors span the entire $1$--$30$~au range of initial orbits, with $61\%$ beginning beyond $3$~au.
Successful systems preferentially have initially eccentric outer orbits and large mutual inclinations (Figure \ref{fig:ekl_simulations_initial_corner} and Appendix~\ref{app:ekl_method}). 
Although these constraints apply to the initial orbits because we do not perform numerical integrations, they also make predictions for the systems observed today. The outer eccentricity should change little because the outer orbit carries most of the angular momentum. The final mutual-inclination distribution should also be broad, and possibly bimodal, if initial inclinations are drawn isotropically \citep{FabryckyTremaine2007,NaozFarrRasio2012, WeldonNaozHansen2025}.
Together, these results show that the observed stellar companions could produce most HJs through EKL, even if only a minority of cold Jupiters undergo migration.

\subsection{Comparison with Ngo et al. (2016)} \label{subsec:ekl_ngo_comparison}

Our conclusion that stellar EKL is dynamically feasible
for nearly all HJs with resolved companions differs from the interpretation of \citet{Ngo2016}.
Using the measured stellar, planetary, and companion properties, they similarly compared $t_{\rm quad}$ and $t_{\rm GR}$ while adopting $e_{\rm out}=0.5$ and broad mutual inclinations. 
They found capable companions around $16\pm6\%$, $34\pm7\%$, and $47\pm7\%$ of all HJ hosts for proto-HJs beginning at $1$, $2.5$, and $5$~au, respectively. The $5$~au result is roughly their inferred wide-companion fraction, indicating that essentially their full companion population could satisfy the GR criterion at that initial separation. 
Their headline estimate of $16\pm5\%$ followed only after weighting progenitors from $1$--$5$~au with the \citet{Cumming2008} cold-Jupiter occurrence law, which gave $32\pm7\%$, and multiplying by $0.49$ under the assumption that every additional long-period giant planet suppresses stellar KL evolution \citep{Knutson2014}.

Our analysis differs from \citet{Ngo2016} in three main respects. First, they restricted proto-HJs to $1$--$5$~au and weighted them using the \citet{Cumming2008} occurrence law, which was fitted only to $\approx3$~au.
Recent RV surveys show that cold Jupiters commonly reside at $1$--$10$~au, with $\sim50\%$ at $5$--$10$~au and a tail extending to $30$~au \citep{Fulton2021,VanZandt2026}. For these wider progenitors, EKL torques from the stellar companion are stronger. 
Thus, by excluding the $>5$~au population, \citet{Ngo2016} underestimated the fraction of EKL-capable HJ progenitors. 
In our sample, $56/71$ and $64/71$ companions satisfy $t_{\rm quad}\dot{\omega_{\rm GR}}<1$ for proto-HJs at $5$~au and $10$~au, respectively, adopting $e_{\rm out}=0.5$.

Second, their calculation assessed only the initial $t_{\rm quad}\dot{\omega_{\rm GR}}<1$ criterion, whereas we sample orbital configurations and calculate the octupole-level descent time to the present-day planetary orbit \citep{Weldon2024}. 
Among suitable configurations, the median $t_{\rm desc}<10$~Gyr in $95\%$ of systems, showing that the required eccentricity growth for EKL-driven migration can occur within the system lifetime for nearly all HJs with companions. 
Third, \citet{Ngo2016} assumed that long-period giant planets universally suppress stellar EKL. \citet{Knutson2014} inferred that $51\pm10\%$ of HJs host a $1$--$13~{\rm M_{\rm J}}$ companion at $1$--$20$~au\footnote{
The inferred outer-giant fraction remains uncertain because $11/14$ reported FOHJ companions were constrained only as linear RV trends, which can also arise from stellar activity \citep[e.g.,][]{Lovis2011,Costes2021}. Follow-up measurements find that three systems also show strong correlations between their RVs and Ca~II H\&K activity indicators \citep[p.~145]{Rosenthal2022Thesis}.}, hence \citet{Ngo2016} reduced the fraction of EKL-capable systems by $\sim50\%$. We do not apply this universal EKL suppression because the effect of an additional planet depends sensitively on its mass and orbit \citep{PuLai2018,Denham2019,Zhang2024}. In some architectures, an additional outer planet can facilitate rather than suppress migration by cascading the stellar companion's secular torque inward, exciting large eccentricity oscillations and tidal migration of the inner giant \citep[e.g.,][]{Takeda2008,Hamers2017, PuLai2018}. 

To summarize, our result contrasts with the conclusion of \citet{Ngo2016} that most observed stellar companions were \emph{``unlikely to cause Kozai--Lidov migration.''} They placed a $16\pm5\%$ upper limit on the contribution of stellar EKL to the HJ population. 
We instead find that nearly all observed companions are dynamically capable of driving EKL migration (Figures~\ref{fig:ekl_q_separation} and \ref{fig:ekl_tdesc_by_system}).
Together with their high intrinsic occurrence, our companion census suggests that stellar EKL could contribute substantially to HJ formation.

\section{Conclusions}
\label{sec:conclusions}

By combining homogeneous adaptive-optics imaging with \textit{Gaia} common proper-motion pairs,
we construct a homogeneous census of resolved stellar companions to $147$ northern HJs with measured obliquities (Figure \ref{fig:sample_cmd_architectures}). 
We then model the survey selection function to infer intrinsic demographic properties of stellar companions to HJs. Finally, we discuss the implications of our results for HJ formation and obliquity evolution.
Our main conclusions are summarized as follows:

\begin{enumerate}

    \item {\it New candidates:} We perform high-resolution AO imaging for $78$ HJ systems, most without previously identified stellar companions, using the PHARO instrument. The search identifies $8$ new stellar companion candidates (Figure \ref{fig:new_pharo_detections}) around CoRoT-19, HAT-P-49, HAT-P-50, HAT-P-70, HATS-02, HATS-14, WASP-60, and WASP-084.

    \item {\it Companion census:} Combining our uniform high-resolution imaging with {\it Gaia} and previous works reveals that $71/147 = 48 \pm 4\%$ of HJs have resolved (observed) stellar companion candidates.

    \item {\it Intrinsic Demographics:} By modeling the selection function of our imaging survey, previous imaging surveys, and {\it Gaia} astrometry (Figure \ref{fig:contrast_curves}), we infer the completeness-corrected demographics of stellar companions to HJ hosts. We find an intrinsic
    \begin{enumerate}
        
        \item {\it present-day binary fraction} of $f_{\rm bin}=62 \pm 5\%$ for stellar companions with mass ratios $0.1\leq q\leq1$ and separations $50\leq s/{\rm au} \leq 50{,}000$, roughly $3$--$4\times$ larger than field stars (Figure \ref{fig:binary_fraction_comparison}).
        Accounting for white dwarf companions gives an even larger {\it birth} companion fraction (Appendix~\ref{app:wd_comps});

        \item {\it mass ratio distribution} that decreases toward higher $q_\star$ over $0.1$--$1$ (Figure \ref{fig:intrinsic_q_s_dist}, left); and

        \item {\it separation distribution} consistent with being log-uniform from $50$--$2000$~au, ${\rm d}N/{\rm d}\log s \propto s^{0}$ (Figure~\ref{fig:intrinsic_q_s_dist}, right).

    \end{enumerate}

    \item {\it Companion--obliquity association:} 
    HJs with resolved companions at $50$--$2{,}000$~au are nearly twice as likely to be misaligned as systems without detected companions (Figure \ref{fig:misalignment_single_panel}): $24/52=46\%$ compared with $21/88=24\%$ ($p=0.009$).

    \item {\it Kraft Break:} 
    HJs with stellar companions have a higher misaligned fraction on both sides of the Kraft Break (Figure \ref{fig:teff_lambda}), although neither trend is individually significant with the current sample.
    Below $6500$~K, $16/57$ companion hosts are misaligned, compared with $9/59$ systems without a known companion ($p=0.116$). Above 6500~K, the corresponding counts are $10/11$ and $10/15$ ($91\%$ versus $67\%$; $p=0.197$).

    \item {\it Implications for tides:} The misaligned fraction rises steadily from $5\%$ in the coolest temperatures to $40\%$ immediately below the Kraft Break and $80\%$ in the hottest bin. By contrast, the intrinsic companion fraction remains steadily large with no clear temperature dependence (Figure~\ref{fig:teff_fractions}). 
    These trends are consistent with HJs beginning with a broad obliquity distribution, followed by tidal realignment that becomes gradually less efficient toward the Kraft Break (Figure~\ref{fig:tidal_models}).

    \item {\it Implications for eccentric Kozai-Lidov evolution:} 
    Most detected stellar companions are massive and close enough to excite EKL oscillations in a proto-HJ (Figure~\ref{fig:ekl_q_separation}). In $64/71$ systems, the median time for EKL to drive the planet to its observed orbit is less than 3~Gyr (Figure~\ref{fig:ekl_tdesc_by_system}). Stellar EKL can therefore produce the eccentricity growth required for HJ migration in nearly all systems with resolved companions, under suitable orbits.
    
    A simple analytic forward model converts approximately $14.4\%$ of cold Jupiters in wide binaries into HJs, similar to the $12.2\%$ obtained by full three-body integrations \citep{Weldon2026}. These efficiencies predict an HJ occurrence of $0.56\%$--$0.66\%$ per FGK star, comparable to the observed occurrence of $0.57^{+0.14}_{-0.12}\%$ \citep{Petigura2018}. 
    Together, these results suggest that stellar EKL could be a dominant pathway for HJ formation.

\end{enumerate}

\section*{Acknowledgments} \label{acknowledgments}
We are grateful to Advait Mehla for obtaining the NIRC2 observation of WASP-60. 
We thank Luke Handley, Andrew Howard, and Smadar Naoz for useful discussions.
S.W. thanks Adam Kraus and Catherine Clark for helpful discussions.
C.S. acknowledges support from the Department of Energy Computational Science Graduate Fellowship. This material is based upon work supported by the U.S. Department of Energy, Office of Science, Office of Advanced Scientific Computing Research, under Award Number DE-SC0026073. 
This research was supported in part by NSF grant AST-2540180.

M.R. acknowledges support from Heising-Simons Foundation Grant \#2023-4478 and NASA Exoplanets Research Program NNH23ZDA001N-XRP (grant No. 80NSSC24K0153). X.Y.W. acknowledges support from the Sullivan Prize Fellowship.
J.W.X. acknowledges support from the Heising-Simons Foundation 51 Pegasi b Fellowship (grant \#2025-5887).  
S.W. was supported in part by the NASA Exoplanets Research Program NNH23ZDA001N-XRP (Grant No. 80NSSC24K0153), the NASA TESS General Investigator Program, Cycle 7, NNNH23ZDA001N-TESS (Grant No. 80NSSC25K7912), and the Heising-Simons Foundation (Grant No. 2023-4050). 
S.W. also acknowledges support through grant JWST-GO-09025.010-A, provided by NASA via the Space Telescope Science Institute under the JWST General Observers Program \#9025, and gratefully acknowledges support from the John and A-Lan Reynolds Faculty Research Fund.

We are grateful to the Palomar Observatory staff for their support on the PHARO observations.
This work presents results from the European Space Agency (ESA) space mission {\it Gaia}. {\it Gaia} data are being processed by the {\it Gaia} Data Processing and Analysis Consortium (DPAC). Funding for the DPAC is provided by national institutions, in particular the institutions participating in the {\it Gaia} MultiLateral Agreement (MLA). The {\it Gaia} mission website is \url{https://www.cosmos.esa.int/Gaia}. The {\it Gaia} archive website is \url{https://archives.esac.esa.int/Gaia}.

\software{This work made use of \texttt{OverCite} \citep{Shariat2026}, an in-editor citation tool for \LaTeX. This work made use of the following software packages: \texttt{Jupyter} \citep{kluyver2016jupyter}, \texttt{matplotlib} \citep{Hunter:2007}, \texttt{numpy} \citep{numpy}, \texttt{pandas} \citep{pandas_17806077}, \texttt{python} \citep{python}, \texttt{scipy} \citep{scipy_17467817}, \texttt{Cython}.
Software citation information aggregated using \texttt{\href{https://www.tomwagg.com/software-citation-station/}{The Software Citation Station}} \citep{software-citation-station-paper,software-citation-station-zenodo}.
}

\appendix
\twocolumngrid

\section{White dwarf companions} \label{app:wd_comps}

The main demographic analysis considers only present-day main-sequence companions less massive than the HJ host ($0.1 \leq q_\star \leq 1$). 
To estimate the companion fraction \emph{at birth}, we also model binaries in which the HJ host began as the lower-mass star.
In this case, $q_\star=M_{\rm host}/M_{\rm prog}$, where $M_{\rm prog}$ is the initial mass of the WD progenitor. We draw $q_\star$ from the same 
mass-ratio distribution of \citet{ElBadry2019Twins} and weight $M_{\rm prog}$ by the Kroupa initial mass function \citep{Kroupa2001} convolved with the mass-dependent wide-binary fraction $f_{\rm wide~bin} \propto M^{0.4}$ \citep[e.g.,][]{ElBadry2019Twins}. This method of weighted sampling ensures that the population of trial companions for each host is consistent with field wide binaries.

To quantify the fraction of stellar companions that are now WDs, we estimate the age of each HJ host in our sample \citep[][see their Section~3.1 for details]{Golonka2026}. In brief, we interpolate MIST isochrones \citep{Dotter2016,Choi2016} using the \texttt{isochrones} package \citep{Morton2015b} and the reported $T_{\rm eff}$, $\log g$, $[{\rm Fe/H}]$, {\it Gaia} parallax, and $G$, $G_{\rm BP}$, and $G_{\rm RP}$ photometry for each HJ host. The photometry is dereddened using the \citet{Edenhofer2024} dust map.
Nested sampling is applied through \texttt{MultiNest} to sample the posterior distribution of the stellar properties \citep{Feroz2009}.  
Seven hosts have $T_{\rm eff}>7500$~K\footnote{HAT-P-69 \citep{Zhou2019}, KELT-9 \citep{Borsa2019}, KELT-20 \citep{Distler2026}, HAT-P-70 \citep{Zhou2019}, KELT-21 \citep{Johnson2018}, Kepler-13 \citep{Morton2016}, and WASP-189 \citep{Lendl2020}.} 
and one is a pre-main-sequence star \citep[K2-33;][]{Mann2016}.
For these sources, we adopt the reported ages.
For the full $N=147$ sample, the median formal $1\sigma$ age uncertainty is $17\%$. 

We sample these age distributions in the forward model and approximate the time to WD formation as $t_{\rm WD}=10~{\rm Gyr}(M_{\rm prog}/{\rm M_\odot})^{-2.5}$. If the sampled age exceeds $t_{\rm WD}$, we assign the companion a WD mass using the initial--final mass relation of \citet{Cummings2018} and its brightness at a cooling age of $t_{\rm age}-t_{\rm WD}$ using the Montreal pure-hydrogen models \citep{Bedard2020}. We then evolve the orbit through mass loss and the WD kicks following \citet{Fuller2026}. Unbound companions have a recovery probability of zero, and bound WDs pass through the same selection function as main-sequence companions (set by the contrast curves, Figure~\ref{fig:contrast_curves}).

This calculation raises the inferred companion fraction from $62\%$ for present-day main-sequence companions to $69 \pm 5\%$ after WD cooling and to $71 \pm 6\%$ after including mass loss and kicks. Of the modeled birth companions, $7.0\%$ become unbound WDs and $8.5\%$ remain bound but are missed. If HJ hosts have an intrinsic correlation with WD companions \citep[e.g.,][]{Stephan2024}, then the reported fractions are underestimates.

\section{Dependence on the Kraft break}
\label{app:kraft_break}

\begin{figure*}
    \centering
     \includegraphics[width=0.9\linewidth]{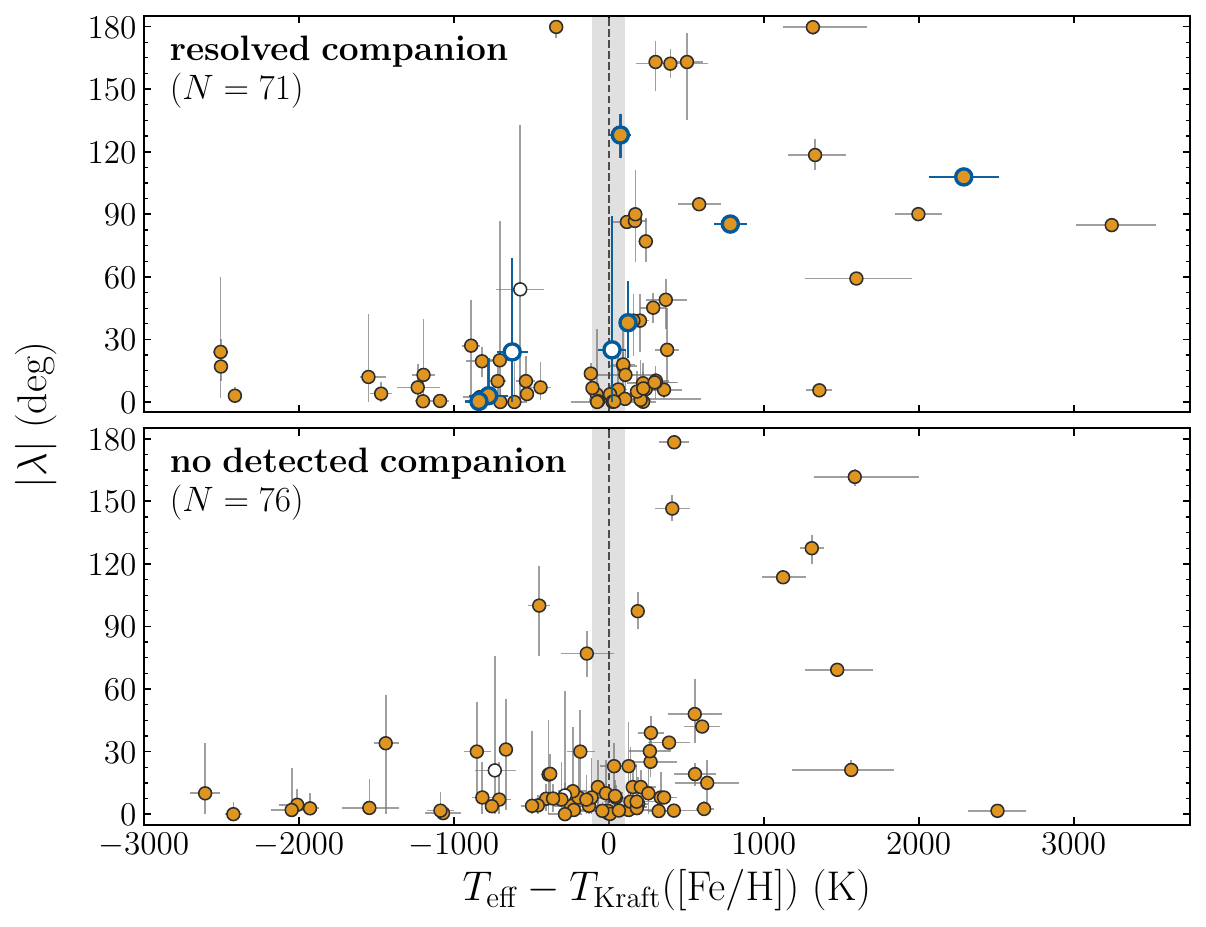}
    \caption{Projected obliquity versus $T_{\rm eff}-T_{\rm Kraft}([\mathrm{Fe/H}])$ for HJ hosts with (top) and without (bottom) a known resolved companion. Positive values correspond to hosts hotter than their metallicity-dependent Kraft-break temperature. Formatting is the same as in Figure~\ref{fig:teff_lambda}.}
    \label{fig:teff_lambda_kraft}
\end{figure*}

\begin{figure}
    \centering
    \includegraphics[width=\linewidth]{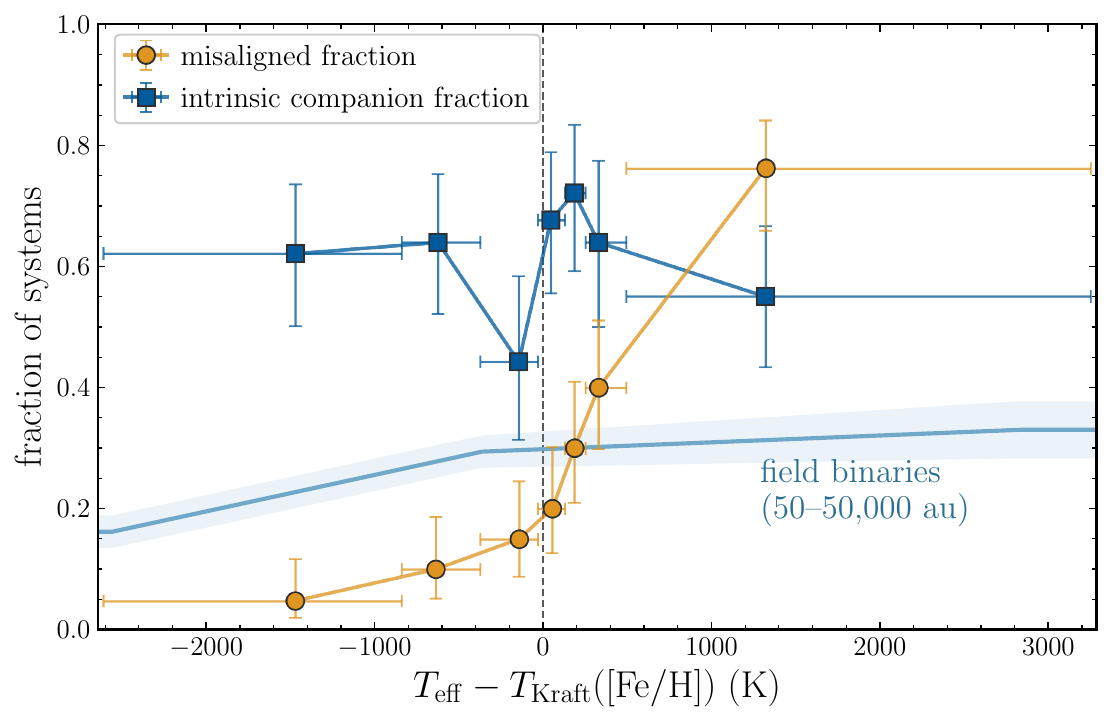}
    \caption{Misalignment fraction as a function of the host star's effective temperature relative to its metallicity-dependent Kraft-Break temperature. The formatting is identical to Figure \ref{fig:teff_fractions}. The misaligned fraction rises steadily with host-star temperature, even when shown as a difference from its Kraft Break, while the intrinsic companion fraction remains roughly constant across the same temperature range.}
    \label{fig:teff_fractions_kraft}
\end{figure}

Metallicity changes the effective temperature at which a main-sequence star's outer convective envelope becomes thin, so a single temperature boundary does not describe the same structural transition in every host star. In particular, \citet{SpaldingWinn2022} predict that this temperature decreases from approximately $6300$ to $6000$~K as $[\mathrm{Fe/H}]$ increases from $-0.3$ to $+0.3$; we therefore test whether the trends in Figures~\ref{fig:teff_lambda} and \ref{fig:teff_fractions} persist after accounting for this dependence.

Following \citet{SpaldingWinn2022}, we use the temperature at which the convective-envelope mass fraction reaches $M_{\rm env}/M_\star=0.003$ as a metallicity-dependent proxy for the Kraft Break. We derive this temperature from the stellar-evolution models of \citet{Amard2019}, adopting the nonrotating $1$~Gyr tracks. At each model metallicity, we interpolate logarithmically in $M_{\rm env}/M_\star$ between the masses bracketing this threshold and then interpolate linearly in $[\mathrm{Fe/H}]$ between the resulting temperatures. We define
\begin{equation}
    \Delta T_{\rm Kraft}
    \equiv T_{\rm eff}-T_{\rm Kraft}([\mathrm{Fe/H}]).
\end{equation}
The resulting solar-metallicity threshold is $T_{\rm Kraft}\simeq6140$~K. This structural threshold is lower than the empirical rotation break near $6550$~K \citep{Beyer2024}: the former is defined by an envelope-mass threshold, whereas the latter is derived from an observed rotation distribution. Repeating the calculation with model ages of $0.1$–$2$ Gyr shifts the Kraft-break temperature by less than $30$ K across the adopted metallicity grid, consistent with the minimal evolution expected for stellar envelopes of main-sequence stars on timescales shorter than the main-sequence lifetime.
At $1$~Gyr, adopting the rotating rather than nonrotating \citet{Amard2019} models changes the thresholds by less than $2$~K.

We use the adopted host-star temperatures and published metallicities compiled for the host stars in our census. The model grid spans $-0.5\leq[\mathrm{Fe/H}]\leq+0.3$; for the 22 hosts with $[\mathrm{Fe/H}]>0.3$, we linearly extrapolate the slope between the two highest-metallicity grid points. All $147$ census hosts therefore have an assigned $\Delta T_{\rm Kraft}$, and $142$ have sufficiently precise obliquity measurements ($\sigma_\lambda < 50^\circ$). For Figure~\ref{fig:teff_fractions_kraft}, we divide the 142 retained obliquities into seven roughly equal-count bins ($21,20,20,20,20,20,21$) and repeat the companion-fraction inference from Equation~\eqref{eq:binary_fraction_likelihood} using all census hosts in each bin and their individual $50$--$50{,}000$~au selection functions. The calculation does not propagate uncertainties in the stellar models.

Figures~\ref{fig:teff_lambda_kraft} and \ref{fig:teff_fractions_kraft} show the results as a function of $\Delta T_{\rm Kraft}$. In general, we find the same qualitative behavior as when they are plotted directly against $T_{\rm eff}$. Systems far below their metallicity-dependent Kraft Break are predominantly aligned, whereas the obliquity distribution broadens toward and above $\Delta T_{\rm Kraft}=0$. Notably, we still do not find evidence for a sharp obliquity transition at or near the Kraft break.
The binned misaligned fraction increases gradually from $1/21$ ($4.8\%$) in the coolest bin to $16/21$ ($76\%$) in the hottest bin, while the inferred companion fraction stays relatively constant. This result remains consistent with tidal realignment becoming less efficient as stellar convective envelopes become thinner \citep{Lai2012Tides,LinOgilvie2017,AndersonWinnPenev2021,SpaldingWinn2022}.

\section{Planet properties and cool-star obliquity}
\label{app:planet_mass_ratio_obliquity}

\begin{figure*}
    \centering
    \includegraphics[width=\linewidth]{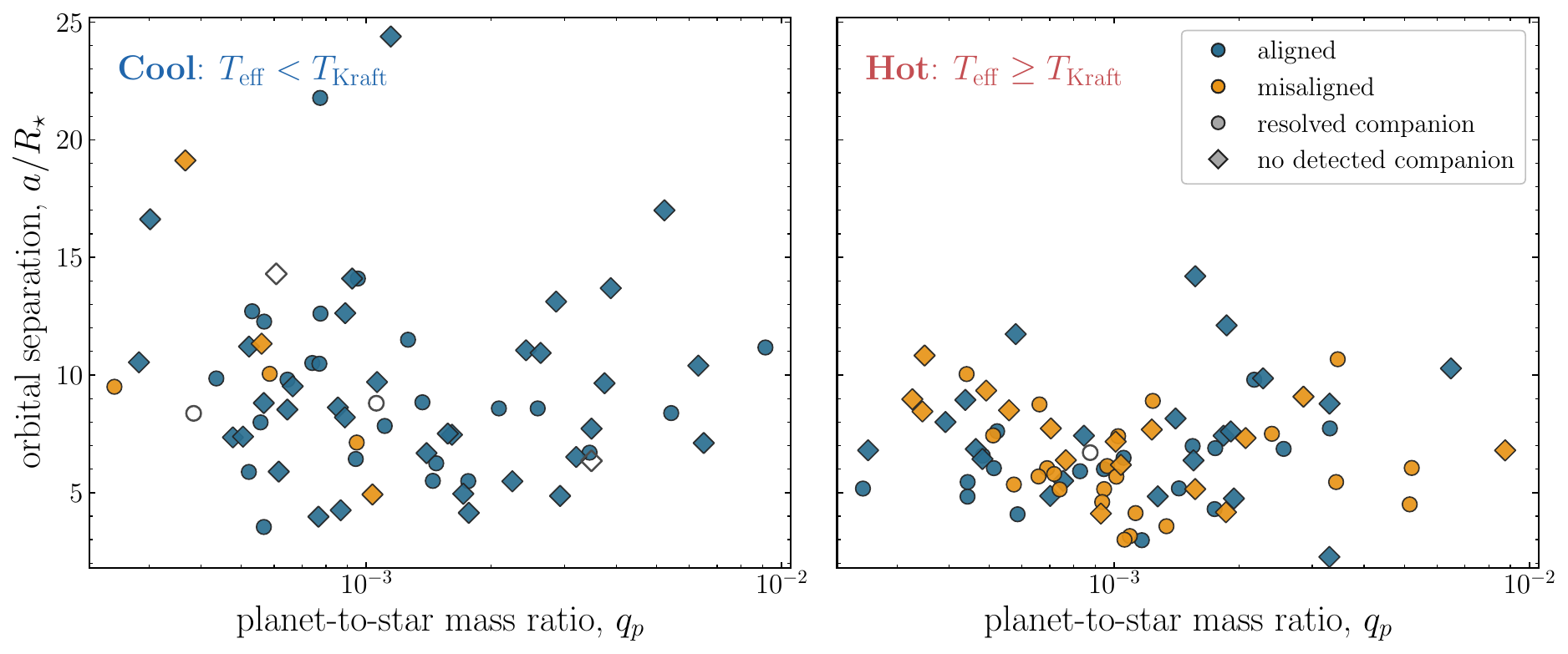}
    \caption{Planet-to-star mass ratio and scaled orbital separation for HJs in our sample around cool (left) and hot (right) host stars. The temperature boundary is determined by each host star's metallicity-dependent Kraft break ($T_{\rm Kraft}$).
    Blue and orange markers denote aligned and misaligned systems, respectively. Circles and diamonds denote systems with and without a known resolved companion; hollow points have ambiguous $\lambda$ measurements. Misaligned HJs around cold stars are concentrated toward lower planet-to-star mass ratios, while the hot sample is predominantly misaligned.}
    \label{fig:planet_tidal_parameters}
\end{figure*}

Beyond their dependence on host-star temperature and stellar companions, HJ obliquities also vary with planet-to-star mass ratio ($q_p=M_p/M_\star$) and orbital separation ($a/R_\star$), which regulate the strength and timescale of tidal realignment \citep{Albrecht2012,Lai2012Tides,AndersonWinnPenev2021,AlbrechtDawsonWinn2022,Zanazzi2024,Rusznak2025}.
Figure~\ref{fig:planet_tidal_parameters} illustrates the sensitivity of HJ misalignment and binarity to $q_p$ and $a/R_\star$ around hot and cool hosts, defined by each host star's metallicity-dependent Kraft Break (see Section \ref{app:kraft_break}).

Below the Kraft break, misaligned systems tend to be less massive (lower $q_{\rm p}$) than aligned systems.
The $58$ aligned and $6$ misaligned systems have median $q_p=1.08\times10^{-3}$ and $5.73\times10^{-4}$, respectively ($p=0.028$).
This trend is expected if more massive planets realign cool host stars more efficiently. 
Only six systems below the Kraft Break are misaligned, split equally between hosts with and without detected stellar companions. Neither companions subgroup separately shows a signficant mass-ratio difference, so the interpretation is limited by small-sample statistics.

Across all $142$ hosts, HJs with resolved companions are closer to their host stars than those without detected companions: the median $a/R_\star$ is $6.64$ versus $7.73$ ($p=0.028$). This difference disappears below the Kraft break, where both medians are $\sim8.7$ ($p=0.891$). Moreover, smaller $a/R_\star$ should strengthen tidal realignment and make companion hosts more aligned, opposite to the observed trend. Differences in planetary separation therefore do not explain why systems with resolved companions are more often misaligned.

\section{EKL capability calculation}
\label{app:ekl_method}

\subsection{Initial EKL and GR timescales}

We denote the host, planet, and stellar-companion masses by $M_\star$, $M_p$, and $M_3=q_\star M_\star$. The planetary and outer orbits have semimajor axes $a_{\rm in}$ and $a_{\rm out}$, eccentricities $e_{\rm in}$ and $e_{\rm out}$, and periods $P_{\rm in}$ and $P_{\rm out}$. At quadrupole order, the characteristic EKL timescale is \citep{FabryckyTremaine2007}
\begin{equation}
t_{\rm KL}=\frac{2P_{\rm out}^2}{3\pi P_{\rm in}}
\frac{M_\star+M_p+M_3}{M_3}(1-e_{\rm out}^2)^{3/2},
\label{eq:ekl_timescale}
\end{equation}
and the initial GR apsidal-precession rate is
\begin{equation}
\dot{\omega}_{\rm GR}= \frac{2\pi}{t_{\rm GR}}=\frac{3G^{3/2}(M_\star+M_p)^{3/2}}
{a_{\rm in}^{5/2}c^2(1-e_{\rm in}^2)}.
\label{eq:gr_precession_rate}
\end{equation}
The gray curves in Figure~\ref{fig:ekl_q_separation} solve $t_{\rm KL}\omega_{\rm GR}=1$ as a function of $q_\star$ and $s$, matching the choice in \citet{Ngo2016}. Quadrupole Kozai oscillations can persist at favorable mutual inclinations for $1<t_{\rm KL}\dot{\omega}_{\rm GR}<3$ \citep{FabryckyTremaine2007}, making the displayed curves conservative. For these curves only, we set $M_\star=1~{\rm M_\odot}$, $M_p=1~{\rm M_{\rm J}}$, $e_{\rm in}=0$, $e_{\rm out}=0.5$, and $a_{\rm out}=s$.

\subsection{Sampling orbits}

Stellar companion masses are measured or photometrically inferred for $69$ systems. For the white dwarfs WASP-71 B and WASP-136 B, we adopt $M_3=0.6~{\rm M_\odot}$. We hold the measured stellar, planetary, and companion properties fixed and draw $10^4$ orbital configurations per system. The proto-HJ semimajor axis is drawn from the mass-dependent occurrence rates in the $1$--$3$, $3$--$10$, and $10$--$30$~au bins of \citet{VanZandt2026}, with uniform sampling in $\ln a_{\rm in}$ within each bin. The proto-HJ eccentricity is drawn from the completeness-corrected histograms of \citet{Blunt2026}: planets below $1000~{\rm M_\oplus}=3.15~{\rm M_{\rm J}}$ use their $30$--$1000$~${\rm M_\oplus}$ distribution, and more massive planets use their $1000$--$6000$~${\rm M_\oplus}$ distribution. We apply these distributions to all systems, motivated by the fact that single- and multiple-star systems show no significant differences in the masses or orbital properties of RV-detected giant planets \citep{Ngo2017}.

We draw the outer eccentricity from the separation-dependent {\it Gaia} wide-binary distribution
\begin{equation}
p(e_{\rm out}\mid s)=[1+\alpha(s)]e_{\rm out}^{\alpha(s)},
\label{eq:hwang_eccentricity_prior}
\end{equation}
using Equation~18 of \citet{HwangTingZakamska2022}, with $\alpha(s)=1.25\tanh[(\log_{10}(s/{\rm au})-1.87)/0.88]+0.12$ and $\alpha=0$ below $\log_{10}(s/{\rm au})=1.78$. Mutual inclinations are isotropic. We infer $a_{\rm out}$ from the measured projected separation by drawing a random mean anomaly, argument of pericenter, and viewing orientation. For eccentric anomaly $E$, true anomaly $f$, and sky inclination $i_{\rm sky}$,
\begin{equation}
\frac{s}{a_{\rm out}}=(1-e_{\rm out}\cos E)
\left[1-\sin^2 i_{\rm sky}\sin^2(\omega_{\rm out}+f)\right]^{1/2}.
\label{eq:projected_separation_deprojection}
\end{equation}

We retain configurations that satisfy the Mardling--Aarseth stability criterion \citep{Mardling2001,FabryckyTremaine2007},
\begin{equation}
\begin{aligned}
\frac{a_{\rm out}}{a_{\rm in}}>{}&2.8
\left(1+\frac{M_3}{M_\star+M_p}\right)^{2/5}\\
&\times\frac{(1+e_{\rm out})^{2/5}}{(1-e_{\rm out})^{6/5}}
\left(1-\frac{0.3i_{\rm mut}}{\pi}\right),
\end{aligned}
\label{eq:mardling_aarseth}
\end{equation}
where $i_{\rm mut}$ is in radians. We also require initial GR precession to be slower than quadrupole evolution ($t_{\rm GR}=2\pi/\dot{\omega}_{\rm GR}\geq t_{\rm quad}$) and restrict the calculation to the tested domain of the general-mass descent-time fit: $0.006\leq\epsilon_{\rm oct}\leq0.06$, $e_{\rm out}\geq0.1$, and $50\arcdeg\leq i_{\rm mut}<89.5\arcdeg$ \citep{Weldon2024}. The inclination cut is the range over which the analytic fit of  \citet{Weldon2024} was calibrated. The octupole strength is \citep[e.g.,][]{Naoz2016}
\begin{equation}
\epsilon_{\rm oct}=\left|\frac{M_\star-M_p}{M_\star+M_p}\right|
\frac{a_{\rm in}}{a_{\rm out}}
\frac{e_{\rm out}}{1-e_{\rm out}^2}.
\label{eq:octupole_strength}
\end{equation}

\subsection{Descent time}

For each accepted configuration, we use the general-mass expression of \citet{Weldon2024},
\begin{equation}
t_{\rm desc}=t_{\rm quad}+\Upsilon t_{\rm oct}
\left[\mathcal{F}(x_{\rm start})-\mathcal{F}(x_{\rm target})\right],
\label{eq:weldon_descent_time}
\end{equation}
where $x=r_p/a_{\rm in}$, $x_{\rm target}=r_{p,{\rm obs}}/a_{\rm in}$, $r_{p,{\rm obs}}=a_{\rm obs}(1-e_{\rm obs})$, and
\begin{align}
t_{\rm quad}&=\frac{16}{15\sqrt{G}}
\frac{a_{\rm out}^3}{a_{\rm in}^{3/2}}
\frac{\sqrt{M_\star+M_p}}{M_3}(1-e_{\rm out}^2)^{3/2},\\
t_{\rm oct}&=\frac{64}{15\sqrt{G}}
\frac{a_{\rm out}^4}{a_{\rm in}^{5/2}}
\frac{(M_\star+M_p)^{3/2}}{(M_\star-M_p)M_3}
\frac{(1-e_{\rm out}^2)^{5/2}}{e_{\rm out}},\\
\Upsilon&=\frac{1.87}{G_1/G_2+|\cos i_{\rm mut}|}.
\end{align}
Here $G_1/G_2$ is the ratio of the inner and outer orbital angular momenta,
\begin{equation}
\frac{G_1}{G_2}
=
\frac{\mu_1}{\mu_2}
\left[
\frac{(M_\star+M_p)a_{\rm in}(1-e_{\rm in}^2)}
{(M_\star+M_p+M_3)a_{\rm out}(1-e_{\rm out}^2)}
\right]^{1/2},
\end{equation}
where
\begin{equation}
\mu_1=\frac{M_\star M_p}{M_\star+M_p},
\qquad
\mu_2=\frac{M_3(M_\star+M_p)}
{M_\star+M_p+M_3}.
\end{equation}
We calculate $x_{\rm start}$ from the first quadrupole eccentricity maximum,
\begin{equation}
x_{\rm start}=1-\left[1-\frac{5}{3}
(1-e_{\rm in}^2)\cos^2 i_{\rm mut}\right]^{1/2},
\end{equation}
and use \citep{Weldon2024}
\begin{align}
\mathcal{F}(x)={}&\frac{\sqrt{2}}{7}x^{1/2}
+\frac{79}{294\sqrt{2}}x^{3/2}
+\frac{9507}{27440\sqrt{2}}x^{5/2}\nonumber\\
&+\frac{526955}{1075648\sqrt{2}}x^{7/2}.
\label{eq:weldon_f}
\end{align}

For each system, the plotted value (Figure \ref{fig:ekl_tdesc_by_system}) is the median $t_{\rm desc}$ among accepted configurations and the 16th--84th percentile range from marginalizing over the orbits. All $71$ systems have at least $130$ accepted draws. We use the current pericenter as the criterion for descent, which is generally different from the tidal capture boundary. The analytic expression estimates the duration of a single octupole descent; several such descents may be required for high-$e$ migration. The simple calculation performed here does not include tidal dissipation or planetary disruption, which is discussed in \citet{Weldon2026}.

\begin{figure*}
    \centering
    \includegraphics[width=0.8\linewidth]{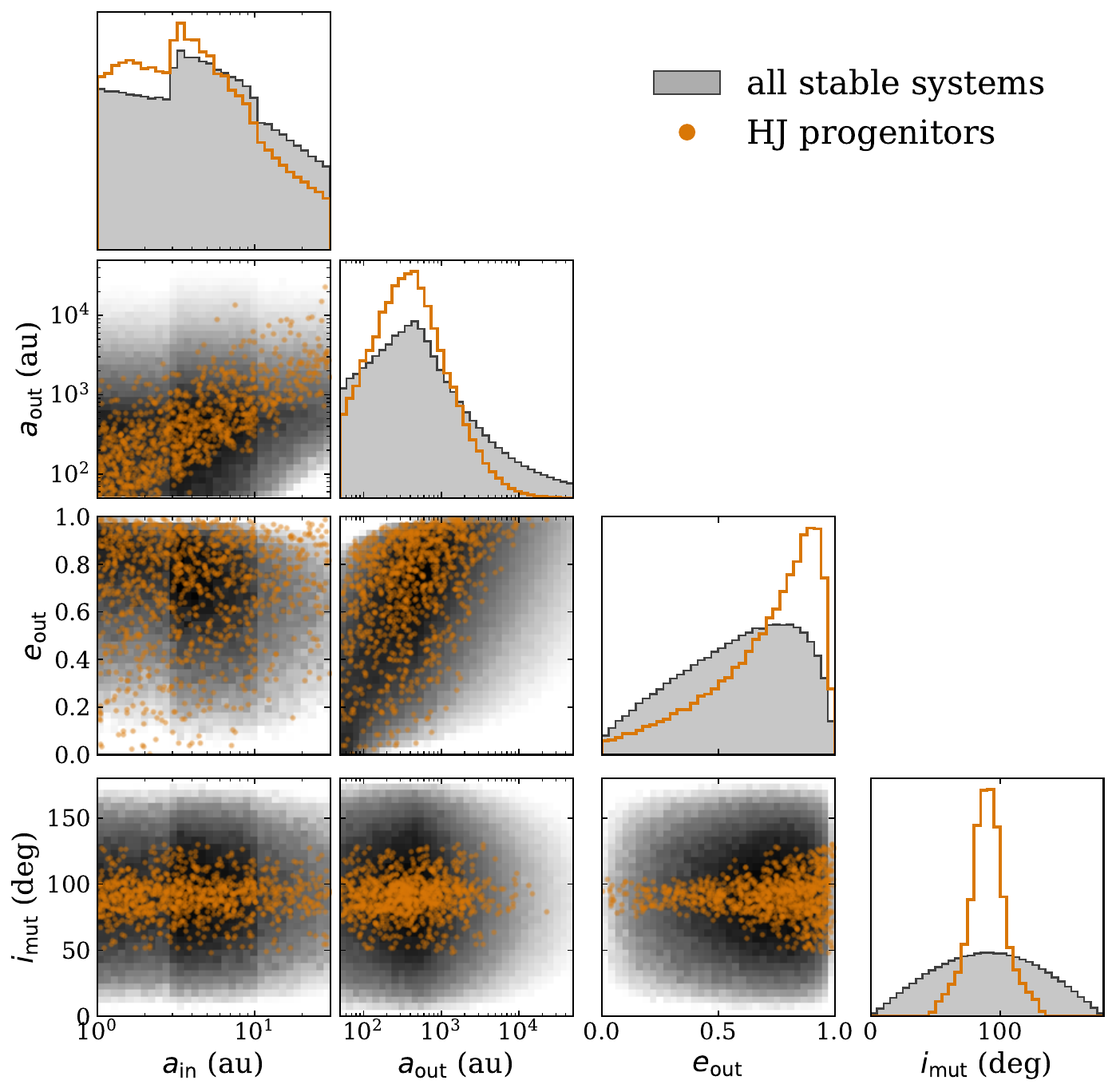}
    \caption{
    Initial architectures in the \emph{simplified} stellar-EKL population synthesis. We show the initial planetary semimajor axis $a_{\rm in}$, stellar-companion semimajor axis $a_{\rm out}$, outer eccentricity $e_{\rm out}$, and mutual inclination $i_{\rm mut}$. Gray densities show the full initially stable population, while orange points show the subsample of systems that form HJs. The diagonal panels show normalized 1D histograms, and the lower panels show binned 2D densities. The HJ-forming systems are concentrated near larger mutual inclinations and $e_{\rm out}$ than the stable population, while their initial planetary semimajor axes ($a_{\rm in}$) span the full adopted $1$--$30$~au range. The relative deficit at large $a_{\rm in}$ is caused by complete disruption, which becomes more common as stronger EKL forces drive planets beyond the HJ survival window.
    }
    \label{fig:ekl_simulations_initial_corner}
\end{figure*}

\section{Companion properties and EKL results}
\label{app:companion_tables}

Table~\ref{tab:resolved_companion_properties} lists each accepted companion component in the census. Table~\ref{tab:ekl_architectures_tdesc} gives the one-perturber architecture adopted for each companion host and the resulting EKL descent-time interval.

Figure~\ref{fig:ekl_simulations_initial_corner} shows the initial architectures used in the population synthesis and the subset that are HJ progenitors.  The initial planetary semimajor axis is sampled uniformly in $\ln a_{\rm in}$ within the three occurrence bins of \citet{VanZandt2026}, $1$--$3$, $3$--$10$, and $10$--$30$~au, causing the step-like features in the initial $a_{\rm in}$ distribution. Of the $10^6$ proposal draws, $856{,}408$ satisfy the stability criterion. Among these, $14.4\%$ form HJs. The HJ-forming systems are concentrated near polar mutual inclinations and have higher outer eccentricities than the stable population, with a weighted median $e_{\rm out}=0.75$ compared with $0.61$. 
Their initial planetary semimajor axes ($a_{\rm in}$) span the full adopted range, with $39\%$, $44\%$, and $17\%$ originating in the $1$--$3$, $3$--$10$, and $10$--$30$~au bins, respectively. 
Interestingly, HJ progenitors are not concentrated towards larger 
$a_{\rm in}$, where EKL is stronger, because such planets are also more likely to be completely tidally disrupted \citep[e.g.,][]{MuNozLaiLiu2016,Weldon2026}.
We remark that the reported fractions are intended as an illustrative population estimate based on analytic estimates. They do not model the full tidal evolution and planetary survival, which can change the HJ conversion efficiency \citep[e.g.,][and references therein]{Weldon2026}.

\startlongtable
\begin{deluxetable*}{lllccccccr}
\tabletypesize{\scriptsize}
\setlength{\tabcolsep}{2.2pt}
\tablewidth{0pt}
\tablecaption{Resolved stellar companions in the HJ sample.\label{tab:resolved_companion_properties}}
\tablehead{
\colhead{System} & \colhead{Companion} & \colhead{Source} & \colhead{Band} & \colhead{$\theta$ ($\arcsec$)} & \colhead{$\Delta m$ (mag)} & \colhead{$s$ (au)} & \colhead{$M_{\rm comp}$ ($M_\odot$)} & \colhead{$q_\star$} & \colhead{Refs.}}
\startdata
CoRoT-02 & Gaia DR3 4287820852697823872 & Gaia & $G$ & 4.080 & 3.23 & 870 & 0.498 & 0.519 & -- \\
CoRoT-11 & Gaia DR3 4285511294152417408 & Gaia & $G$ & 2.535 & 2.18 & 1653 & 0.801 & 0.631 & -- \\
CoRoT-19 & CoRoT-19 B & PHARO & $K_s$ & 0.792 & $4.29^{+0.18}_{-0.21}$ & 632 & $0.312^{+0.038}_{-0.032}$ & 0.258 & [1] \\
HAT-P-01 & HAT-P-1 A & Gaia & $G$ & 11.268 & -0.57 & 1807 & 1.231 & 1.069 & -- \\
HAT-P-03 & Gaia DR3 1510191594552968960 & Gaia & $G$ & 9.800 & 5.66 & 1321 & 0.224 & 0.211 & -- \\
HAT-P-04 & Gaia DR3 1291119606434912384 & Gaia & $G$ & 91.779 & 0.38 & 29539 & 1.167 & 0.935 & -- \\
HAT-P-07 & HAT-P-7 B & Gaia & $G$ & 3.857 & 8.21 & 1286 & 0.259 & 0.166 & -- \\
HAT-P-08 & HAT-P-8 B & Previous & $K_s$ & 1.040 & 5.84 & 236 & 0.175 & 0.138 & [5,6] \\
HAT-P-08 & HAT-P-8 C & Previous & $K_s$ & 1.045 & $6.31^{+0.06}_{-0.06}$ & 238 & $0.139^{+0.011}_{-0.011}$ & 0.109 & [5,6] \\
HAT-P-14 & HAT-P-14 B & Previous & $K_s$ & 0.857 & $5.65^{+0.03}_{-0.03}$ & 176 & $0.211^{+0.010}_{-0.010}$ & 0.080 & [5] \\
HAT-P-16 & HAT-P-16 B & Previous & $K_s$ & 0.690 & $5.46^{+0.04}_{-0.04}$ & 162 & $0.185^{+0.025}_{-0.025}$ & 0.152 & [5] \\
HAT-P-16 & HAT-P-16 C & Gaia & $G$ & 23.347 & 2.94 & 5193 & 0.739 & 0.607 & [7] \\
HAT-P-20 & HAT-P-20 B & Gaia & $G$ & 6.989 & $2.12^{+0.01}_{-0.01}$ & 499 & 0.528 & 0.698 & [7,8] \\
HAT-P-22 & Gaia DR3 846946625690867328 & Gaia & $G$ & 9.163 & 3.45 & 747 & 0.541 & 0.479 & -- \\
HAT-P-24 & Gaia DR3 3167323048322924672 & Gaia & $G$ & 4.944 & 6.45 & 2043 & 0.368 & 0.269 & -- \\
HAT-P-27 & HAT-P-27 B & Previous & $K$ & 0.654 & $3.52^{+0.05}_{-0.05}$ & 133 & $0.323^{+0.048}_{-0.048}$ & 0.353 & [9] \\
HAT-P-30 & Gaia DR3 3096441729861715968 & Gaia & $G$ & 3.834 & 4.93 & 798 & 0.517 & 0.334 & -- \\
HAT-P-32 & HAT-P-32 B & Previous & $K_s$ & 2.936 & $3.62^{+0.06}_{-0.06}$ & 831 & $0.409^{+0.010}_{-0.010}$ & 0.361 & [5] \\
HAT-P-33 & HAT-P-33 B & Previous & $K_s$ & 0.307 & $3.68^{+0.07}_{-0.07}$ & 119 & $0.525^{+0.022}_{-0.022}$ & 0.288 & [5] \\
HAT-P-41 & Gaia DR3 4290415081653653376 & Gaia & $G$ & 3.613 & 3.71 & 1269 & 0.694 & 0.271 & -- \\
HAT-P-49 & HAT-P-49 B & PHARO & $K_s$ & 0.798 & $5.56^{+0.11}_{-0.23}$ & 273 & $0.375^{+0.018}_{-0.013}$ & 0.169 & [1] \\
HAT-P-50 & HAT-P-50 B & PHARO & $K_s$ & 0.659 & $2.97^{+0.07}_{-0.32}$ & 300 & $0.533^{+0.046}_{-0.010}$ & 0.419 & [1] \\
HAT-P-67 & Gaia DR3 1358614983131339904 & Gaia & $G$ & 9.100 & 6.74 & 3387 & 0.488 & 0.282 & -- \\
HAT-P-70 & HAT-P-70 B & PHARO & $K_s$ & 0.628 & $4.21^{+0.10}_{-0.12}$ & 198 & $0.462^{+0.023}_{-0.014}$ & 0.244 & [1] \\
HATS-02 & HATS-2 B & PHARO & $K_s$ & 1.225 & $3.71^{+0.11}_{-0.19}$ & 409 & $0.228^{+0.027}_{-0.009}$ & 0.252 & [1] \\
HATS-14 & HATS-14 B & PHARO & $K_s$ & 1.373 & $5.20^{+0.06}_{-0.06}$ & 690 & $0.135^{+0.004}_{-0.004}$ & 0.139 & [1] \\
HD\_189733 & HD 189733 B & Gaia & $G$ & 11.442 & 5.77 & 226 & 0.202 & 0.256 & -- \\
K2-029 & Gaia DR3 150054788545735296 & Gaia & $G$ & 4.308 & 1.80 & 763 & 0.614 & 0.654 & -- \\
K2-267 & K2-267 B & Previous & $i'$ & $5.899^{+0.001}_{-0.001}$ & $5.98^{+0.10}_{-0.10}$ & 2175 & $0.450^{+0.040}_{-0.040}$ & 0.350 & [8] \\
KELT-09 & Gaia DR3 2064327205637457024 & Gaia & $G$ & 12.888 & 10.96 & 2671 & 0.190 & 0.075 & -- \\
KELT-18 & Gaia DR3 1612165353792906112 & Gaia & $G$ & 3.420 & 5.42 & 1082 & 0.584 & 0.383 & -- \\
KELT-19 & KELT-19 B & Previous & $K_s$ & $0.640^{+0.030}_{-0.030}$ & $2.04^{+0.03}_{-0.03}$ & 188 & 0.500 & 0.320 & [10] \\
KELT-19 & KELT-19 C & Gaia & $G$ & 38.813 & 7.25 & 11429 & 0.387 & 0.248 & -- \\
KELT-21 & KELT-21 C & Previous & $K_s$ & $1.214^{+0.014}_{-0.014}$ & $7.30^{+0.06}_{-0.06}$ & 504 & $0.110^{+0.010}_{-0.010}$ & 0.075 & [11] \\
KELT-21 & KELT-21 B & Previous & $K_s$ & $1.261^{+0.012}_{-0.012}$ & $7.00^{+0.06}_{-0.06}$ & 523 & $0.130^{+0.020}_{-0.010}$ & 0.089 & [11] \\
KELT-24 & Gaia DR3 1076970406752355584 & Gaia & $G$ & 2.067 & 5.05 & 200 & 0.546 & 0.431 & -- \\
Kepler-013 & Kepler-13 BC & Gaia & $G$ & 1.156 & 0.19 & 569 & 2.390 & 1.239 & [7,12] \\
Qatar-1 & Gaia DR3 2244877876891663872 & Gaia & $G$ & 181.839 & 0.54 & 34059 & 0.718 & 0.857 & -- \\
Qatar-6 & Gaia DR3 1265513389372846720 & Gaia & $G$ & 4.804 & 5.03 & 485 & 0.236 & 0.287 & -- \\
TrES-1 & Gaia DR3 2098964884220974592 & Gaia & $G$ & 13.161 & 8.69 & 2101 & 0.102 & 0.099 & -- \\
TrES-4 & Gaia DR3 4609062381822880384 & Gaia & $G$ & 1.560 & 4.90 & 793 & 0.604 & 0.559 & -- \\
WASP-001 & WASP-1 B & Previous & $K_s$ & 4.582 & 4.70 & 1587 & 0.296 & 0.239 & [5] \\
WASP-003 & WASP-3 B & Previous & $K_s$ & 1.191 & 6.55 & 299 & 0.109 & 0.067 & [5] \\
WASP-003 & WASP-3 C & Gaia & $G$ & 18.332 & 3.15 & 4230 & 0.772 & 0.477 & [7] \\
WASP-011 & HAT-P-10 B & Previous & $K_s$ & 0.355 & 2.76 & 42.5 & 0.353 & 0.249 & [5] \\
WASP-011 & HAT-P-10 C & Gaia & $G$ & 16.405 & 4.98 & 2131 & 0.250 & 0.176 & [7] \\
WASP-012 & WASP-12 B & Previous & $K_s$ & 1.059 & 3.30 & 462 & 0.558 & 0.421 & [5,6] \\
WASP-012 & WASP-12 C & Previous & $K_s$ & 1.068 & $3.38^{+0.03}_{-0.03}$ & 466 & $0.555^{+0.029}_{-0.029}$ & 0.419 & [5,6] \\
WASP-014 & WASP-14 B & Previous & $K_s$ & $1.473^{+0.011}_{-0.011}$ & $4.76^{+0.05}_{-0.05}$ & 237 & $0.250^{+0.040}_{-0.040}$ & 0.154 & [5,8] \\
WASP-014 & WASP-14 C & Gaia & $G$ & 11.540 & 7.65 & 1855 & 0.233 & 0.144 & [7] \\
WASP-020 & WASP-20 B & SPHERE & $K_s$ & $0.259\pm0.003$ & $0.86\pm0.06$ & 74.6 & $0.88^{+0.08}_{-0.07}$ & 0.733 & [15] \\
WASP-024 & WASP-24 B (EB) & Gaia & $G$ & 21.834 & 5.47 & 7199 & 0.740 & 0.517 & [7] \\
WASP-026 & Gaia DR3 2416782705960292608 & Gaia & $G$ & 15.386 & 2.91 & 3888 & 0.695 & 0.808 & -- \\
WASP-033 & WASP-33 C & Gaia & $G$ & 48.972 & 3.24 & 5955 & 0.900 & 0.602 & [7] \\
WASP-049 & Gaia DR3 2991284162905572992 & Gaia & $G$ & 2.264 & 5.64 & 441 & 0.315 & 0.342 & -- \\
WASP-060 & WASP-60 B & PHARO & $K_{\rm cont}$ & 0.313 & $4.98^{+0.34}_{-0.32}$ & 144 & $0.204^{+0.026}_{-0.028}$ & 0.171 & [1] \\
WASP-071 & WASP-71 B & Gaia & $G$ & 6.366 & $8.60^{+0.01}_{-0.01}$ & 2261 & $0.600^{\dagger}$ & 0.385 & [7] \\
WASP-076 & WASP-76 B & Previous & $K$ & 0.443 & 2.65 & 53.0 & 0.712 & 0.488 & [9] \\
WASP-077 & WASP-77 B & Gaia & $G$ & 3.277 & 1.73 & 346 & 0.771 & 0.808 & [7] \\
WASP-078 & WASP-78 B & Gaia & $G$ & 42.129 & 3.02 & 30416 & 0.806 & 0.431 & -- \\
WASP-084 & WASP-84 B & PHARO & Br$\gamma$ & 0.714 & $4.34^{+0.11}_{-0.12}$ & 71.3 & $0.163^{+0.010}_{-0.008}$ & 0.186 & [1] \\
WASP-085 & Gaia DR3 3909745223886018560 & Gaia & $G$ & 1.466 & 0.93 & 207 & 0.819 & 0.844 & -- \\
WASP-103 & WASP-103 close source & PHARO & $K_s$ & 0.227 & $2.37^{+0.04}_{-0.04}$ & 80.3 & 0.519 & 0.425 & [1] \\
WASP-140 & Gaia DR3 5094154336332482176 & Gaia & $G$ & 7.235 & 2.35 & 845 & 0.607 & 0.675 & -- \\
WASP-180 & WASP-180 B & Gaia & $G$ & 4.862 & 0.86 & 1223 & 1.053 & 0.900 & [7,13] \\
WASP-189 & Gaia DR3 6339097675623770496 & Gaia & $G$ & 9.417 & 7.83 & 932 & 0.428 & 0.211 & -- \\
XO-2 & XO-2 S & Gaia & $G$ & 31.207 & -0.04 & 4677 & 0.944 & 1.005 & -- \\
KELT-2 A & Gaia DR3 3438059442854472832 & Gaia & $G$ & 2.381 & 3.39 & 320 & 0.758 & 0.577 & -- \\
KELT-23 A & Gaia DR3 1644692068838995840 & Gaia & $G$ & 4.542 & 5.35 & 577 & 0.372 & 0.395 & -- \\
KELT-3 & Gaia DR3 806492023788937216 & Gaia & $G$ & 3.745 & 4.21 & 783 & 0.658 & 0.515 & -- \\
KELT-4 A & KELT-4 B & Previous & $K_s$ & $1.5692\pm0.0018$ & $2.02\pm0.36$ & 342 & $0.776\pm0.081$ & 0.646 & [14] \\
KELT-4 A & KELT-4 C & Previous & $K_s$ & $1.5616\pm0.0020$ & $3.10\pm0.15$ & 341 & $0.584\pm0.029$ & 0.486 & [14] \\
TIC 46432937 & Gaia DR3 2984391358868786560 & Gaia & $G$ & 39.751 & 1.09 & 3604 & 0.396 & 0.704 & -- \\
TOI-1259 A & Gaia DR3 2294170834291960832 & Gaia & $G$ & 13.906 & 7.41 & 1654 & 0.561 & 0.752 & [16] \\
TOI-1333 & Gaia DR3 1978027912379523712 & Gaia & $G$ & 2.870 & 3.30 & 563 & 0.780 & 0.533 & -- \\
TOI-1789 & Gaia DR3 646124645103549312 & Gaia & $G$ & 79.845 & 0.44 & 17845 & 1.322 & 0.877 & -- \\
TOI-3714 & Gaia DR3 178924390476838784 & Gaia & $G$ & 2.670 & 4.56 & 302 & 1.070 & 2.019 & [17] \\
TOI-5293 A & Gaia DR3 2640121482094497024 & Gaia & $G$ & 3.570 & 2.90 & 579 & 0.196 & 0.364 & -- \\
WASP-136 & Gaia DR3 2441013811933003136 & Gaia & $G$ & 5.077 & 9.75 & 1364 & $0.600^{\dagger}$ & 0.426 & [18] \\
WASP-54 & Gaia DR3 3661983846370746624 & Gaia & $G$ & 5.722 & 8.71 & 1436 & 0.190 & 0.176 & [15] \\
\enddata
\tablecomments{Sources give the origin of $\theta$ and $\Delta m$: PHARO measurements from this work [1], {\it Gaia} DR3 [2], or previous imaging. {\it Gaia} binaries are from [3] unless cited otherwise. Main-sequence masses are published values or are inferred from [4]. Unless published, $s$ and $q_\star=M_{\rm comp}/M_\star$ are calculated from the {\it Gaia} distance and adopted host mass. $\dagger$: assumed fiducial WD mass. Separations are projected. PHARO $\theta$ values use the nominal $0.025\arcsec$ pixel scale and lack calibrated astrometric errors; their mass errors propagate only $\Delta m$. Table~\ref{tab:ekl_architectures_tdesc} gives the adopted aggregate architecture for systems with multiple components. References: [1] this work; [2] \citep{Gaia2023}; [3] \citep{ElBadry2021}; [4] \citep{PecautMamajek2013}; [5] \citep{Ngo2015}; [6] \citep{Bechter2014}; [7] \citep{Michel2024}; [8] \citep{Schlagenhauf2024}; [9] \citep{Ngo2016}; [10] \citep{Siverd2018}; [11] \citep{Johnson2018}; [12] \citep{Santerne2012}; [13] \citep{Temple2019}; [14] \citep{Ngo2017}; [15] \citep{Bohn2020}; [16] \citep{Martin2021}; [17] \citep{Canas2022}; [18] \citep{Mugrauer2022}.}
\end{deluxetable*}

\startlongtable
\begin{deluxetable*}{llccccccr}
\tablecaption{Companion architectures and EKL descent times.\label{tab:ekl_architectures_tdesc}}
\tablehead{
\colhead{System} & \colhead{Perturber} & \colhead{$|\lambda|$ ($\arcdeg$)} & \colhead{$M_p$ ($M_{\rm J}$)} & \colhead{$s$ (au)} & \colhead{$M_3$ ($M_\odot$)} & \colhead{$q_\star$} & \colhead{$t_{\rm desc}$ (Gyr)} & \colhead{Refs.}}
\startdata
CoRoT-02 & Gaia companion & 10.0 & 3.47 & 870 & 0.498 & 0.519 & $0.091^{+0.747}_{-0.082}$ & [1,2] \\
CoRoT-11 & Gaia companion & 0.1 & 2.33 & 1653 & 0.801 & 0.631 & $0.25^{+1.77}_{-0.23}$ & [3,4] \\
CoRoT-19 & CoRoT-19 B & $25.0^{\dagger}$ & 1.11 & 632 & 0.312 & 0.258 & $0.093^{+0.722}_{-0.084}$ & [1,5] \\
HAT-P-01 & Gaia companion & 3.6 & 0.525 & 1807 & 1.23 & 1.069 & $0.21^{+1.32}_{-0.20}$ & [1,6] \\
HAT-P-03 & Gaia companion & 19.5 & 0.650 & 1321 & 0.224 & 0.211 & $0.76^{+4.24}_{-0.70}$ & [1,7] \\
HAT-P-04 & Gaia companion & 6.0 & 0.671 & 29539 & 1.17 & 0.935 & $42.1^{+86.2}_{-37.9}$ & [1,8] \\
HAT-P-07 & Gaia companion & 163.0 & 1.84 & 1286 & 0.259 & 0.166 & $0.69^{+3.59}_{-0.63}$ & [1,7] \\
HAT-P-08 & HAT-P-8 BC & 17.4 & 1.28 & 237 & 0.314 & 0.247 & $0.016^{+0.143}_{-0.014}$ & [1,7] \\
HAT-P-14 & HAT-P-14 B & 163.0 & 3.44 & 176 & 0.211 & 0.080 & $0.010^{+0.116}_{-0.009}$ & [1,7] \\
HAT-P-16 & HAT-P-16 B & 18.0 & 4.22 & 162 & 0.185 & 0.152 & $0.0093^{+0.0953}_{-0.0084}$ & [1,9] \\
HAT-P-20 & HAT-P-20 B & 4.0 & 7.25 & 499 & 0.528 & 0.698 & $0.029^{+0.278}_{-0.027}$ & [1,10] \\
HAT-P-22 & Gaia companion & 0.0 & 2.47 & 747 & 0.541 & 0.479 & $0.090^{+0.611}_{-0.082}$ & [1,7] \\
HAT-P-24 & Gaia companion & 9.0 & 0.750 & 2043 & 0.368 & 0.269 & $0.97^{+5.14}_{-0.89}$ & [1,7] \\
HAT-P-27 & HAT-P-27 B & 20.0 & 0.620 & 133 & 0.323 & 0.353 & $0.0039^{+0.0335}_{-0.0034}$ & [1,11] \\
HAT-P-30 & Gaia companion & 77.0 & 0.830 & 798 & 0.517 & 0.334 & $0.13^{+1.00}_{-0.12}$ & [1,7] \\
HAT-P-32 & HAT-P-32 B & 86.3 & 0.680 & 831 & 0.409 & 0.361 & $0.16^{+1.10}_{-0.15}$ & [1,12] \\
HAT-P-33 & HAT-P-33 B & 5.9 & 0.920 & 119 & 0.525 & 0.288 & $0.0023^{+0.0237}_{-0.0020}$ & [13,7] \\
HAT-P-41 & Gaia companion & 9.4 & 1.19 & 1269 & 0.694 & 0.271 & $0.30^{+2.03}_{-0.28}$ & [1,7] \\
HAT-P-49 & HAT-P-49 B & 85.3 & 2.20 & 273 & 0.375 & 0.169 & $0.021^{+0.232}_{-0.019}$ & [14,7] \\
HAT-P-50 & HAT-P-50 B & 38.0 & 1.35 & 300 & 0.533 & 0.419 & $0.016^{+0.123}_{-0.015}$ & [1,15] \\
HAT-P-67 & Gaia companion & 6.2 & 0.450 & 3387 & 0.488 & 0.282 & $1.90^{+10.75}_{-1.78}$ & [1,16] \\
HAT-P-70 & HAT-P-70 B & 107.9 & 6.78 & 198 & 0.462 & 0.244 & $0.0060^{+0.0566}_{-0.0054}$ & [17,18] \\
HATS-02 & HATS-2 B & 3.0 & 1.37 & 409 & 0.228 & 0.252 & $0.064^{+0.493}_{-0.057}$ & [19] \\
HATS-14 & HATS-14 B & $24.0^{\dagger}$ & 1.07 & 690 & 0.135 & 0.139 & $0.30^{+2.16}_{-0.27}$ & [1,20] \\
HD\_189733 & Gaia companion & 0.3 & 1.13 & 226 & 0.202 & 0.256 & $0.018^{+0.160}_{-0.016}$ & [1,7] \\
K2-029 & Gaia companion & 10.0 & 0.730 & 763 & 0.614 & 0.654 & $0.073^{+0.557}_{-0.065}$ & [1,21] \\
K2-267 & K2-267 B & 1.5 & 3.00 & 2175 & 0.450 & 0.350 & $0.66^{+3.98}_{-0.61}$ & [22] \\
KELT-09 & Gaia companion & 84.8 & 2.88 & 2671 & 0.190 & 0.075 & $3.79^{+18.25}_{-3.44}$ & [23,24] \\
KELT-18 & Gaia companion & 94.8 & 1.18 & 1082 & 0.584 & 0.383 & $0.19^{+1.26}_{-0.17}$ & [25,26] \\
KELT-19 & KELT-19 B & 179.7 & 4.07 & 188 & 0.500 & 0.320 & $0.0044^{+0.0412}_{-0.0040}$ & [27,28] \\
KELT-21 & KELT-21 BC & 5.6 & 3.91 & 667 & 0.240 & 0.165 & $0.12^{+0.99}_{-0.11}$ & [29] \\
KELT-24 & Gaia companion & 49.0 & 4.59 & 200 & 0.546 & 0.431 & $0.0036^{+0.0415}_{-0.0032}$ & [1,30] \\
Kepler-013 & Kepler-13 BC & 59.2 & 9.28 & 569 & 2.39 & 1.239 & $0.012^{+0.100}_{-0.011}$ & [31,32] \\
Qatar-1 & Gaia companion & 7.0 & 1.29 & 34059 & 0.718 & 0.857 & $66.3^{+170.4}_{-50.1}$ & [1,33] \\
Qatar-6 & Gaia companion & 0.5 & 0.668 & 485 & 0.236 & 0.287 & $0.061^{+0.562}_{-0.055}$ & [1,34] \\
TrES-1 & Gaia companion & 27.0 & 0.840 & 2101 & 0.102 & 0.099 & $2.83^{+16.36}_{-2.54}$ & [1,7] \\
TrES-4 & Gaia companion & 10.2 & 0.780 & 793 & 0.604 & 0.559 & $0.085^{+0.553}_{-0.078}$ & [1,7] \\
WASP-001 & WASP-1 B & 86.8 & 0.854 & 1587 & 0.296 & 0.239 & $0.77^{+4.18}_{-0.71}$ & [1,35] \\
WASP-003 & WASP-3 B & 9.4 & 2.43 & 299 & 0.109 & 0.067 & $0.085^{+0.755}_{-0.076}$ & [1,7] \\
WASP-011 & HAT-P-10 B & 13.0 & 0.790 & 42.5 & 0.353 & 0.249 & $0.00029^{+0.00344}_{-0.00026}$ & [1,7] \\
WASP-012 & WASP-12 BC & 39.0 & 1.47 & 464 & 1.11 & 0.840 & $0.022^{+0.150}_{-0.020}$ & [1,33] \\
WASP-014 & WASP-14 B & 45.2 & 8.84 & 237 & 0.250 & 0.154 & $0.017^{+0.171}_{-0.015}$ & [1,7] \\
WASP-020 & WASP-20 B & 13.6 & 0.311 & 74.6 & 0.880 & 0.733 & $0.00051^{+0.00432}_{-0.00046}$ & [1,62,63] \\
WASP-024 & WASP-24 B (EB) & 3.7 & 1.24 & 7199 & 0.740 & 0.517 & $5.34^{+19.61}_{-4.99}$ & [1,7] \\
WASP-026 & Gaia companion & 0.4 & 0.850 & 3888 & 0.695 & 0.808 & $1.35^{+7.05}_{-1.22}$ & [1,7] \\
WASP-033 & WASP-33 C & 118.4 & 2.09 & 5955 & 0.900 & 0.602 & $3.46^{+12.68}_{-3.15}$ & [1,36] \\
WASP-049 & Gaia companion & $54.0^{\dagger}$ & 0.370 & 441 & 0.315 & 0.342 & $0.051^{+0.401}_{-0.046}$ & [37,7] \\
WASP-060 & WASP-60 B & 128.0 & 0.550 & 144 & 0.204 & 0.171 & $0.0076^{+0.0636}_{-0.0068}$ & [1,7] \\
WASP-071 & WASP-71 B & 1.0 & 1.39 & 2261 & $0.600^{\ddagger}$ & 0.385 & $0.66^{+4.27}_{-0.60}$ & [1,7] \\
WASP-076 & WASP-76 B & 25.0 & 0.894 & 53.0 & 0.712 & 0.488 & $0.00030^{+0.00263}_{-0.00026}$ & [1,38] \\
WASP-077 & WASP-77 B & 3.8 & 1.67 & 346 & 0.771 & 0.808 & $0.015^{+0.122}_{-0.014}$ & [1,39] \\
WASP-078 & Gaia companion & 0.0 & 1.11 & 30416 & 0.806 & 0.431 & $62.2^{+206.9}_{-57.5}$ & [1,7] \\
WASP-084 & WASP-84 B & 0.3 & 0.692 & 71.3 & 0.163 & 0.186 & $0.0014^{+0.0145}_{-0.0012}$ & [1,40] \\
WASP-085 & Gaia companion & 7.0 & 1.26 & 207 & 0.819 & 0.844 & $0.0036^{+0.0274}_{-0.0032}$ & [1,41] \\
WASP-103 & WASP-103 close source & 13.0 & 1.49 & 80.3 & 0.519 & 0.425 & $0.0013^{+0.0099}_{-0.0012}$ & [1,42] \\
WASP-140 & Gaia companion & 2.3 & 2.44 & 845 & 0.607 & 0.675 & $0.085^{+0.760}_{-0.076}$ & [1,43] \\
WASP-180 & WASP-180 B & 162.2 & 0.900 & 1223 & 1.05 & 0.900 & $0.12^{+0.81}_{-0.11}$ & [1,44] \\
WASP-189 & Gaia companion & 90.1 & 1.99 & 932 & 0.428 & 0.211 & $0.22^{+1.49}_{-0.20}$ & [45,46] \\
XO-2 & Gaia companion & 0.0 & 0.629 & 4677 & 0.944 & 1.005 & $1.38^{+7.24}_{-1.29}$ & [1,47] \\
KELT-2 A & Gaia companion & 0.1 & 1.524 & 320 & 0.758 & 0.577 & $0.011^{+0.096}_{-0.010}$ & [48,49] \\
KELT-23 A & Gaia companion & 179.8 & 0.938 & 577 & 0.372 & 0.395 & $0.079^{+0.562}_{-0.072}$ & [1,50] \\
KELT-3 & Gaia companion & 6.5 & 1.477 & 783 & 0.658 & 0.515 & $0.083^{+0.621}_{-0.076}$ & [1,51] \\
KELT-4 A & KELT-4 BC & 90.0 & 0.902 & 341 & 1.360 & 1.132 & $0.0075^{+0.0602}_{-0.0067}$ & [1,52] \\
TIC 46432937 & Gaia companion & 3.0 & 3.200 & 3604 & 0.396 & 0.704 & $1.38^{+7.54}_{-1.24}$ & [53,54] \\
TOI-1259 A & WD companion & 12.0 & 0.444 & 1654 & 0.561 & 0.752 & $0.28^{+1.89}_{-0.26}$ & [1,55] \\
TOI-1333 & Gaia companion & 5.0 & 2.370 & 563 & 0.780 & 0.533 & $0.033^{+0.316}_{-0.030}$ & [56,57] \\
TOI-1789 & Gaia companion & 0.2 & 0.700 & 17845 & 1.322 & 0.877 & $12.0^{+42.9}_{-10.7}$ & [48,58] \\
TOI-3714 & WD companion & 17.0 & 0.700 & 302 & 1.070 & 2.019 & $0.0063^{+0.0541}_{-0.0057}$ & [1,59] \\
TOI-5293 A & Gaia companion & 24.0 & 0.540 & 579 & 0.196 & 0.364 & $0.11^{+0.89}_{-0.10}$ & [1,60] \\
WASP-136 & WD companion & 39.0 & 1.510 & 1364 & $0.600^{\ddagger}$ & 0.426 & $0.25^{+1.57}_{-0.23}$ & [1,61] \\
WASP-54 & Gaia companion & 6.7 & 0.590 & 1436 & 0.190 & 0.176 & $0.86^{+5.50}_{-0.80}$ & [1,7] \\
\enddata
\tablecomments{Values are medians and 16th--84th percentiles. $\dagger$: excluded from obliquity statistics. $\ddagger$: assumed fiducial WD mass. Table~\ref{tab:resolved_companion_properties} gives the component measurements. Row references source $|\lambda|$ and the system parameters; $t_{\rm desc}$ is calculated here as described in Appendix~\ref{app:ekl_method}. References: [1] \citep{WangWangBatygin2026}; [2] \citep{Gillon2010}; [3] \citep{Gandolfi2012}; [4] \citep{Gandolfi2010}; [5] \citep{Guenther2012}; [6] \citep{Nikolov2014}; [7] \citep{Stassun2017}; [8] \citep{Torres2008}; [9] \citep{Bonomo2017}; [10] \citep{Bakos2010}; [11] \citep{Brown2012}; [12] \citep{Wang2019}; [13] \citep{Bourrier2023}; [14] \citep{Balkoova2026}; [15] \citep{Hartman2015}; [16] \citep{Wang2025}; [17] \citep{BelloArufe2022}; [18] \citep{Zhou2019}; [19] \citep{Biagiotti2024}; [20] \citep{Mancini2015b}; [21] \citep{Santerne2016}; [22] \citep{Yu2018}; [23] \citep{Stephen2022}; [24] \citep{Borsa2019}; [25] \citep{Rubenzahl2024}; [26] \citep{McLeod2017}; [27] \citep{Kawai2024}; [28] \citep{Siverd2018}; [29] \citep{Johnson2018}; [30] \citep{Giovinazzi2024}; [31] \citep{Howarth2017}; [32] \citep{Esteves2015}; [33] \citep{Collins2017}; [34] \citep{Alsubai2018}; [35] \citep{Maciejewski2014}; [36] \citep{Sengupta2019}; [37] \citep{Wyttenbach2017}; [38] \citep{Ehrenreich2020}; [39] \citep{Noguer2024}; [40] \citep{Maciejewski2023}; [41] \citep{Mocnik2016}; [42] \citep{Gillon2014}; [43] \citep{Hellier2017}; [44] \citep{Temple2019}; [45] \citep{Prinoth2024}; [46] \citep{Lendl2020}; [47] \citep{Knutson2014}; [48] \citep{Polanski2026}; [49] \citep{Beatty2012}; [50] \citep{Johns2019}; [51] \citep{Pepper2013}; [52] \citep{Eastman2016}; [53] \citep{Weisserman2026}; [54] \citep{Hartman2024}; [55] \citep{Veldhuis2026}; [56] \citep{Knudstrup2026TOI1333}; [57] \citep{Rodriguez2021}; [58] \citep{Khandelwal2022}; [59] \citep{Canas2022}; [60] \citep{Canas2023}; [61] \citep{Knudstrup2024}; [62] \citep{Anderson2015WASP20}; [63] \citep{Bohn2020}.}
\end{deluxetable*}

\section{All PHARO Images}
\label{app:image_atlas}
\begin{figure*}[!p]
\centering
\includegraphics[width=0.98\textwidth]{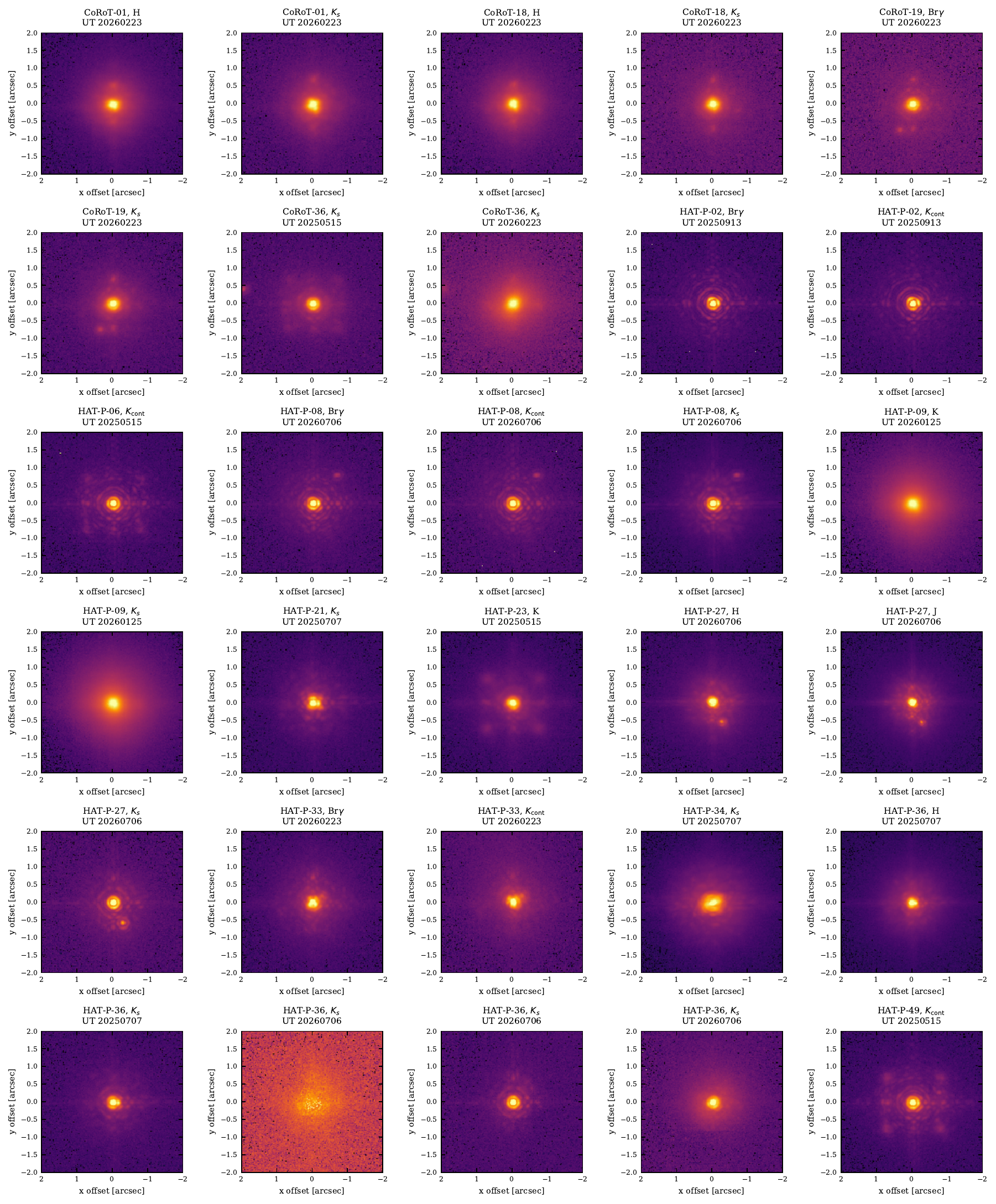}
\caption{PHARO image atlas, page 1 of 6.}
\label{fig:pharo_atlas_01}
\end{figure*}

\begin{figure*}[!p]
\centering
\includegraphics[width=0.98\textwidth]{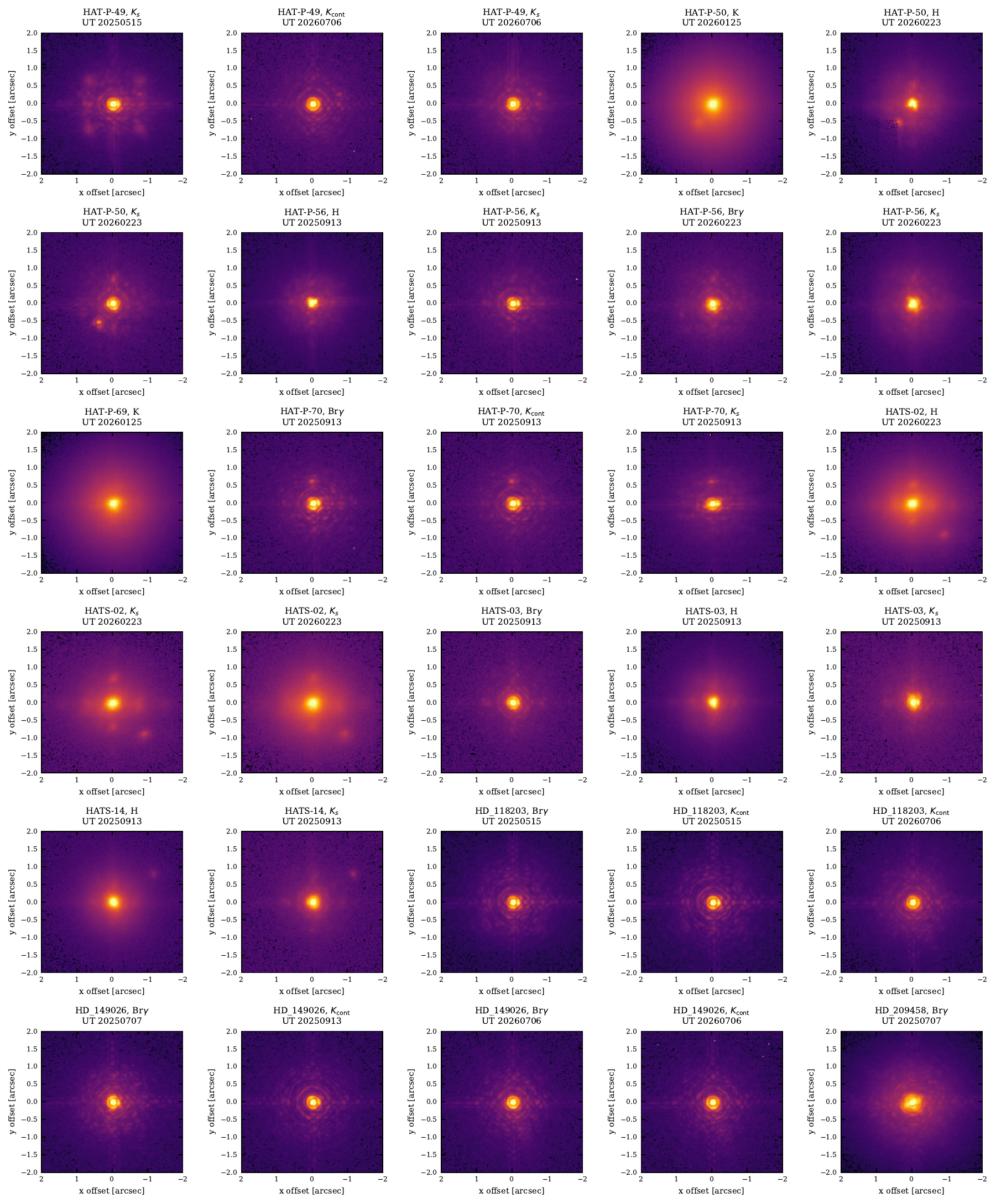}
\caption{PHARO image atlas, page 2 of 6.}
\label{fig:pharo_atlas_02}
\end{figure*}

\begin{figure*}[!p]
\centering
\includegraphics[width=0.98\textwidth]{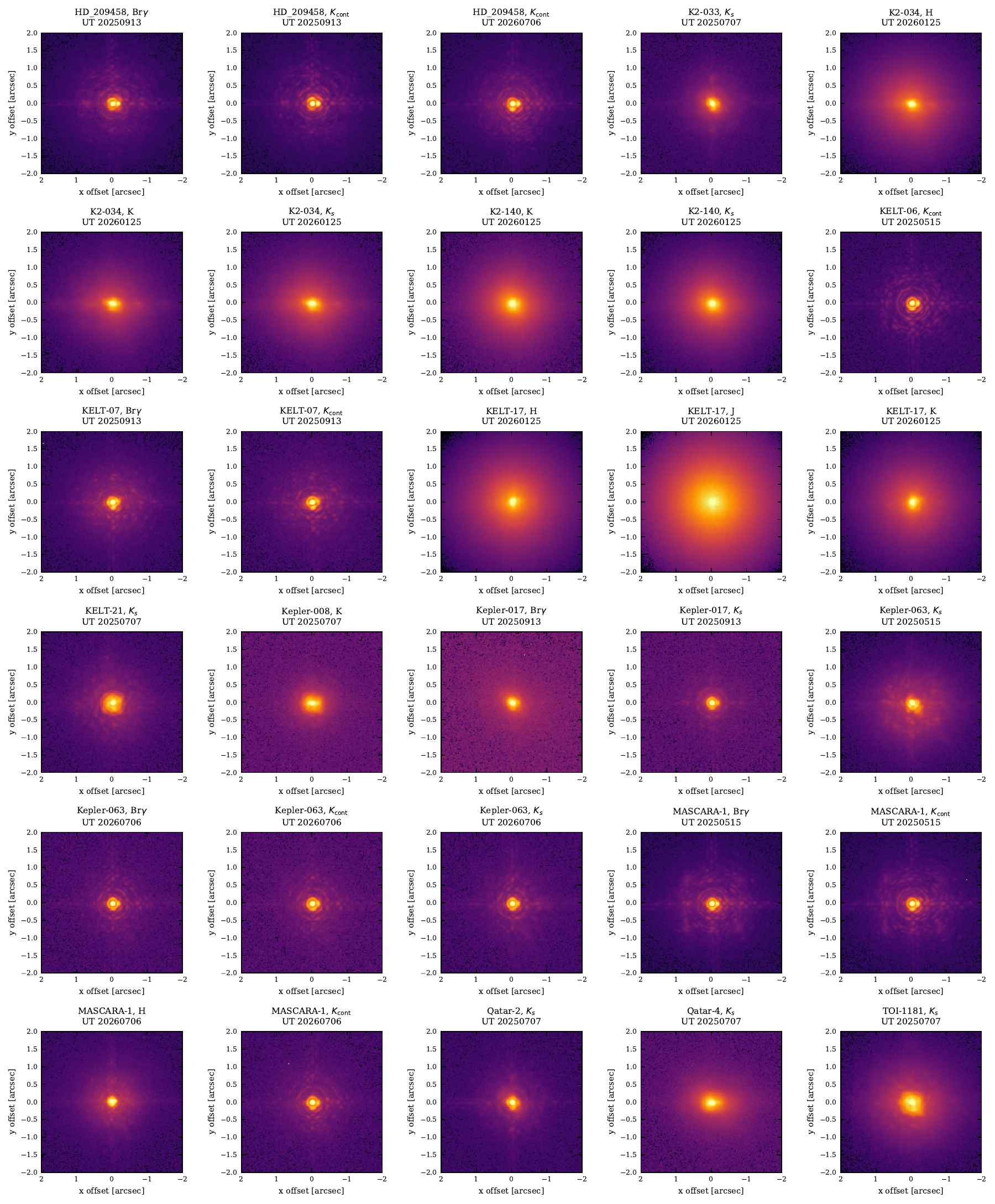}
\caption{PHARO image atlas, page 3 of 6.}
\label{fig:pharo_atlas_03}
\end{figure*}

\begin{figure*}[!p]
\centering
\includegraphics[width=0.98\textwidth]{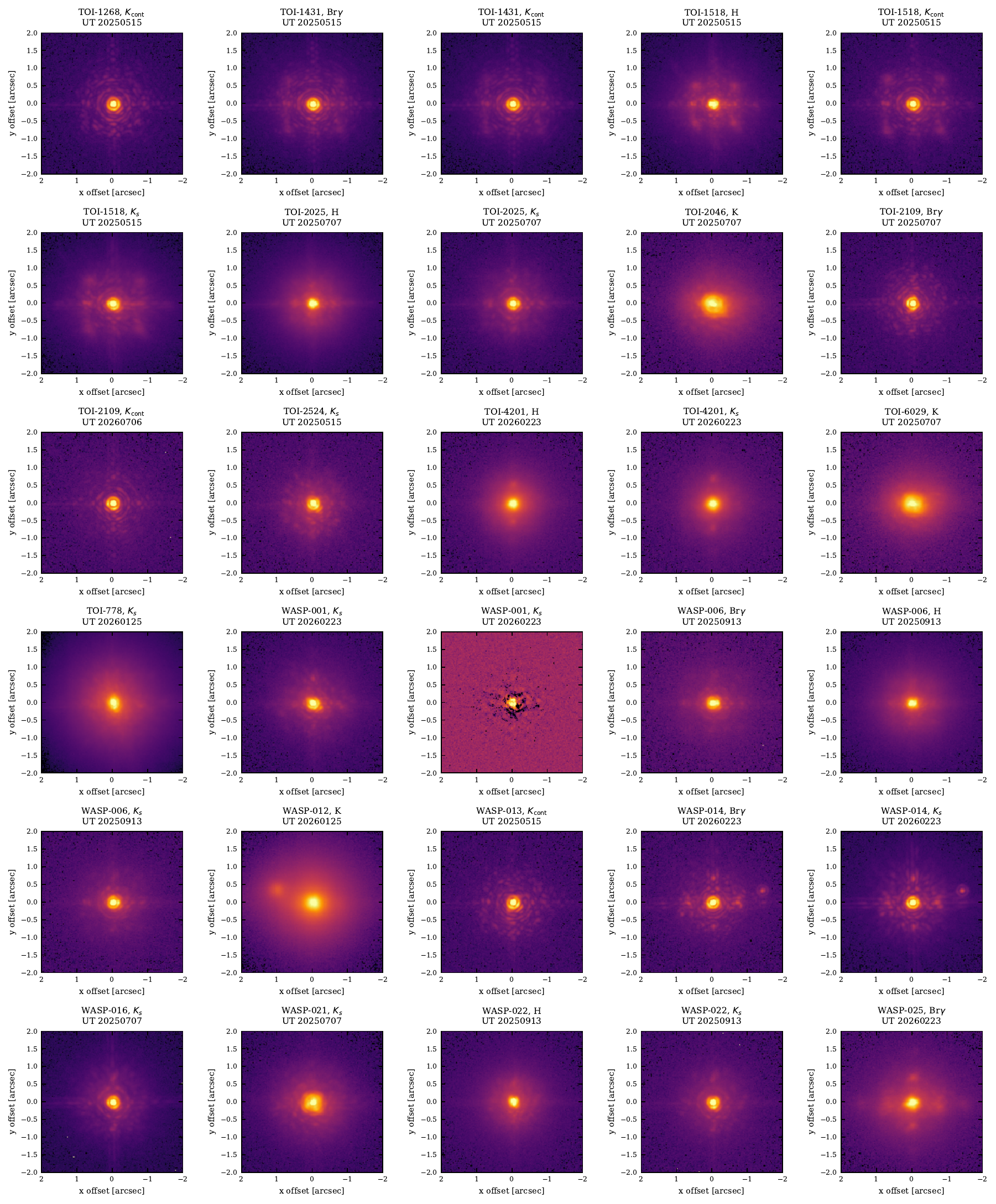}
\caption{PHARO image atlas, page 4 of 6.}
\label{fig:pharo_atlas_04}
\end{figure*}

\begin{figure*}[!p]
\centering
\includegraphics[width=0.98\textwidth]{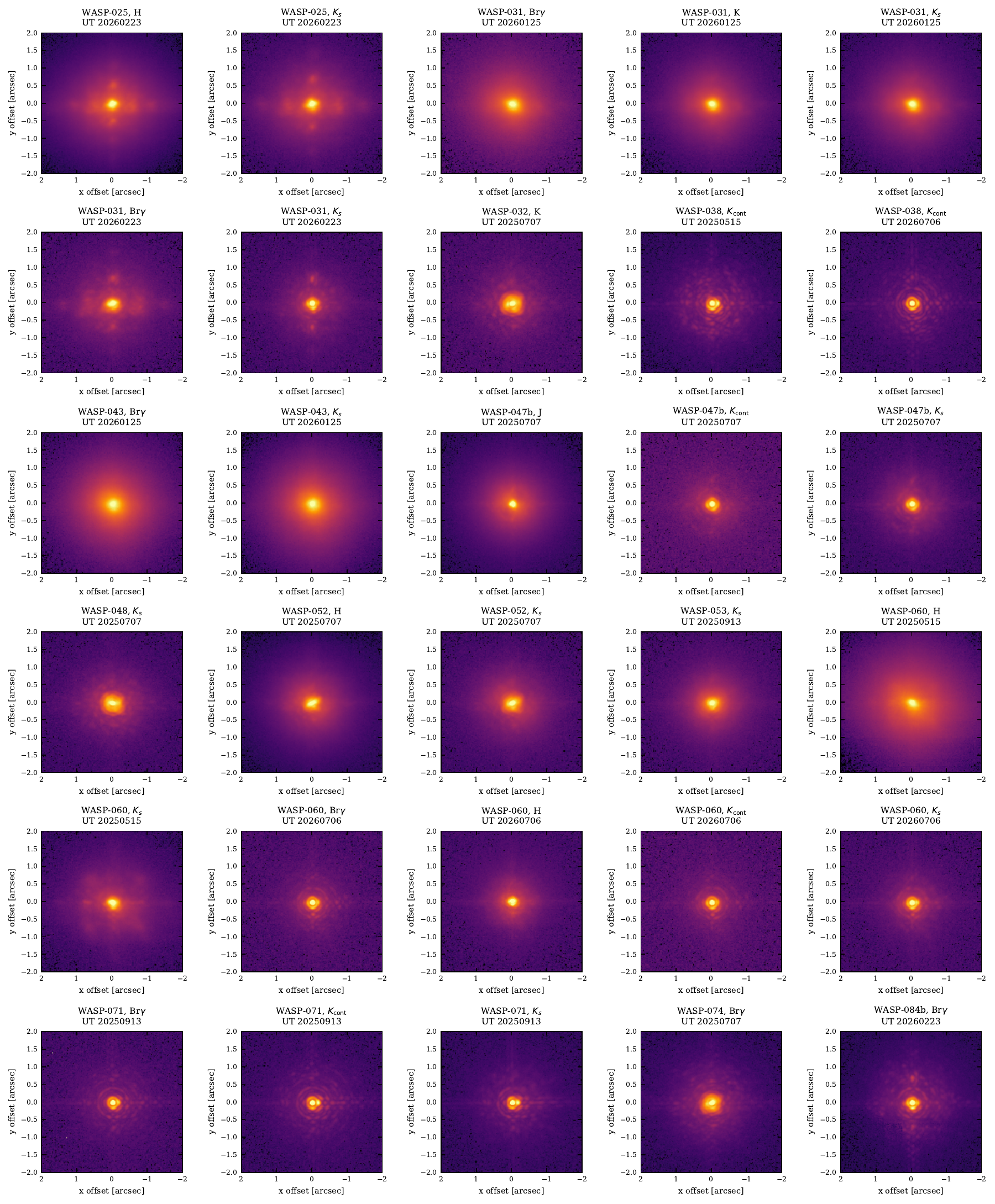}
\caption{PHARO image atlas, page 5 of 6.}
\label{fig:pharo_atlas_05}
\end{figure*}

\begin{figure*}[!p]
\centering
\includegraphics[width=0.98\textwidth]{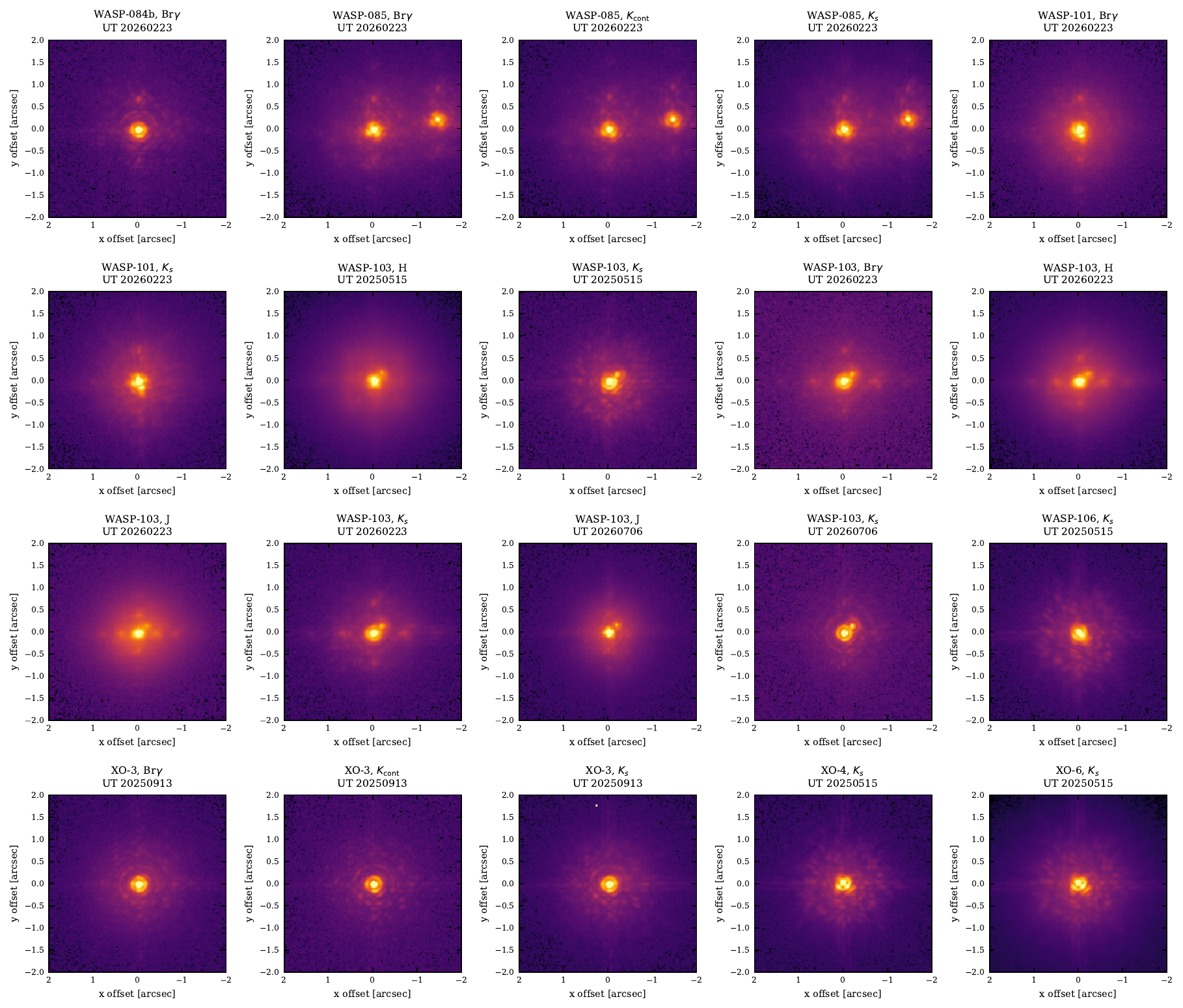}
\caption{PHARO image atlas, page 6 of 6.}
\label{fig:pharo_atlas_last}
\end{figure*}

Figures~\ref{fig:pharo_atlas_01}--\ref{fig:pharo_atlas_last} show all reduced stacked PHARO images.  Each $4\arcsec\times4\arcsec$ panel uses a fourth-root stretch from the 1st to 99.9th percentile. 

\clearpage

\bibliographystyle{aasjournal}
\bibliography{references}

\end{document}